\documentclass[a4paper,twocolumn,showpacs,superscriptaddress,floatfix,nofootinbib,prd]{revtex4-1}
\usepackage{graphicx}
\usepackage{epsf}
\usepackage{epsfig}
\usepackage{amssymb,amsmath}
\usepackage[usenames]{color}
\usepackage{amssymb}
\usepackage{float}
\usepackage{times}
\usepackage{hyperref}
\hypersetup{
  colorlinks=true,        
  linkcolor=blue,         
  citecolor=cyan,         
}
\usepackage{subfigure}
\newcommand{\cf}{cf.,~}
\newcommand{\ie}{i.e.,~}
\newcommand{\eg}{e.g.,~}
\newcommand{\ms}{\,{\rm ms}}
\newcommand{\m}{{\rm m}}
\newcommand{\km}{{\rm km}}
\newcommand{\erg}{{\rm erg}}
\newcommand{\cm}{{\rm cm}}
\newcommand{\gr}{{\rm gr}}
\newcommand{\G}{{\rm G}}

\newcommand{\Msun}{M_{\odot}}

\begin{document}
\title[GRMHD simulations of binary neutron stars]{General-relativistic
  resistive-magnetohydrodynamic simulations of binary neutron stars}
\author{Kyriaki~Dionysopoulou} 
\affiliation{School of Mathematics,
  University of Southampton, SO17 1BJ, Southampton, United Kingdom}
\affiliation{Max-Planck-Institut f\"ur Gravitationsphysik,
  Albert-Einstein-Institut, D-14476, Golm, Germany}
\author{Daniela~Alic}
\affiliation{Institut f{\"u}r Theoretische Physik,
  Max-von-Laue-Stra{\ss}e 1, 60438, Frankfurt, Germany}
\affiliation{Max-Planck-Institut f\"ur Gravitationsphysik,
Albert-Einstein-Institut, D-14476, Golm, Germany}
\author{Luciano~Rezzolla}
\affiliation{Institut f{\"u}r Theoretische Physik,
  Max-von-Laue-Stra{\ss}e 1, 60438, Frankfurt, Germany}
\date{29 October 2015}

\begin{abstract}
  We have studied the dynamics of an equal-mass magnetized
neutron-star binary within a resistive magnetohydrodynamic (RMHD)
approach in which the highly conducting stellar interior is matched to
an electrovacuum exterior. Because our analysis is aimed at assessing
the modifications introduced by resistive effects on the dynamics of
the binary after the merger and through to collapse, we have carried
out a close comparison with an equivalent simulation performed within
the traditional ideal magnetohydrodynamic approximation. We have found
that there are many similarities between the two evolutions but also
one important difference: the survival time of the hypermassive
neutron star increases in a RMHD simulation. This difference is due to
a less efficient magnetic-braking mechanism in the resistive regime,
in which matter can move across magnetic-field lines, thus reducing
the outward transport of angular momentum. Both the RMHD and the ideal
magnetohydrodynamic simulations carried here have been performed at
higher resolutions and with a different grid structure than those in
previous work of ours [L. Rezzolla, B. Giacomazzo, L. Baiotti,
J. Granot, C. Kouveliotou, and M. A. Aloy, Astrophys. J. Letters 732,
L6 (2011)], but confirm the formation of a low-density funnel with an
ordered magnetic field produced by the black hole--torus system. In
both regimes the magnetic field is predominantly toroidal in the
highly conducting torus and predominantly poloidal in the nearly
evacuated funnel. Reconnection processes or neutrino annihilation
occurring in the funnel, none of which we model, could potentially
increase the internal energy in the funnel and launch a relativistic
outflow, which, however, is not produced in these simulations.
\end{abstract}

\maketitle

\section{Introduction}

With the rapid progress made in upgrading and testing a series of
advanced interferometric gravitational-wave detectors such as
LIGO~\citep{Harry2010}, Virgo~\citep{Accadia2011_etal}, and
KAGRA~\citep{Aso:2013}, there are now great expectations that in the next
five years we will witness the first direct detection of gravitational
waves. Prime sources for such a detection are binary systems of compact
objects, namely, binary systems comprising either two black holes, a
black hole and a neutron star, or two neutron stars. The latter
configuration, in particular, is potentially a very interesting one, as
it will represent the most common source, with a realistic expected
detection rate of $\sim 40\, \mathrm
{yr}^{-1}$~\cite{Abadie:2010_etal}. A detection of gravitational waves
from binary neutron stars would yield a wealth of information about the
chirp mass, the orientation, and the localization of the binary but also
possibly the mass, spin and radius of the individual
stars~\cite{Andersson:2009yt,Weinstein2012}. In turn, this information
could set constraints on the equation of state (EOS) of the matter in
their interior. Indeed, a number of recent investigations have revealed
that it is possible to set serious constraints on the properties of the
neutron-star structure and EOS, either when using the inspiral signal
only \cite{Read2013,DelPozzo2013}, or when exploiting the rich spectral
features of the postmerger signal \cite{Bauswein2011, Bauswein2012,
  Hotokezaka2013c, Takami2014, Takami2015}.

At the same time, the merger of a binary system containing at least one
neutron star represents arguably the most attractive scenario to explain
the complex phenomenology associated with short gamma-ray bursts (SGRBs),
although many alternatives exist [see Ref.~\cite{Berger2013b} for a recent
  review]. While such a scenario was suggested already 30 years
ago~\cite{Eichler89,Narayan92}, numerical simulations (see, \eg
Refs.~\cite{Shibata99d, Baiotti08, Anderson2007, Bernuzzi2011,
  Paschalidis2014}) and new observations~\cite{Giacomazzo2012b,
  Berger2013b} have put this scenario on firmer grounds. In particular,
what these simulations have shown is that the merger of a binary system
of neutron stars inevitably leads to the formation of a metastable
object, which we dub the binary-merger product or BMP. Depending on the
total mass and mass ratio of the binary, the BMP can either be a
supramassive neutron star (SMNS), that is, a star with mass above the maximum
mass for nonrotating configurations $M_{_{\rm TOV}}$ but below the
maximum mass for uniformly rotating configurations $M_{\rm max}$, with
$M_{\rm max} \simeq (1.15-1.20)\,M_{_{\rm TOV}}$~\cite{Lasota1996}; a
hypermassive neutron star (HMNS), that is, a star more massive than a
SMNS; or a black hole\footnote{In principle, the BMP can also be a stable
  neutron star, but this would require that the stars have mass $M
  \lesssim M_{_{\rm TOV}}/2$. Since the mass distribution in neutron-star
  binaries is peaked around $1.3-1.4\,\Msun$ \cite{Belczynski2008}, this
  implies that $M_{_{\rm TOV}} \gtrsim 2.8\,\Msun$. Although this cannot
  be excluded, there is also no observational evidence that such massive
  neutron stars exist.}.

The general-relativistic hydrodynamical modelling of binary neutron
stars has seen very considerable progress over the last decade (see,
\eg Refs.~\cite{Shibata99d, Baiotti08, Anderson2007}), and it has now
reached a rather mature state. In fact, it is presently possible to
calculate inspiral waveforms having a phase accuracy comparable to that
of binary black-hole simulations thanks to the use of high-order
methods with high-order convergence rates \cite{Radice2012a,
  Radice2013b, Radice2015} or of very high-resolution and long
inspirals \cite{Kyutoku2014, Bernuzzi2015}. At the same time, the
space of parameters is also being carefully investigated, both in
terms of the variety of the EOSs considered~\cite{Hotokezaka2013c,
  Takami2014, Takami2015}, and of the treatment of radiative losses
via neutrino cooling~\cite{Sekiguchi2011, Neilsen2014}.

When magnetic fields are present, on the other hand, the bulk of work
carried out so far is considerably more limited. Investigations of the
impact that magnetic fields have on the dynamics of the binary have in
fact started with the first short-inspiral works in
Refs.~\cite{Anderson2008,Liu:2008xy}, which were complemented by the longer
simulations in Ref.~\cite{Giacomazzo:2009mp}. The latter work, together with
Ref.~\cite{Giacomazzo:2010}, also investigated the possibility that magnetic
fields could have an impact on the gravitational-wave signals emitted by
these systems during the inspiral. The conclusions reached were that, for
realistic strengths $B \lesssim 10^{12}\,\G$, the presence of
magnetic fields could not be revealed by detectors such as advanced LIGO
or advanced Virgo. The astrophysical implications of the merger of a
magnetized neutron-star binary were explored in Ref.~\cite{Rezzolla:2011},
where it was shown that instabilities in the torus orbiting the black
hole amplify the magnetic field by 3 orders of magnitude and generate
a magnetic-jet structure characterized by an ordered poloidal magnetic
field along the black-hole spin axis. The broad consistency with the
observations in terms of black-hole spin, torus mass and accretion rate,
and magnetic-field topology offered the first evidence that the merger
of magnetized neutron stars can provide the basic conditions for the
central engine of SGRBs.

Considerable effort has also been dedicated to investigating the
properties of the HMNS under more controlled conditions. For instance,
using ultrahigh spatial resolutions but axisymmetric initial data,
Ref. ~\cite{Siegel2013} has provided the first evidence from
three-dimensional global simulations that a magnetorotational instability
(MRI)~\cite{Velikhov1959, Chandrasekhar1960} is likely to develop during
the lifetime of the HMNS (see also Refs.~\cite{Duez:2006qe, Duez2006a} for
earlier work in two dimensions). In addition, again using axisymmetric
initial data and different magnetic field configurations, it has been
shown that a magnetically driven wind can be launched from the outer
layers of the HMNS as a result of its differential rotation
\cite{Kiuchi2012b,Siegel2014}. These works have also highlighted that for
realistic magnetic field topologies the wind is baryon loaded and
quasi-isotropic, with bulk velocities of $\sim 0.1\,c$
\cite{Siegel2014}. More recently, instead, the use of subgrid modeling
as an effective way to describe the turbulent dynamics that develops in
the shear layer between the two neutron stars at merger, has suggested
that amplifications of up to 5 orders of magnitude are possible
\cite{Giacomazzo:2014b}, although these amplifications are not produced
in direct simulations \cite{Giacomazzo:2010,Rezzolla:2011}, even at very
high resolution \cite{Kiuchi2014}. Finally, progress has taken place also
on the derivation of improved numerical techniques, such as those in
Ref. \cite{Etienne2012a}, where the significant advantages of a
vector-potential approach and of a Lorentz gauge were discussed.

All of the works mentioned above have been carried out within the
ideal-magnetohydrodynamic (IMHD) approximation, in which the electrical
conductivity is assumed to be infinite. Under these conditions, the
magnetic flux is conserved and the magnetic field is frozen into the
fluid, being simply advected with it. This is a very good approximation
for the stellar interior before the merger because it neglects any effect
of resistivity on the dynamics of the plasma. After the merger, however,
there will be spatial regions with very hot plasma where the electrical
conductivity is finite and the resistive effects, most notably, magnetic
reconnection, will take place.

An obvious improvement over the IMHD description is the use of the
general-relativistic resistive-magnetohydrodynamic (RMHD) equations, which provide a
complete magnetohydrodynamic (MHD) description of regions with a high conductivity, such as the
stellar interiors, and of regions with small conductivity, such as the
electrovacuum exterior. Furthermore, when the conductivity is set to
zero, it yields the Maxwell equations in vacuum, thus allowing for the
study of the magnetic-field evolution also well outside the stellar
magnetosphere~\cite{Palenzuela:2008sf}.

Partly because of the increased complexity of the equations and partly
because of the additional difficulties posed by their numerical solution
(the equations easily become stiff in regions of high conductivity), RMHD
simulations have started only rather recently. Most of the work so far
has focussed on problems in flat spacetimes \cite{Dumbser2009,
  Komissarov2007, Palenzuela:2008sf, Zenitani2010, Zanotti2011b,
  Takamoto2011b, Mizuno2013}, but general-relativistic investigations
have also been carried out on fixed spacetimes
\cite{Bucciantini2012a}. Indeed, together with the work carried out in
Refs.~\cite{Palenzuela2013a, Ponce2014}, those reported here are, to the
best of our knowledge, the only RMHD simulations of the dynamics of
binary neutron stars in general relativity. More specifically, we have
followed the inspiral, merger, and collapse to a black hole of a
neutron-star binary in which the stars have the same gravitational mass
of $M=1.625\,\Msun$ and are modelled with a simple ideal-fluid EOS.

Complementing the work reported in Refs.~\cite{Palenzuela2013a,Ponce2014},
which concentrated on the electromagnetic emission during the inspiral
and at the merger, the focus of the simulations reported here is that of
assessing the impact that resistive effects have on the dynamics of the
binary after the merger and through to collapse to a black hole. To this
scope we have carried out a close comparison with an equivalent
simulation performed for the same binary within the traditional IMHD
approximation. In this way it has been possible to determine both the
similarities between the two regimes and the novel features. The most
important of such features is the evidence the survival time of the
HMNS before collapse to a black hole increases in a RMHD simulation. This
difference is associated to a less efficient magnetic braking in the
resistive regime, in which matter is no longer perfectly advected with
the flow, but can move across magnetic-field lines. In turn, this reduces
the transport of angular momentum away from the central regions of the
HMNS, increasing its lifetime. Interestingly, a longer-lived magnetized
HMNS is of help in those models of SGRBs which invoke the existence of a
magnetarlike object produced after the merger \cite{Zhang2001,
  Metzger2008, Metzger:2011, Bucciantini2012, Rezzolla2014b,
  Ciolfi2014}. Another important result of the simulations reported here,
which have been performed at higher resolutions and with a different grid
structure than those in the previous work of Ref. \cite{Rezzolla:2011}, is
the confirmation that a magnetic-jet structure is formed in the
low-density funnel produced by the black hole--torus system. Both in RMHD
and in IMHD, the magnetic field is predominantly toroidal in the highly
conducting torus and predominantly poloidal in the nearly evacuated
funnel. Furthermore, because of the effective decoupling between the
matter and the electromagnetic fields achieved in the RMHD simulations,
the magnetic-jet structure is coherent on the largest scale of our
system. However, as in the IMHD case, also in these RMHD simulations, the
magnetic-jet structure does not lead to an ultrarelativistic
outflow. Reconnection processes or neutrino annihilation occurring in the
funnel could potentially increase the internal energy in the funnel and
launch a relativistic outflow; none of these effects is modeled in our
simulations.

The plan of the paper is as follows. In Sec. \ref{sec:binaries:approach}
we briefly review the mathematical setup of our simulations,
concentrating mostly on the discussion of the general-relativistic RMHD
equations used and on the expression of the generalized Ohm's law we have
employed. Section \ref{sec:numsetup}, on the other hand, is dedicated to
illustrate the numerical strategy employed in the solution of the
combined set of the Einstein and RMHD equations, including the properties
of our computational grid, of our matching to the low-density exterior,
and of our initial data. The core of the paper is represented by
Sec. \ref{sec:binaries:results}, where we present our results. After a
brief overview, we discuss the magnetic-field topology and magnetic-jet
structure produced in the simulations, as well as the comparison with the
IMHD case. Such a comparison goes over a number of aspects, from the
angular-momentum transfer and lifetime of the HMNS, over to the
black hole--torus system, and the electromagnetic luminosities. Finally,
Sec. \ref{sec:binaries:conclusions} contains a conclusive summary of our
results and the prospects for future research. Although this is not the
focus of this work, an illustration of the magnetic-jet structure
obtained in the IMHD simulations is presented in Appendix
\ref{sec:appendix:imhd} for completeness.

We use a spacelike signature $(-,+,+,+)$ and a system of units in which
$c=G=\Msun=1$ unless stated differently.

\section{Mathematical Setup}
\label{sec:binaries:approach}

\subsection{General-relativistic RMHD equations}
\label{sec:ms:rmhd}

Much of the numerical setup used in these simulations has been presented
in greater detail in other
papers~\cite{Giacomazzo:2007ti, Baiotti08, Baiotti:2009gk,
  Giacomazzo:2009mp, Baiotti:2010ka, Dionysopoulou:2012pp}, and for
compactness we will review here only the basic aspects, referring the
interested reader to the papers above for additional
information. However, given its importance here, we will dedicate some
space to a review of our fully general-relativistic RMHD framework, which
was first presented in Ref.~\cite{Dionysopoulou:2012pp} and represents the
extension of the special-relativistic RMHD formalism discussed
in Ref.~\cite{Palenzuela:2008sf}. A similar but independent extension has been
presented recently in Ref.~\cite{Bucciantini2012}, which describes the first
3+1 general-relativistic RMHD implementation in fixed spacetimes.

We start by presenting the augmented Maxwell equations 
\begin{eqnarray}
  \nabla_{\nu} (F^{\mu \nu} + g^{\mu \nu} \psi) &=& I^{\mu} - \kappa
  n^{\mu} \psi \,,
\label{Maxwell1} \\
 \nabla_{\nu} (^*F^{\mu \nu} + g^{\mu \nu} \phi) &=& - \kappa n^{\mu}
 \phi \,,
\label{Maxwell2} 
\end{eqnarray}
where $g^{\mu \nu}$ is the 4-metric, $F^{\mu \nu}$ is the Faraday
tensor, $^{*\!}F^{\mu \nu}$ is the Maxwell tensor, $I^{\mu}$ is the
electric four-current density, and $\phi,\,\psi$ are two auxiliary scalar
variables added to the Maxwell equations to control the constraints for
the magnetic and electric parts, respectively (see below).

After a standard 3+1 splitting of spacetime, the Maxwell and Faraday
tensors can be decomposed in terms of the electric ($E^i$) and magnetic
($B^i$) fields measured by an observer moving along the normal direction
$n^{\nu}$ (\ie normal or Eulerian observer) as
\begin{eqnarray}
F^{\mu\nu} &=& n^{\mu} E^{\nu} - n^{\nu} E^{\mu} +
\epsilon^{\mu\nu\alpha\beta} B_{\alpha} n_{\beta}\,, \\
^{*\!}F^{\mu\nu} &=& n^{\mu} B^{\nu} - n^{\nu} B^{\mu} -
\epsilon^{\mu\nu\alpha\beta} E_{\alpha} n_{\beta}\,,
\end{eqnarray}
with $\epsilon^{\mu\nu\alpha\beta} := \eta^{\mu\nu\alpha\beta} /
{\sqrt{-g}}$, $g$ the determinant of the 4-metric and
$\eta^{\mu\nu\alpha\beta}$ the Levi-Civit\'{a} symbol. The same can be done
for the electric four-current
\begin{equation}
I^{\mu} := q n^{\mu} + J^{\mu}\,,
\label{current1} 
\end{equation}
where $q$ and $J^{\mu}$ are the charge density and the electric current
density for an Eulerian observer, respectively. Using these definitions
and performing a 3+1 decomposition of Eqs.~\eqref{Maxwell1} and
\eqref{Maxwell2} with respect to the normal vector $n^{\mu}$, we arrive
at the following evolution equations,
\begin{align}
&(\partial_t - {\cal L}_{\boldsymbol{\beta}}) E^{i} - \epsilon^{ijk} \nabla_j (\alpha
B_k) + \alpha \gamma^{ij} \nabla_j \psi = \nonumber \\
& \hskip 5.0cm \alpha K E^i - \alpha J^i\,, 
\label{maxwellext_3+1_eq1a}\\
&(\partial_t - {\cal L}_{\boldsymbol{\beta}}) \psi + \alpha \nabla_i E^i   = \alpha
q -\alpha \kappa \psi\,,
\label{maxwellext_3+1_eq1b}\\
&(\partial_t - {\cal L}_{\boldsymbol{\beta}}) B^{i} + \epsilon^{ijk} \nabla_j
(\alpha E_k) + \alpha \gamma^{ij} \nabla_j \phi = \alpha K B^i\,,
\label{maxwellext_3+1_eq1c} \\ 
& (\partial_t - {\cal L}_{\boldsymbol{\beta}}) \phi
+ \alpha \nabla_i B^i = -\alpha \kappa \phi\,,
\label{maxwellext_3+1_eq1d}
\end{align}
where $\gamma_{ij}$ is the spatial 3-metric, $K := K^i_{\ i}$ is the
trace of the extrinsic curvature $K_{ij}$, $\alpha$ is the lapse and
$\boldsymbol{\beta}$ is the shift 4-vector. We recall that
$\mathcal{L}_{\boldsymbol{\beta}}$ denotes the Lie derivative along the
shift vector and that $\epsilon^{ijk}$ is related to the four-dimensional
Levi-Civit\'{a} tensor via $\epsilon^{\nu\kappa\lambda} =
\epsilon^{\mu\nu\kappa\lambda}n_{\mu}$ or alternatively $\epsilon^{ijk} =
\eta^{ijk}/\sqrt{\gamma}$, where $\gamma$ now is the determinant of the
3-metric. The scalar fields $\phi,~\psi$ measure the deviation from
the constrained solution, with $\phi$ driving the solution of
Eq.~\eqref{maxwellext_3+1_eq1d} toward the zero-divergence condition
$\nabla_i B^i=0$, and $\psi$ driving the solution of
Eq.~\eqref{maxwellext_3+1_eq1b} toward the condition $\nabla_i
E^i=q$. This driving is exponentially fast and over a time scale
$1/\kappa$. This approach, named hyperbolic divergence cleaning in the
context of IMHD, was proposed in Ref.~\cite{Dedner:2002} as a simple way of
solving the Maxwell equations and enforcing the conservation of the
divergence-free condition for the magnetic field. This method has been
extended to the resistive relativistic case
in Refs.~\cite{Komissarov2007,Palenzuela:2008sf}.

An obvious consequence of the Maxwell equations in RMHD is the
conservation law associated with the electric charge
\begin{eqnarray}
\nabla_{\mu} I^{\mu} = 0\,,
\end{eqnarray}
which provides an evolution equation for the charge density
\begin{eqnarray}\label{charge}
(\partial_t - \mathcal{L}_{\beta}) q + \nabla_i (\alpha J^i) = \alpha\, K
q\,.
\end{eqnarray}

Combining the MHD and Maxwell equations we obtain the following set of
evolution equations, which we write in a flux-conservative form as
\begin{widetext}
\begin{subequations}
\begin{align}
\begin{split}
\partial_t (\sqrt{\gamma} B^i)+\partial_k(-\beta^k \sqrt{\gamma} B^i
+ \alpha \epsilon^{ikj} \sqrt{\gamma} E_j) =
  -\sqrt{\gamma} B^k
(\partial_k \beta^i) - \alpha \sqrt{\gamma} \gamma^{ij} \partial_j
\phi \,, 
\end{split}
\label{magnetic_evol}
\\
\begin{split}
\partial_t (\sqrt{\gamma} E^i) + \partial_k(-\beta^k \sqrt{\gamma} E^i
- \alpha \epsilon^{ikj} \sqrt{\gamma} B_j) = 
  -\sqrt{\gamma} E^k
(\partial_k \beta^i) - \alpha \sqrt{\gamma} \gamma^{ij} \partial_j \psi - 
\alpha \sqrt{\gamma} J^i\,,
\end{split}
\label{electric_evol}
\\
\begin{split}
\partial_t \phi + \partial_k (-\beta^k \phi + \alpha B^k) =
  -\phi
(\partial_k \beta^k) + B^k (\partial_k \alpha) - \frac{\alpha}{2}
(\gamma^{lm} \partial_k \gamma_{lm}) B^k - \alpha \kappa \phi\,,
\end{split}
\label{psib_evol}
\\
\begin{split}
 \partial_t \psi + \partial_k (-\beta^k \psi + \alpha E^k) =
   -\psi
 (\partial_k \beta^k) + E^k (\partial_k \alpha) - \frac{\alpha}{2}
 (\gamma^{lm} \partial_k \gamma_{lm}) E^k  + \alpha q - \alpha \kappa
 \psi\,,
\end{split}
\label{eq:psie_evol}
\\
\begin{split} 
\partial_t (\sqrt{\gamma} q) + \partial_k [\sqrt{\gamma} (-\beta^k q
 + \alpha J^k)] = 0\,,
\end{split}
\label{charge_evol}
\\
\begin{split} \partial_t (\sqrt{\gamma} D) + \partial_k [\sqrt{\gamma} (-\beta^k D
 + \alpha v^k D)] = 0\,,
\end{split}
\label{rest_mass_evol}
\\
\begin{split} \partial_t (\sqrt{\gamma} \tau) + \partial_k \{\sqrt{\gamma}
 [-\beta^k \tau + \alpha ( S^k - v^k D)]\} =
   \sqrt{\gamma} (\alpha
 S^{lm} K_{lm} - S^k \partial_k \alpha)\,,
\end{split}
\label{energy_density_evol}
\\
\begin{split}
\partial_t (\sqrt{\gamma} S_i) + \partial_k [\sqrt{\gamma}
  (-\beta^k S_i + \alpha  S^k_{\ i}  )] =
    \sqrt{\gamma} \left[\frac{\alpha}{2}
  S^{lm} \partial_i \gamma_{lm}+ S_k \partial_i \beta^k 
- (\tau +
  D) \partial_i \alpha \right]\,. 
\label{momentum_evol}
\end{split}
\end{align}
\label{eq:rmhdconservative}
\end{subequations}
\noindent The fluid variables $D,~U$, $S_{i}$, and $S_{ij}$ are the
conserved rest-mass density, the conserved energy density, the conserved
momentum, and the fully spatial projection of the energy-momentum tensor,
respectively. Their explicit definitions are therefore
\begin{eqnarray}
\label{Tmunu_decomposition2}
   D &:=& \rho W \,, \\
   \tau &:=& U - D = \rho h W^2 - p + \frac{1}{2} (E^2 + B^2) - D\,, \\
   S_{i} &:=& \rho h W^2 v_{i} + \epsilon_{ijk} E^j B^k\,, \\
   S_{ij} &:=& \rho h W^2 v_{i} v_{j} + \gamma_{ij} p 
 -E_i E_j - B_i B_j + \frac{1}{2} \gamma_{ij} (E^2 + B^2)\,.
\end{eqnarray}
\end{widetext}
Here, $W=\alpha u^0 = u_i/v_i$ is the Lorentz factor, where $u^{\mu}$ are
the components of the fluid 4-velocity and $v^i$ are the components of the
3-velocity as measured by the Eulerian observer. Furthermore,
$h:=(e+p)/\rho = 1+\epsilon + p/\rho$ is the enthalpy, with $e = \rho
(1+\epsilon)$ the total energy density, $p$ the pressure, $\epsilon$ the
specific internal energy, and $\rho$ the rest-mass
density~\cite{Rezzolla_book:2013}. The important difference between the
RMHD and IMHD equations is that they involve stiff relaxation terms that
pose serious numerical limitations on the time evolution of the
equations.  For this reason, a distinct class of implicit-explicit
evolution methods has been developed, the RKIMEX
schemes~\cite{pareschi_2005_ier}, which we have presented in detail in
Ref. \cite{Dionysopoulou:2012pp}.

\subsubsection{Generalized Ohm's law}

To close the system of equations presented above, a relation for
the electric current density in terms of the other fields is necessary,
just like Ohm's law provides a prescription for the spatial conduction
current to be proportional to the electric field. A generalized Ohm's law
provides the necessary coupling of the electric current density to the
electromagnetic and matter fields. Previous work toward relativistic
versions of the generalized Ohm's law includes the investigations
in Refs.~\cite{Blackman1994, Khanna1998, Kandus2008, Andersson2012}. In one of
the simpler cases the spatial conduction current can be considered as
proportional to the electric field measured by the comoving
observer. Therefore the electric four-current density can be written as
the superposition of an advective and a conductive current~\cite{Tsamparlisbook}, which takes
the form of the generalized Ohm's law
\begin{equation}
I^{\mu} = \tilde{J}^{\mu}_{\rm adv} + \tilde{J}^{\mu}_{\rm cond} = 
\tilde{q} u^\mu + \tilde{\sigma}^{\mu\nu}e_\nu\,.
\label{eq:localcurrent}
\end{equation}
Here, $\tilde{q}:= -I^{\mu} u_{\mu} = [q + (\tilde{J}_a n^a)]/W$ is the
electric charge density measured in the rest frame comoving with the
fluid, and should be contrasted with ${q}:=-I^{\mu} n_{\mu}$, which is
instead the charge density measured by the Eulerian observer. Similarly,
$e_\nu$ and $\tilde{\sigma}_{\mu\nu}$ are, respectively, the electric field
and the electrical conductivity of the medium (which is a rank-2
symmetric tensor) as measured in the same frame. In collisional plasmas
the current in the comoving frame can be considered to be carried by the
mobile electrons, with charge $e$, and the conductivity tensor becomes
\begin{equation}
\tilde{\sigma}^{\mu\nu}=\sigma (g^{\mu\nu}+\xi^2 b^\mu b^\nu + \xi
\epsilon^{\mu\nu\alpha\beta} u_\alpha u_\beta)\,,
\label{eq:conductivity}
\end{equation}
with $b^{\mu}$ being the magnetic field in the comoving frame,
$\xi:=e\tau_e/m_e$, $\sigma:=n_ee\xi/(1+\xi^2+b^2)$, and $\tau_e$ the
electron collision time scale.

Expressing the four-current density of Eq.~\eqref{eq:localcurrent} in
terms of the fields measured by the Eulerian observer we arrive at
\begin{align}
I^\mu =& q n^\mu+q v^\mu+W\sigma[E^\mu+\epsilon^{\mu\nu\alpha} v_\nu
  B_\alpha - (v_\nu E^\nu)v^{\mu}] + \nonumber \\
\phantom{=} & W\sigma\xi^2(E^\alpha
B_\alpha)[B^\mu-\epsilon^{\mu\nu\alpha}v_\nu E_\alpha-(v_\nu
  B^\nu)v^\mu)]\,
\label{eq:ohm}
\end{align}
In deriving Eq.~\eqref{eq:localcurrent} we have made the implicit
assumption that the collision frequency between particles is much
larger than the typical oscillation frequency of the plasma, which, we
recall, is defined as $\omega_{_{\rm P}} := (4\pi n_e
e^2/m_e)^{1/2}$. This implies that electrons and ions can reach
equilibrium on very short time scales and any correction due to the
mass difference between electrons and protons can be neglected. As a
result, there is no global charge separation and the plasma is
neutral. Note also that the first term in Eq.~\eqref{eq:conductivity}
accounts for an isotropic scalar law for the current, while the rest
represent anisotropies due to the presence of a magnetic field in the
comoving frame.

Ideally, it would be desirable to have a well-defined prescription of
the conductivity tensor as a function of the fluid properties,
$\sigma_{\mu\nu} = \sigma_{\mu\nu} (\rho, \epsilon, b_{\mu})$, which
stems from the microphysical properties of the plasma
(see Ref.~\cite{Andersson2012} for a recent discussion). In practice,
however, we are far from having such a prescription in the extreme
physical conditions characterizing merging neutron stars. However, if
the collision time scale is much smaller than the electron cyclotron
period\footnote{The electron cyclotron period is defined as $P_{c,e} :
  = 2\pi/\omega_{c,e}$, where $\omega_{c,e}:= e B/m_e$ is the
  cyclotron frequency and represents the frequency at which electrons
  gyrate perpendicular to the magnetic-field lines.} or,
equivalently, if $\xi \ll |b^{\mu}b_{\mu}|^{-1}$, electrons do not
have sufficient time to gyrate perpendicular to the magnetic field
lines. Under these conditions, the isotropic part of the conductivity
is the dominant one (electrons essentially slide along the
magnetic-field lines), and expression \eqref{eq:conductivity} can be
approximated as $\sigma_{\mu\nu}\approx\sigma g_{\mu\nu}$. As a
result, the spatial three-current density in the Eulerian frame coming
from the generalized Ohm's law \eqref{eq:ohm} can be simplified so
that
\begin{eqnarray}
\label{eq:ohm_law}
J^i = q v^i + W \sigma [E^i + \epsilon^{ijk} v_j B_k - (v_k E^k) v^i]\,,
\end{eqnarray}
where we used the fact that the four-current density can be also written
as $I^{\mu}=q n^\mu + J^{\mu}$. In our simulations, the conductivity
$\sigma$ is chosen to be either a constant or a function of the rest-mass
density and a discussion will be presented in
Sec.~\ref{sec:rmhdatmo}. The last term in Eq.~\eqref{eq:ohm} represents
the Hall effects, which, however, we set to zero for simplicity.

\section{Numerical Setup}
\label{sec:numsetup}

\subsection{Field equations}
\label{sec:nr:fe}

The evolution of the spacetime (\ie of the 3-metric, extrinsic
curvature, and conformal factor) is obtained using the \texttt{McLachlan}
code, which implements the BSSNOK formulation of the Einstein
equations~\cite{Nakamura87, Shibata95, Baumgarte99} employing
three-dimensional finite-differencing operators for calculating the
fluxes on the right-hand side of the Einstein
equations~\cite{Pollney:2007ss}. The time integration is carried out
using the third-order accurate strong stability preserving
implicit-explicit Runge--Kutta scheme outlined in
Ref. \cite{Dionysopoulou:2012pp}. The time step on each grid is limited by
the Courant--Friedrichs--Lewy (CFL) condition~\cite{Rezzolla_book:2013} and
hence the Courant coefficient is set to be $0.25$ on all refinement
levels. A Kreiss--Oliger type dissipation is added to the spacetime
evolution equations to ensure that any high-frequency noise produced
during the evolution (mostly at the refinement levels boundaries) is
damped. The damping parameter in the evolution equation for the shift was
carefully chosen to be $0.71$ so that the additional term does not become
stiff on the coarser grids.

\subsection{GR-RMHD equations}
\label{sec:nr:rmhd}

The general-relativistic RMHD (GR-RMHD) equations are solved with the
\texttt{WhiskyRMHD} code \cite{Dionysopoulou:2012pp}, which uses
high-resolution shock-capturing schemes~\cite{Toro99,
  Rezzolla_book:2013}. In particular, the reconstruction of the conserved
variables is achieved via the Piecewise Parabolic Method~\cite{Colella84}, while the fluxes are calculated through the
approximate Riemann solver introduced by Harten--Lax--van Leer--Einfeldt~\cite{Harten83} and which requires only knowledge about the
maximum characteristic speeds of the system, \ie the speed of light in
this case.

The system of RMHD equations is closed by describing the fluid as ideal
and with a $\Gamma$-law EOS
\begin{equation}\label{eq:gamma-law}
  p=\rho\epsilon~(\Gamma-1)\,,
\end{equation}
where $\Gamma=2$ and $K=123.6$. This EOS is clearly an idealization. Much
more sophisticated EOSs have been implemented in our numerical
infrastructure~\cite{Galeazzi2013,Takami:2014,Takami2015} and have been
used by other groups in general-relativistic magnetohydrodynamic (GRMHD) simulations~\cite{Neilsen2014}. However,
this idealization is probably adequate at this stage as here we are
mostly interested in assessing the differences introduced in the dynamics
of the BMP by resistive effects. These effects
suffer from even larger uncertainties than those associated to the
different EOSs.

\subsection{Adaptive mesh refinement and symmetries}

Our code makes use of the \texttt{Cactus}~\cite{cactus_url} computational
framework, which allows us to employ a box-in-box vertex-centered
adaptive mesh refinement grid hierarchy that tracks the ``center of
mass'' of the stars as they orbit each other. This was achieved via the
\texttt{BNSTracker} thorn implemented by W. Kastaun and the
\texttt{Carpet} driver~\cite{Schnetter-etal-03b}. The numerical domain
consists of six levels of refinement with the resolution doubling between
adjacent refinements. In addition, we employ moving refinement boxes in
order to track the high-density regions. The outer boundary of the
computational domain is located at $\approx 378\,\km$. The finest
resolution during the inspiral and merger is $\Delta x\approx 296~\m$,
but an extra refinement level is activated right after the merger with a
resolution of $\Delta x\approx 148~\m$. It is important to remark that,
although in a small region of spatial extent $\sim 13.5\,\km$, this
resolution is considerably higher than the one used in
Ref.~\cite{Rezzolla:2011} (\ie $\approx 221\,\m$ with a spatial extent of
$\sim 35.4\,\km$), where the first evidence was given that the merger of
a binary system of magnetized neutron stars can lead to the formation of
a magnetic-jet structure. Hence, although in a resistive framework, the
simulations reported here can be considered as a ``higher-resolution''
counterpart of those presented in Ref.~\cite{Rezzolla:2011}. We also note that
when considering initial data consisting of a binary in quasicircular
orbits (see Sec. \ref{sec:bns:lifetime}) we do not activate the extra
refinement level after the merger, hence keeping a resolution of $\Delta
x\approx 296~\m$ to describe the HMNS. We do this so as to have two
different resolutions to describe the HMNS and hence to study the
numerical consistency of the solution.

To reduce the computational cost associated with the numerical evolution of an equal-mass binary, we use a reflection-symmetry condition across the $z=0$ plane and a $\pi$-symmetry condition across the $x=0$ plane. These conditions respect the symmetries of the scenario we are investigating. The outer
boundary conditions are set by using simple zeroth-order extrapolation
of the hydrodynamic variables. For the electromagnetic and spacetime
variables, we employ simple Sommerfeld radiative boundary conditions
because of the nature of the fields. Given the long time scale over
which our simulations are carried out, reflections due to
imperfections of the boundary conditions experience several domain
crossings before the end of the simulation. We plan to improve on this
in the future by the application of maximally dissipative boundary
conditions that minimize the effects of reflections.

\subsubsection{Exterior matching and atmosphere handling}
\label{sec:rmhdatmo}

Our goal with the use of a RMHD framework is that of modelling the
exterior of the neutron stars as an electrovacuum, where both the
conductivity and the charge density are negligibly small, so that
electromagnetic fields should obey in these regions the Maxwell equations
in vacuum. On the other hand, we want to model the interior of the stars
as highly conducting, so that our equations recover the IMHD limit in
such regions. There are several different ways to achieve this; see, for
example, Refs.~\cite{Lehner2011, Dionysopoulou:2012pp, Palenzuela2013,
  Paschalidis2013c}. Each of them has, in our view, its advantages and
disadvantages. However, because they all try to model the difficult
transition region between two regimes that are intrinsically different,
they all represent an approximation. This difficulty is not specific to
this problem but is a typical feature of physical problems as, for
example, in the transition from an optically thick to an optically thin
regime in radiative-transfer calculations. Our approach is also an
approximation and is similar to the one in Refs.~\cite{Dionysopoulou:2012pp,
  Palenzuela2013} in the sense that the matching from the stellar
interior to the stellar exterior is achieved through a carefully chosen
conductivity profile. More specifically, the conductivity profile adopted
is directly related to the conserved rest-mass density (and hence to the
rest-mass density) and given by
\begin{align}
\sigma:=\sigma_0\max\left[1-\frac{2}{1+\exp\left[2 D_{\rm
        tol}{(D-D_{\rm rel})}/{D_{\rm atm}}\right]},\,0\right]\,,
\label{eq:bns:conductivityprofile}
\end{align}
where $\sigma_0 \gg 1$ corresponds to a uniform scalar conductivity. The
parameters $D_{\rm tol}$, and $D_{\rm rel}$ determine how sharp the
transition to the exterior is. For the simulations reported here, we have
chosen\footnote{We recall that in units in which the speed of light is
  not set to 1, $\tau_d := L^2 \sigma_0/c^2$ represents the Ohmic
  diffusion time scale, with $L$ the typical length scale of the field. For
  $L\sim 10^6{\rm cm}$, $\tau_d \sim 10^2\,\sec$, which is obviously much
  larger than the timescale over which our simulations are carried out,
  \ie $\sim 10^{-2}\,\sec$.}, $\sigma_0 = 10^6 =
2.0\times10^{11}\,\sec^{-1}$, $D_{\rm tol}=0.01$ and $D_{\rm rel}=
100\,D_{\rm atm} = 100\,\rho_{\rm atm}$, where $D_{\rm atm}$ and
$\rho_{\rm atm}$ are the values of the conserved and primitive rest-mass
density in the ``atmosphere.''

We recall that, as in other Eulerian hydrodynamics and MHD codes, also
our \texttt{WhiskyRMHD} code makes use of a very low rest-mass density
fluid to handle the evolution of the MHD equations in regions which
are associated to the exterior of the stars. In such a region, we
follow the same prescription initially implemented in
Ref.~\cite{Baiotti04} and then adopted in essentially all of the
simulations performed with our code, even in its IMHD
incarnation~\cite{Giacomazzo:2007ti}. In essence, we treat as
atmosphere any region in the computational domain which is below
$\rho_{\rm atm}$, which we take here to be $6.17\times 10^{6}\, {\rm
  g\,cm}^{-3}$, that is approximately 8 orders of magnitude smaller
than the maximum rest-mass density.\footnote{In practice,
  to avoid being sensitive to the threshold value, we set to
  atmosphere any cell of which the rest-mass density is below $\rho_{\rm atm}
  + \rho_{\rm tol}$, where $\rho_{\rm tol} \sim 10^{-2}\,\rho_{\rm
    atm}$~\cite{Baiotti08}.} In such a region, we set the fluid
3-velocity to zero, the rest-mass density to a floor value, and
the specific internal energy to the value it assumes for a fluid
following a polytropic EOS~\cite{Rezzolla_book:2013}.\footnote{A
  different prescription is used for the specific internal energy in
  the case of hot, nuclear-physics EOSs (see Ref.~\cite{Galeazzi2013} for
  details).} In addition, and differently from the IMHD
implementation of Ref.~\cite{Giacomazzo:2010}, the use of the
prescription \eqref{eq:bns:conductivityprofile} automatically sets the
conductivity in the atmosphere to zero, so that the Maxwell equations
reduce there to the Maxwell equations in vacuum. Hence, in our
prescription the atmosphere is de facto a cold, static, uniform fluid
in which electromagnetic waves propagate as if in vacuum.

A few remarks should be made about modelling the neutron stars' exterior,
which in reality is expected to be a highly conducting, low-density,
possibly magnetically dominated plasma in very strong magnetic
fields. 

First, with our prescription for the atmosphere and with the conductivity
\eqref{eq:bns:conductivityprofile}, the latter is zero also in regions
that are not at the atmosphere level but close to it, \ie at $D \leq
D_{\rm rel}$. We do this because the strong winds that are produced at
the merger and later on rapidly fill the computational domain. A direct
consequence of this is that the atmosphere is present only close to the
boundaries and therefore an IMHD prescription would be met essentially
everywhere. Yet, we are interested in deviations from perfect-flux
freezing and in particular in regions where plasma is tenuous and the
magnetic fields essentially decouple from the matter. We can effectively
achieve this by setting $\sigma=0$ in any region that is below $D_{\rm
  rel}$. This approach therefore accomplishes our goal of decoupling in
the low-density regions the evolution of the electromagnetic fields from
the dynamics of the plasma. Of course, an electrovacuum prescription where
the rest-mass density is nonzero is, strictly speaking, inconsistent, but
we believe that this is a tolerable inconsistency, given that the
rest-mass density in these regions takes essentially the smallest values
in the whole domain.

Second, albeit somewhat arbitrary, our approach suffers from the same
uncertainties of other approaches suggested in the literature to match
the two different regimes (see, \eg Refs.~\cite{Palenzuela2013,
  Paschalidis2013, Ponce2014}). Ideally, its robustness can be validated
by varying the free parameters $D_{\rm rel}$ and $\rho_{\rm atm}$,
although this is something that admittedly we are not able to do here
because of the computational costs involved. More importantly, we find
that this approach allows us to take an important step beyond the
previous IMHD treatment presented in Ref.~\cite{Giacomazzo2011b}.

Third, while matching to a force-free regime is appealing, it can be
rather dangerous in the physical conditions encountered \emph{after} the
merger (by contrast, the force-free approximation is probably very good
before the merger or in black hole--neutron-star binaries, when only
magnetospheric effects are expected to take place). Our calculations
reveal in fact that the exterior of the conducting matter (either the
HMNS or the torus) is always matter dominated and the plasma beta
parameter, \ie the ratio of the gas-to-magnetic pressure,\footnote{We
  note that our definition of $\beta_{_{\rm P}}$ is the one normally
  used in plasma physics, but the inverse of the one employed in other
  numerical-relativity calculations, \eg Ref.~\cite{Paschalidis2014}.}
$\beta_{_{\rm P}} := 2 p/B^2$, is at least $10^{4}$ and of the order of
$10^{6}$ in the polar regions. This is far from the condition of
$\beta_{_{\rm P}} \ll 1$, where the matter inertia can be neglected and
the force-free approximation is a good one.

Finally, despite the fact that our implementation offers a control over
the amount of resistivity in different parts of the flow, the choice of
realistic values for the resistivity is far from trivial (see, \eg
Ref.~\cite{Uzdensky2011}) and not addressed at all in these simulations.

\subsubsection{Miscellanea}

As already mentioned in Sec.~\ref{sec:ms:rmhd}, in order to ensure that
the magnetic field is essentially divergence free, we have employed the
divergence cleaning method of Ref.~\cite{Dedner:2002}, which, however, requires
choosing a suitable value for the constant $\kappa$ [\cf
  Eqs.~\eqref{maxwellext_3+1_eq1b} and \eqref{maxwellext_3+1_eq1d}]. In
the simulations presented here, we have set $\kappa = 0.075$. This value
does not lead to stiffness problems in the coarser grids and at the same
time provides a rapid damping of the constraint violation on a time scale
$\tau \sim (B^i B_i)^{1/2}/\nabla_i B^i$. 

When after the merger the HMNS collapses to a black hole, steep gradients
appear in the rest-mass density and cover just few grid points of
the finest grid. If the resolution is sufficiently low, these gradients
are simply dissipated numerically and the evolution can proceed without
problems. However, for the rather high resolutions used here for the
finest grid, \ie $\approx 148\,\m$, the numerical dissipation is smaller
and the gradients are not removed, leading to failures in the
conversion of the conserved variables to the primitive ones. To counter
this problem we reset to atmosphere the hydrodynamical variables in a
mask inside which the lapse function goes below a threshold, \eg $\alpha
< \alpha_{\rm thr}=0.1$. This reset is done only for
the hydrodynamical variables, while the spacetime and electromagnetic
ones are evolved as usual. Furthermore, the reset is applied in practice
to a handful of cells, well inside the apparent horizon and thus not
influencing the matter dynamics.

\subsection{Initial data}
\label{sec:id}

The initial data consist of a magnetized binary neutron-star system of
total Arnowitt--Deser--Misner (ADM) mass $M_{_{\rm ADM}} = 3.25\,M_{\odot}$
and an initial orbital separation of $45\,\km$. Each star has a baryon
mass equal to $1.625\,M_{\odot}$ and an equatorial radius of $R_{\rm eq}
= 13.68\,{\rm km}$, so that the initial separation corresponds to
approximately $3.3\,R_{\rm eq}$. The initial orbital velocity is
$\Omega_0=1.85\,{\rm rad}\,\ms^{-1}$, and the maximum rest-mass density is
$5.91~\times10^{14}\,\gr\,\cm^{-3}$. Lacking self-consistent initial data
for magnetized binaries, our initial data are generated by the
\texttt{LORENE} library as an unmagnetized irrotational binary in
equilibrium on a quasicircular
orbit~\cite{Gourgoulhon-etal-2000:2ns-initial-data, lorene41}.  The
magnetic field is then superimposed on the unmagnetized
constraint-satisfying solution. Following Refs.~\cite{Giacomazzo:2009mp,
  Giacomazzo:2010}, the initial magnetic field is fully contained inside
the stars and purely poloidal. This is achieved after prescribing the
toroidal vector potential $A_\phi$ to have the form
\begin{equation}
A_\phi = A_b \left[ \max(P - P_{\mathrm{cut}},0) \right]^2\,,
\label{eq:rmhdA_b}
\end{equation}
where $P_{\rm cut}=0.04~P_{\rm max}$ determines the point at which the
magnetic field goes to zero (typically before it reaches the
surface). The resulting (poloidal) magnetic field is just the curl of the
vector potential and leads to a maximum magnetic-field strength of
$1.97\times 10^{12}\,{\rm G}$ at the pressure maximum. Because the
initial data constructed in this way are not a solution of the full
Einstein--Euler--Maxwell system, they will introduce an increased violation of
the constraint equations, which amount to $\sim 2.5\times
10^{-6}/M^2_{_{\rm ADM}}$ and $\sim 1.7~\times 10^{-6}/M^2_{_{\rm ADM}}$
in the $L_2$-norm, of the Hamiltonian and momentum constraints,
respectively. These values should be compared with those before the
introduction of the magnetic field and which are about 1 order of
magnitude smaller, \ie $5\times10^{-7}/M^2_{_{\rm ADM}}$ and $1.5\times
10^{-7}/M^2_{_{\rm ADM}}$.

The perturbations introduced by the addition of the magnetic fields are
small enough so as not to have a significant effect on the dynamics of
the binary. Indeed, our experience is that the $L_2$-norm of the
constraints relaxes to values comparable to those of simulations of
unmagnetized binaries after about one crossing time or, equivalently, one
orbit. Obvious ways to improve this approach, and which could become
important for an accurate modelling of the gravitational-wave emission
during the inspiral, exist. Among them are the use of consistent initial
data for magnetized binaries or the simulation of binaries with much
larger separations, so that the system has several orbits to reach a more
consistent MHD equilibrium.

In addition, because the main focus of this work is the assessment of
resistive effects on the postmerger dynamics and hence a comparison
between IMHD and RMHD simulations, our interest in an accurate treatment
of the initial data is rather limited here. As a result, and once again
to reduce the computational costs, we accelerate the inspiral as first
suggested in  Ref.~\cite{Kastaun2013}. More specifically, for most of our
simulations we modify the initial linear momenta by adding an initial
inward radial velocity which is $\sim 20\%$ of the orbital velocity. This
reduces the number of orbits at this separation from $\sim 3.5$ to only
$\sim 1.5$. Of course, the constraint violations introduced in this way
are even larger than those discussed above with the introduction of
magnetic fields\footnote{For completeness, we can compare the violations
  in the $L_2$-norm of the Hamiltonian and momentum constraints for
  binaries with reduced initial momenta with the corresponding violations
  for binaries on quasicircular orbits. The latter amount to $\sim
  2\times10^{-7}/M^2_{_{\rm ADM}}$ for the Hamiltonian constraint and to
  $\sim 4\times10^{-8}/M^2_{_{\rm ADM}}$ for the average of the momentum
  constraints.}, but we have also verified that this does not introduce
qualitative differences by comparing the results of these grazing
collisions with those obtained from the corresponding binaries in
quasicircular orbits.

\section{Results}
\label{sec:binaries:results}

The dynamics of the same neutron-star binary considered here has been
previously investigated in hydrodynamic simulations \cite{Baiotti08}, in
IMHD simulations \cite{Giacomazzo:2010}, and, more recently, also in RMHD
simulations \cite{Ponce2014}. In all cases, it was shown that after the
merger this specific model forms a rapidly rotating HMNS with a high
degree of differential rotation. The transient object collapses later to
form a rotating black hole of mass $\sim 2.9-3.0\,M_{\odot}$. The
magnetic-field strengths chosen here and in previous works
\cite{Giacomazzo:2010, Ponce2014} are not sufficiently high to affect the
bulk dynamics of the binary during the inspiral~\cite{Giacomazzo:2009mp},
which can be considered to be equivalent to the purely hydrodynamical
case for all practical purposes.

In the following subsections, we focus on the results obtained from the
application of our RMHD implementation. We start by providing a general
description of the basic features of the RMHD dynamics
(Sec.~\ref{sec:bns:features}) and then move to make a detailed comparison
with an IMHD simulation of the same neutron-star binary
(Sec.~\ref{sec:bns:comparison}). We note that our focus here is different
from the one in Ref. \cite{Ponce2014}, which matched the resistive
description to a force-free one to study the interaction of the two
stellar magnetospheres before the merger. Here, on the other hand, we are
mostly interested in the postmerger object and on the effects that
resistivity has on the dynamics of the HMNS and subsequent
black hole--torus system. Because of this, and because the scenario we
are investigating is polluted by the large baryonic winds produced after
merger, our resistive matching is made to an electrovacuum exterior. In
this sense, the work carried out here and in Ref. \cite{Ponce2014} provide a
complementary description of the dynamics of magnetized binary neutron
stars in full general relativity.

\begin{figure*}
\centering
\includegraphics[width=0.32\textwidth]{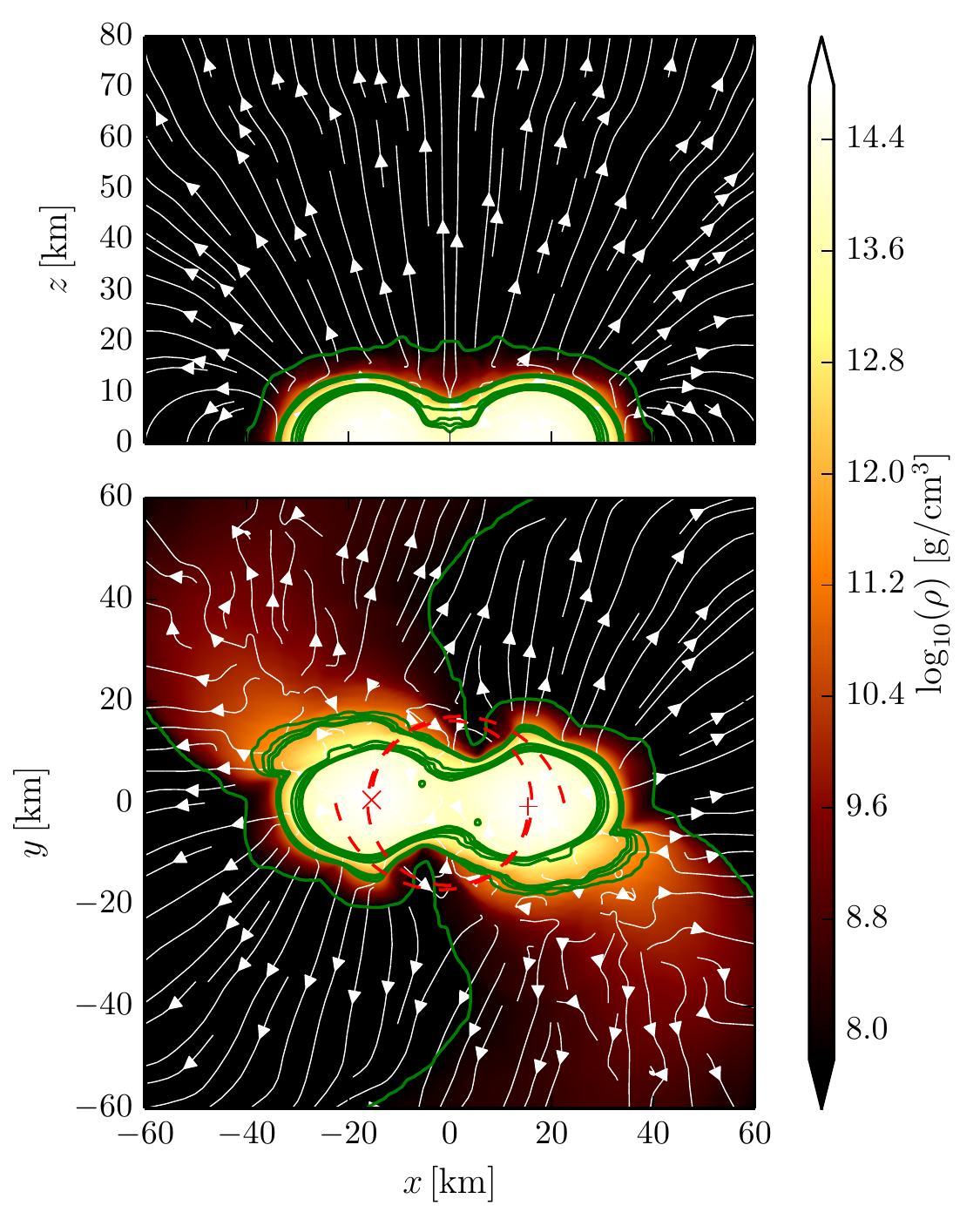}
\includegraphics[width=0.32\textwidth]{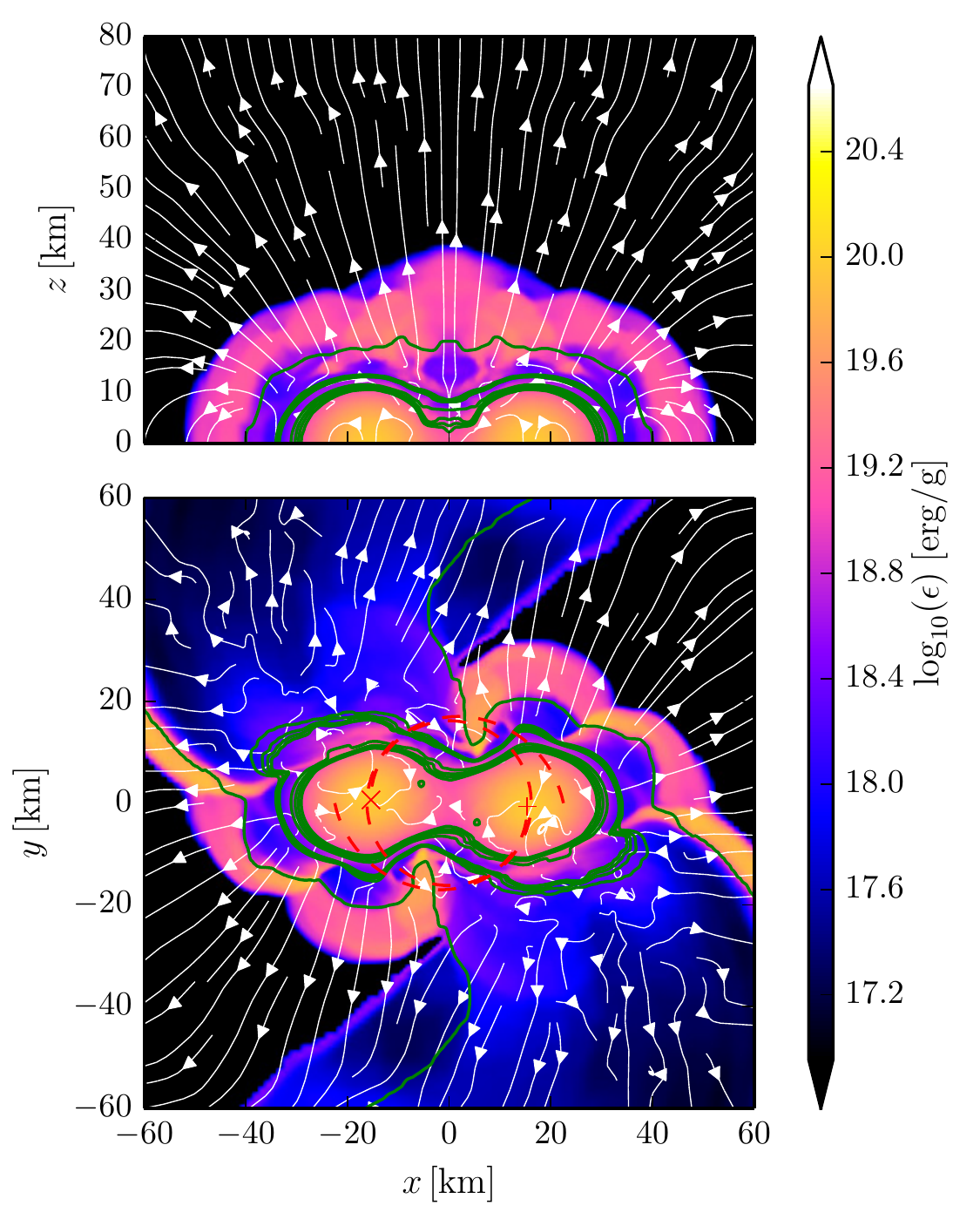}
\includegraphics[width=0.32\textwidth]{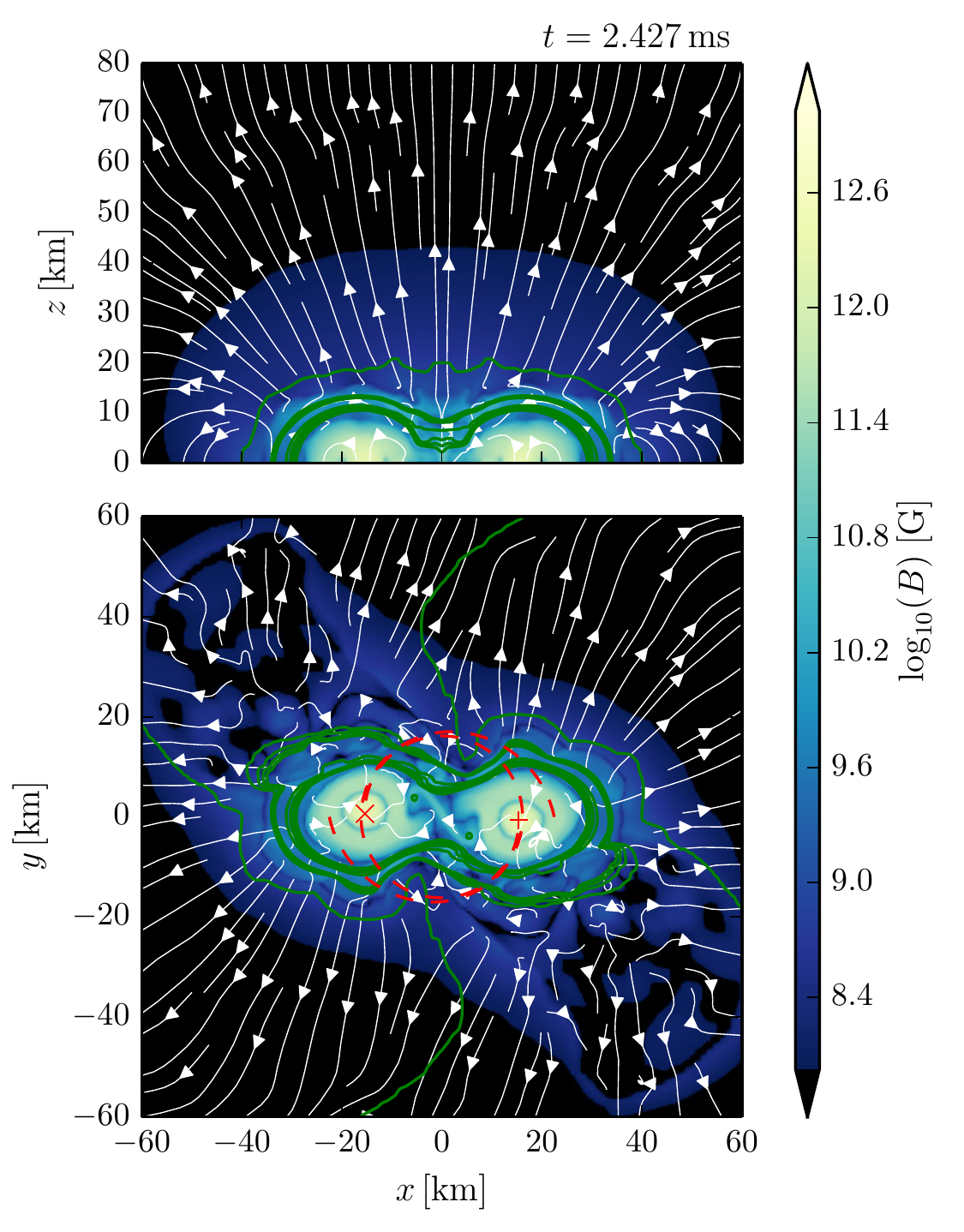}
\includegraphics[width=0.32\textwidth]{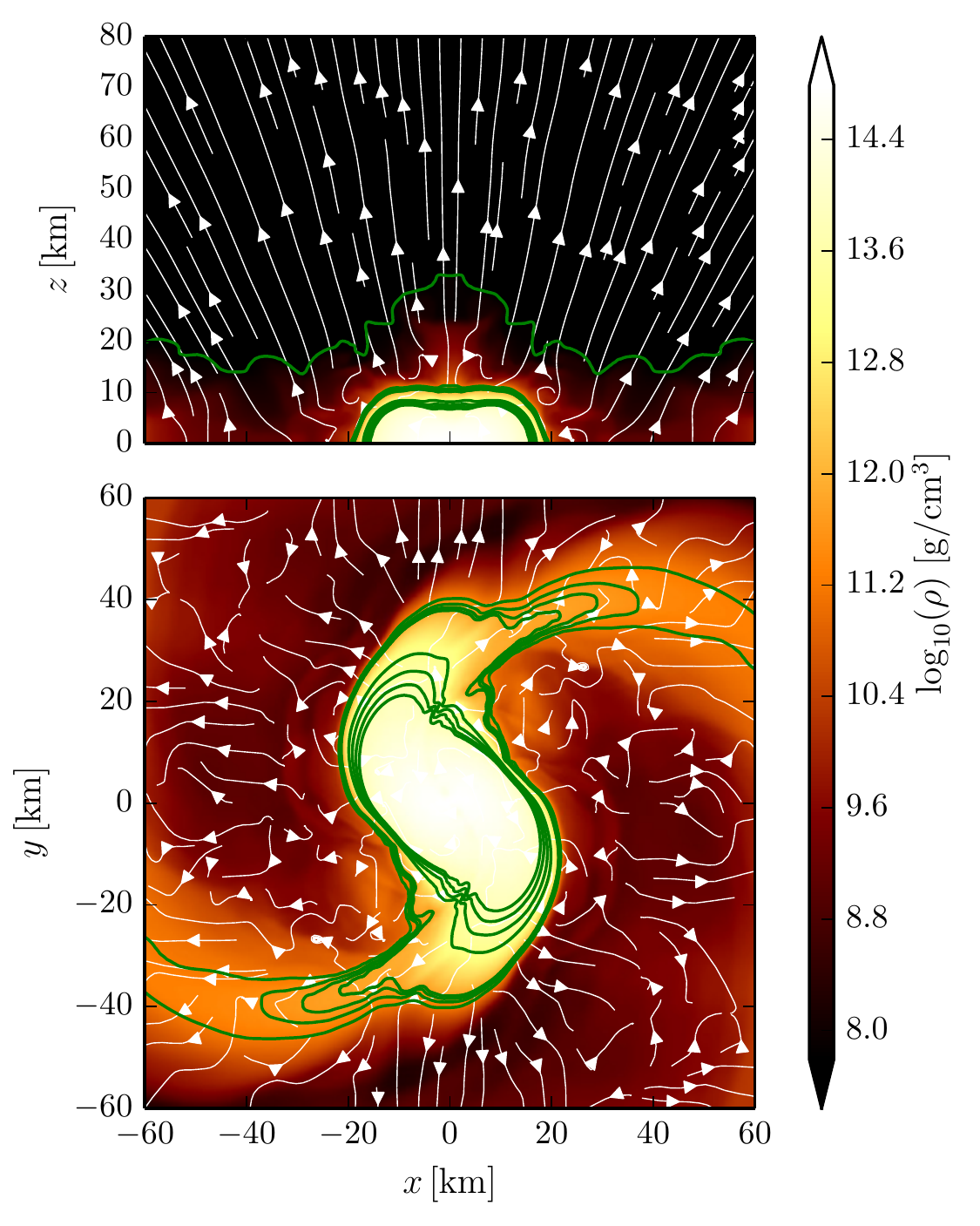}
\includegraphics[width=0.32\textwidth]{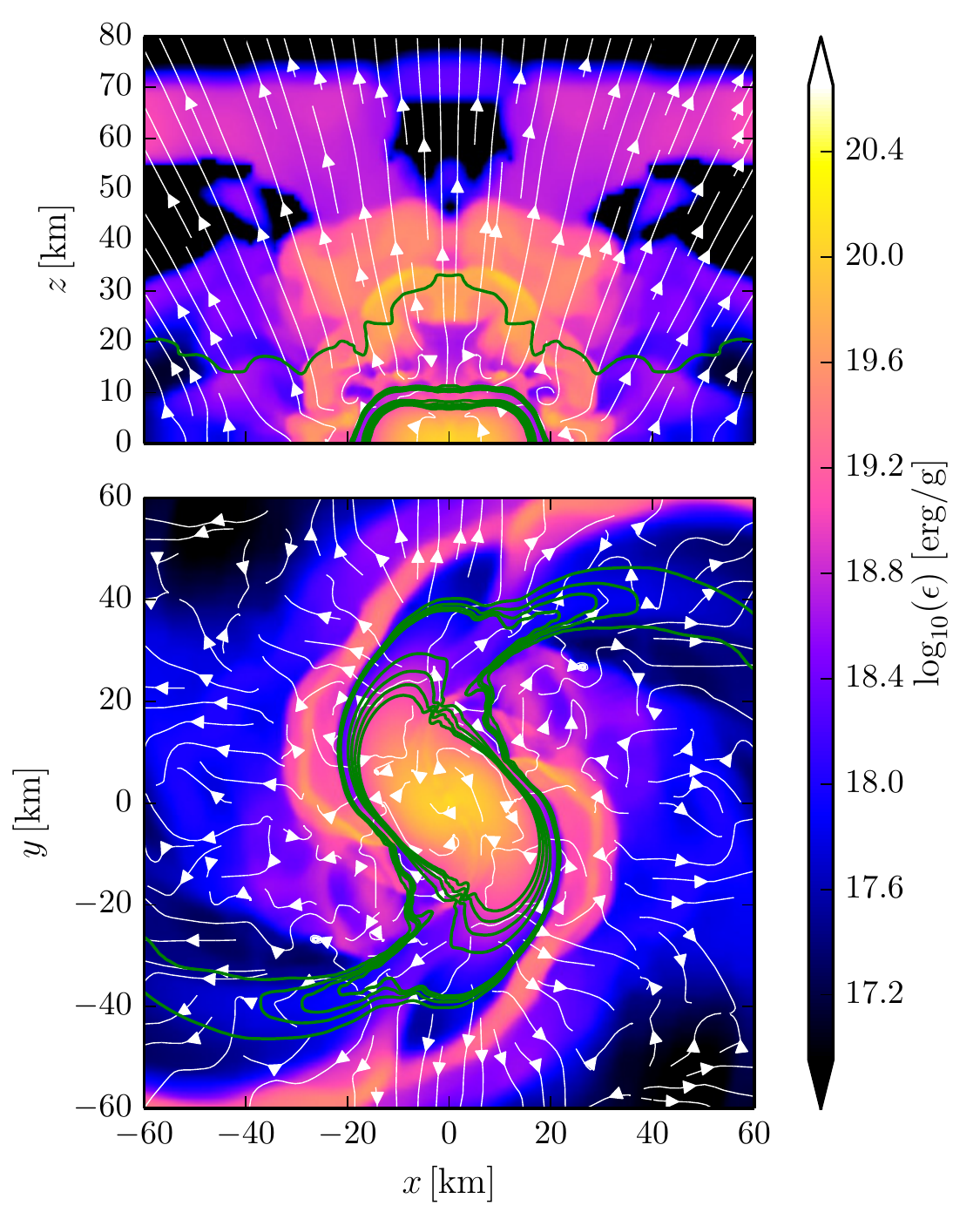}
\includegraphics[width=0.32\textwidth]{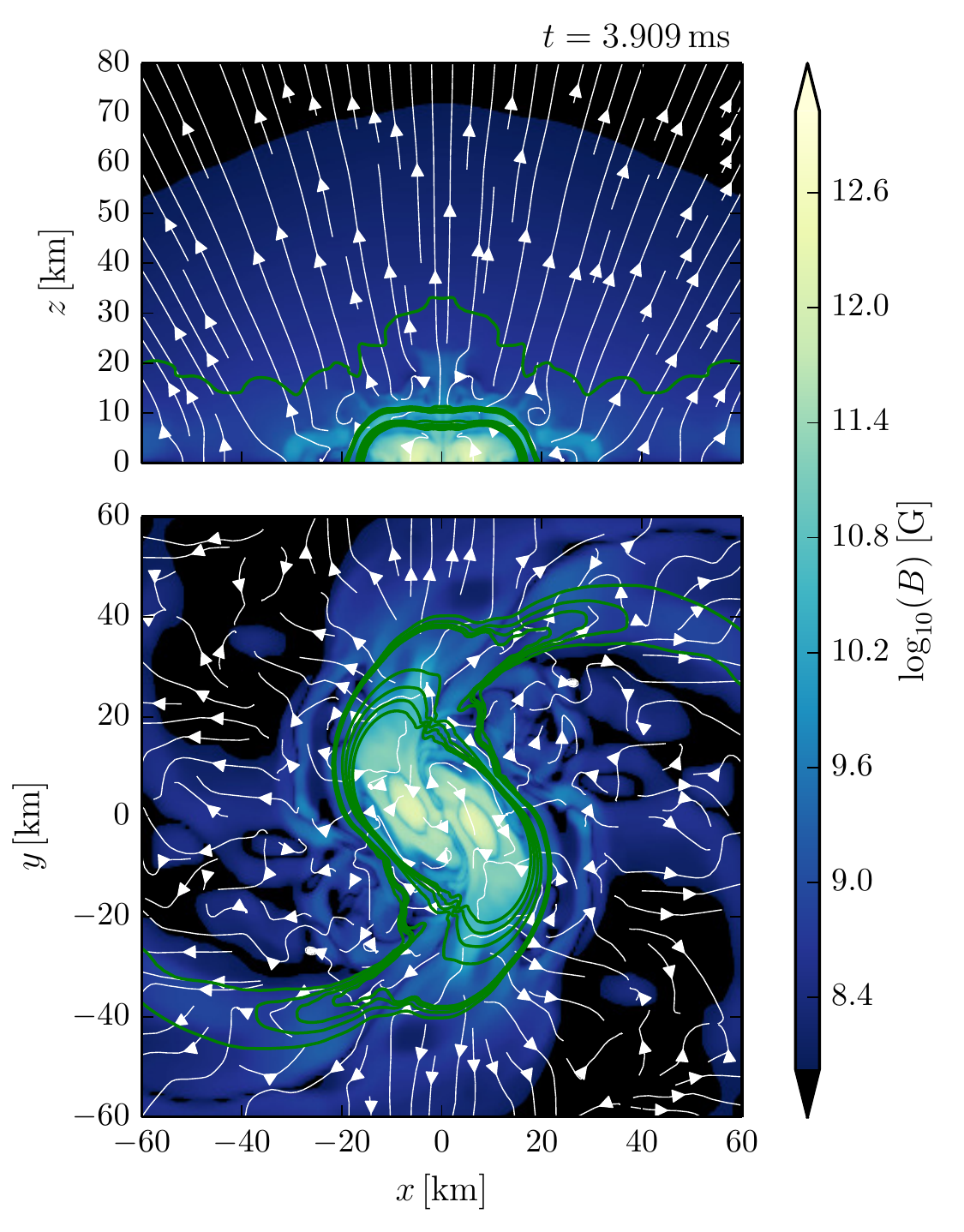} 
\includegraphics[width=0.32\textwidth]{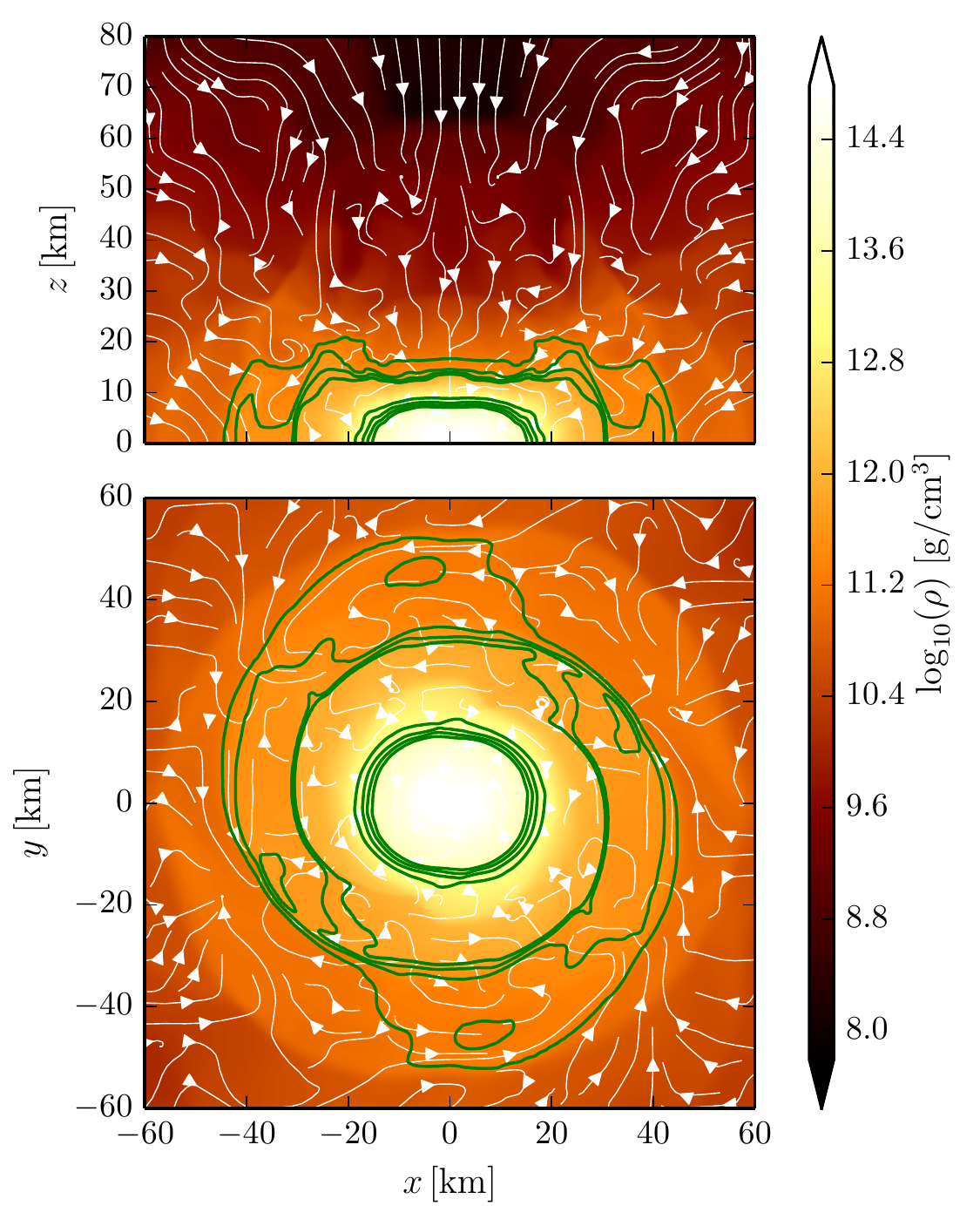}
\includegraphics[width=0.32\textwidth]{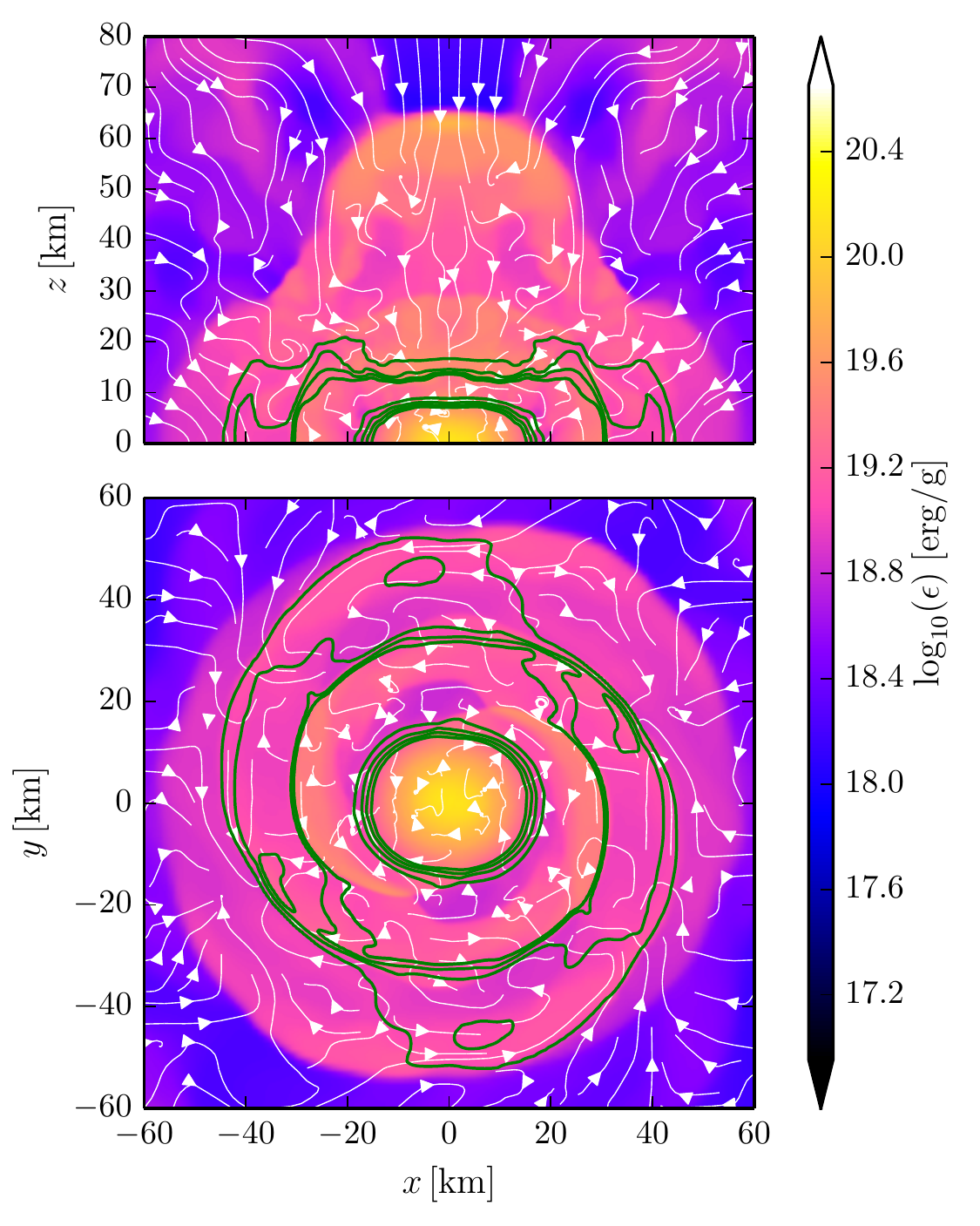}
\includegraphics[width=0.32\textwidth]{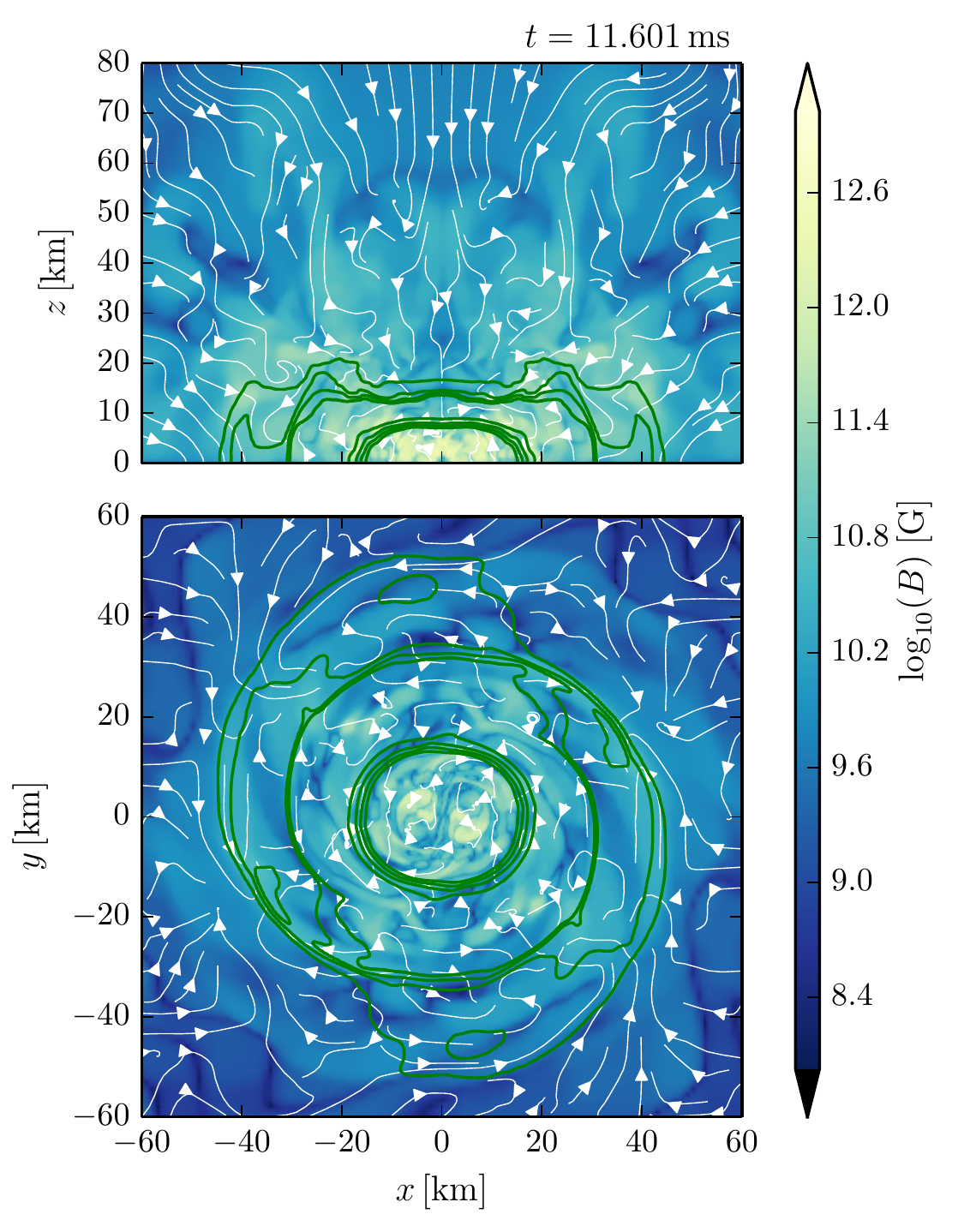} 
\caption{Snapshots of two-dimensional cuts in the $(x,y)$ and $(x,z)$
  planes of the rest-mass density $\rho$ (left panel), of the specific
  internal energy $\epsilon$ (middle panel), and of the modulus of the
  magnetic field $|B|:=(BiB_i)^{1/2}$ (right panel). From the top, the
  snapshots correspond to times $t=2.43\,\ms$ {(top row)}, $t=3.91\,\ms$
  {(middle row)}, and $t=11.60\,\ms$ {(bottom row)}. Shown with white
  lines are the projection of the magnetic-field lines on the different
  planes, while marked with green solid lines are the isocontours of the
  rest-mass density at $\rho=\{6.2\times 10^7,~1.2\times
  10^{11},~2.5\times 10^{11},~3.7\times 10^{11},~4.9\times
  10^{11},~6.2\times 10^{11},~1.3\times 10^{13},~2.5\times
  10^{13},~3.7\times 10^{13},~4.9\times 10^{13},\allowbreak~6.2\times
  10^{13}\}~\gr\,\cm^{-3}$. The positions of the two stars are marked
  with $\times$ and $+$ symbols, and the stars' trajectories are marked with red
  dashed lines. The different panels represent the evolution of the HMNS
  and highlight that no ordered magnetic-field topology emerges; this will
  change when the HMNS collapses to a black hole (see also
  Figs. \ref{fig:bns:rhoB6}--\ref{fig:bns:magtorpol}).}
\label{fig:bns:rhoB}
\end{figure*}
\begin{figure*}[!ht]
\center
\includegraphics[width=0.32\textwidth]{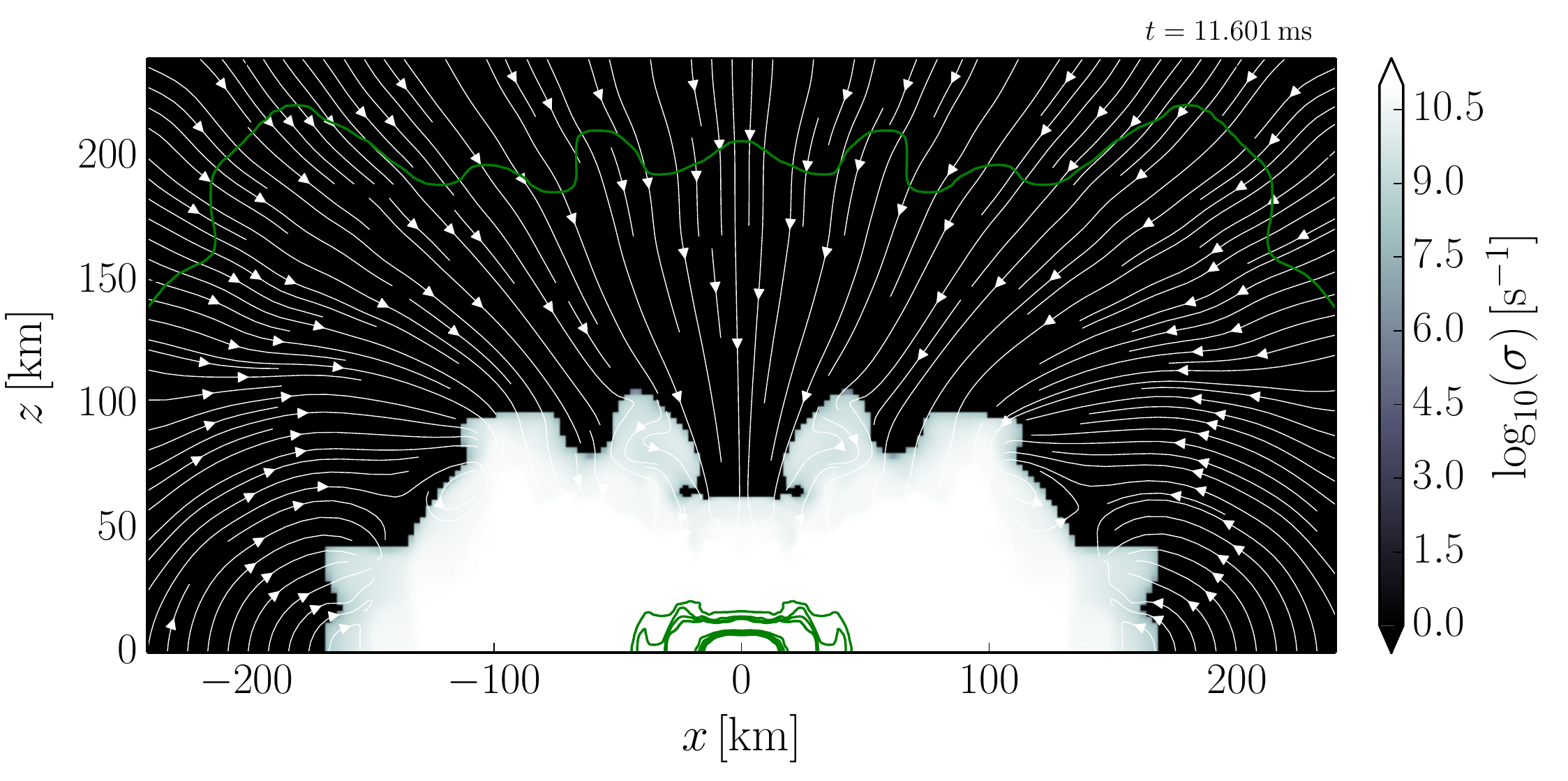}
\hskip 0.2cm
\includegraphics[width=0.32\textwidth]{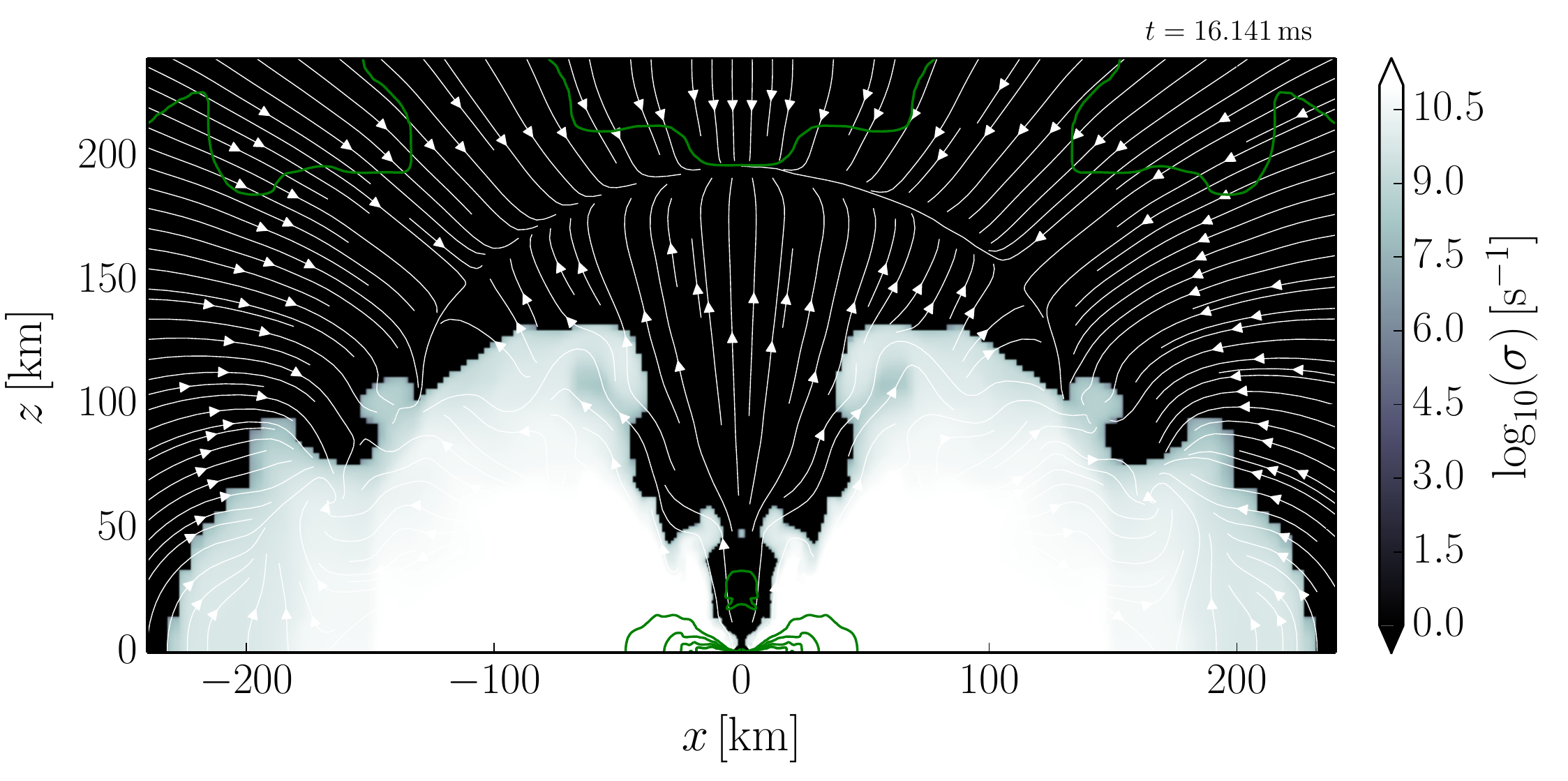}
\hskip 0.2cm
\includegraphics[width=0.32\textwidth]{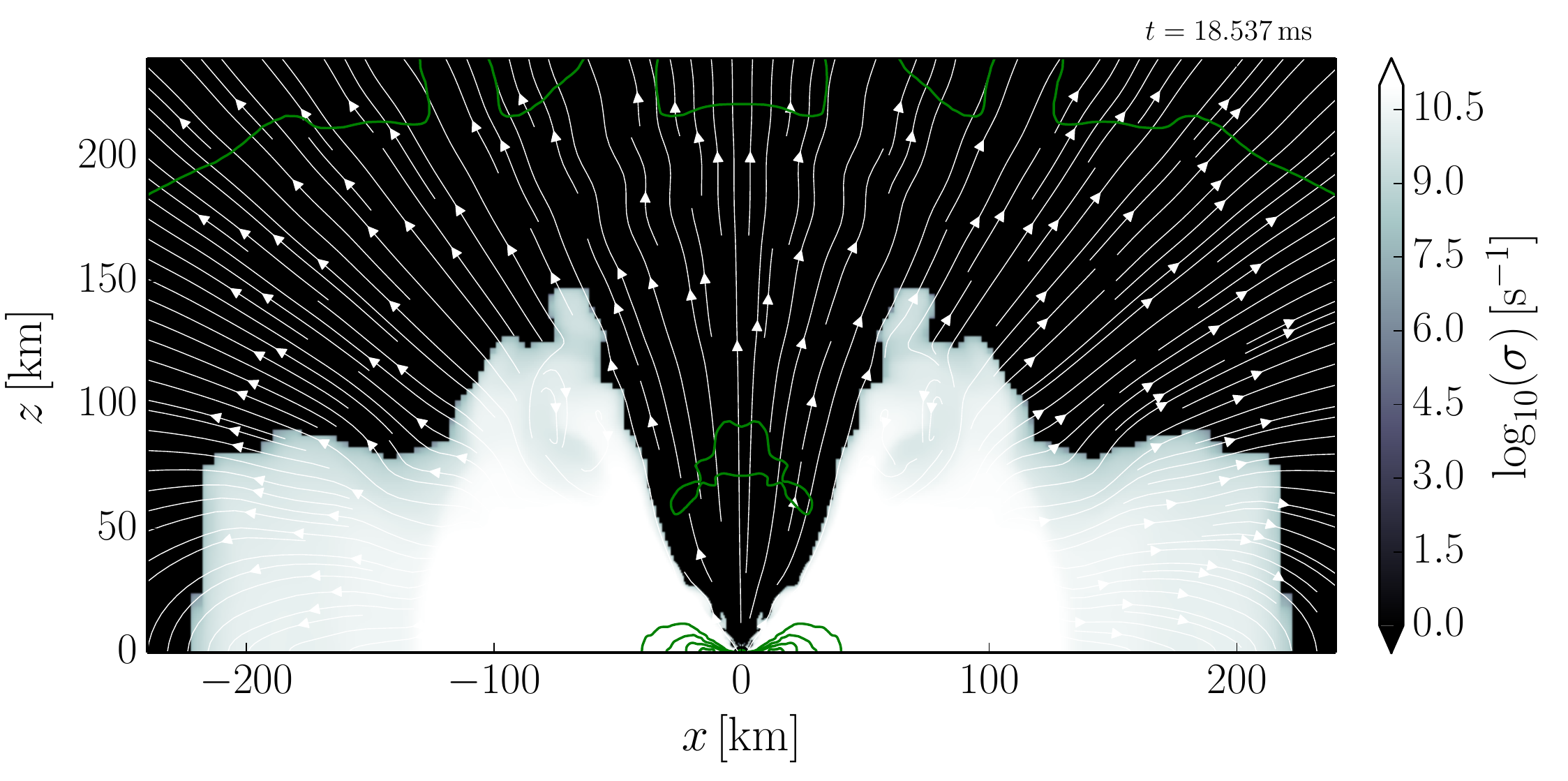}
\caption{Two-dimensional cuts on the $(x,z)$ plane of the electrical
  conductivity distribution at different times
  $t=\{11.601,~16.141,~18.537\}\,\ms$ (left to right). Note that, the
  conductivity is very large in the HMNS (left panel) and torus (middle
  and right panel), where the IMHD is recovered. On the other hand, the
  conductivity is outside the HMNS/torus, where the electrovacuum limit
  is reached. Note also that the region in black does not
  coincide with the atmosphere, but is filled with tenuous as can be seen in the left columns of Figs.~\ref{fig:bns:rhoB}
  and \ref{fig:bns:rhoB6}.}
\label{fig:bns:sigma}
\end{figure*}

\subsection{Rapid overview}
\label{sec:bns:features}
 
\begin{figure*}
\centering
\includegraphics[width=0.32\textwidth]{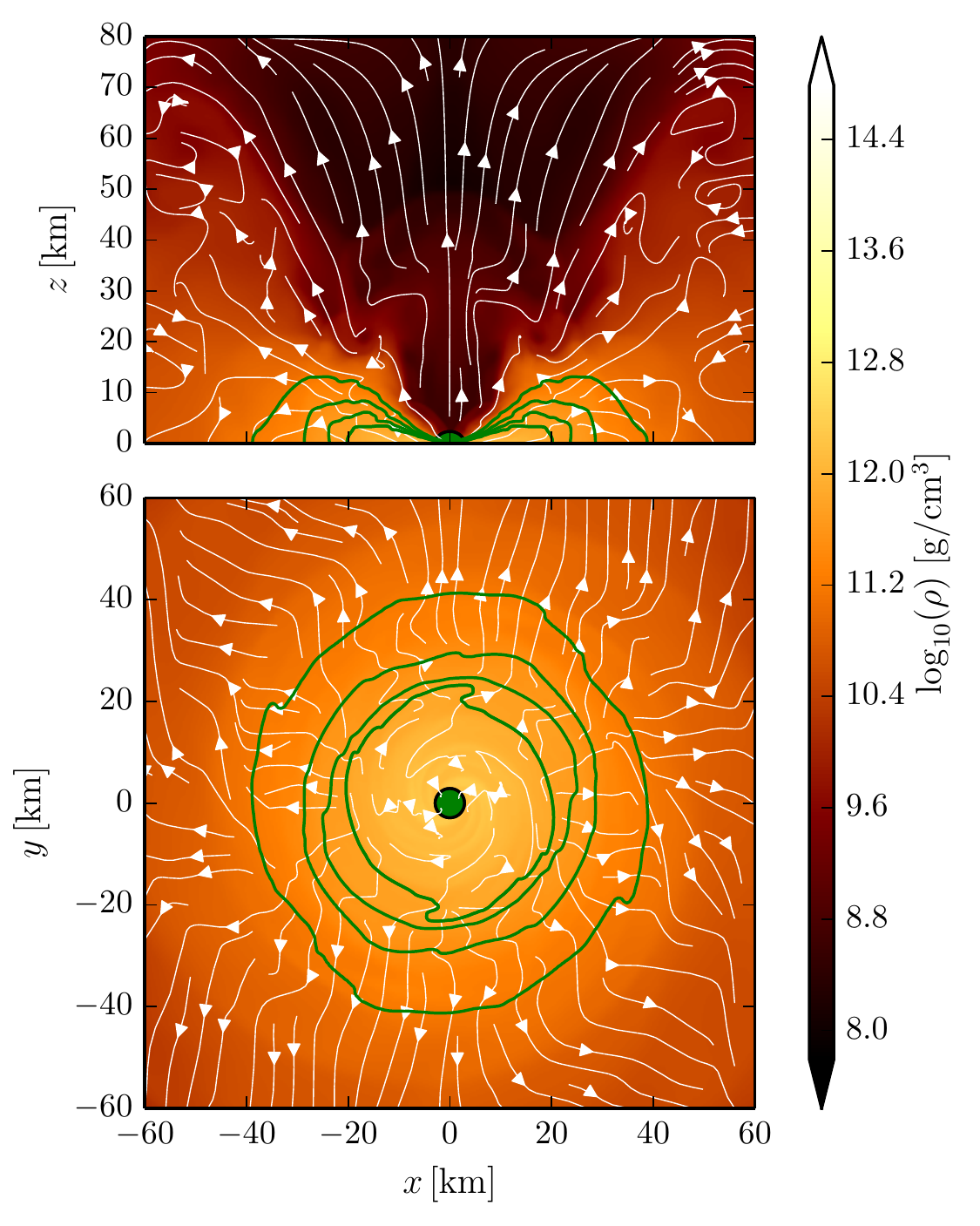}
\includegraphics[width=0.32\textwidth]{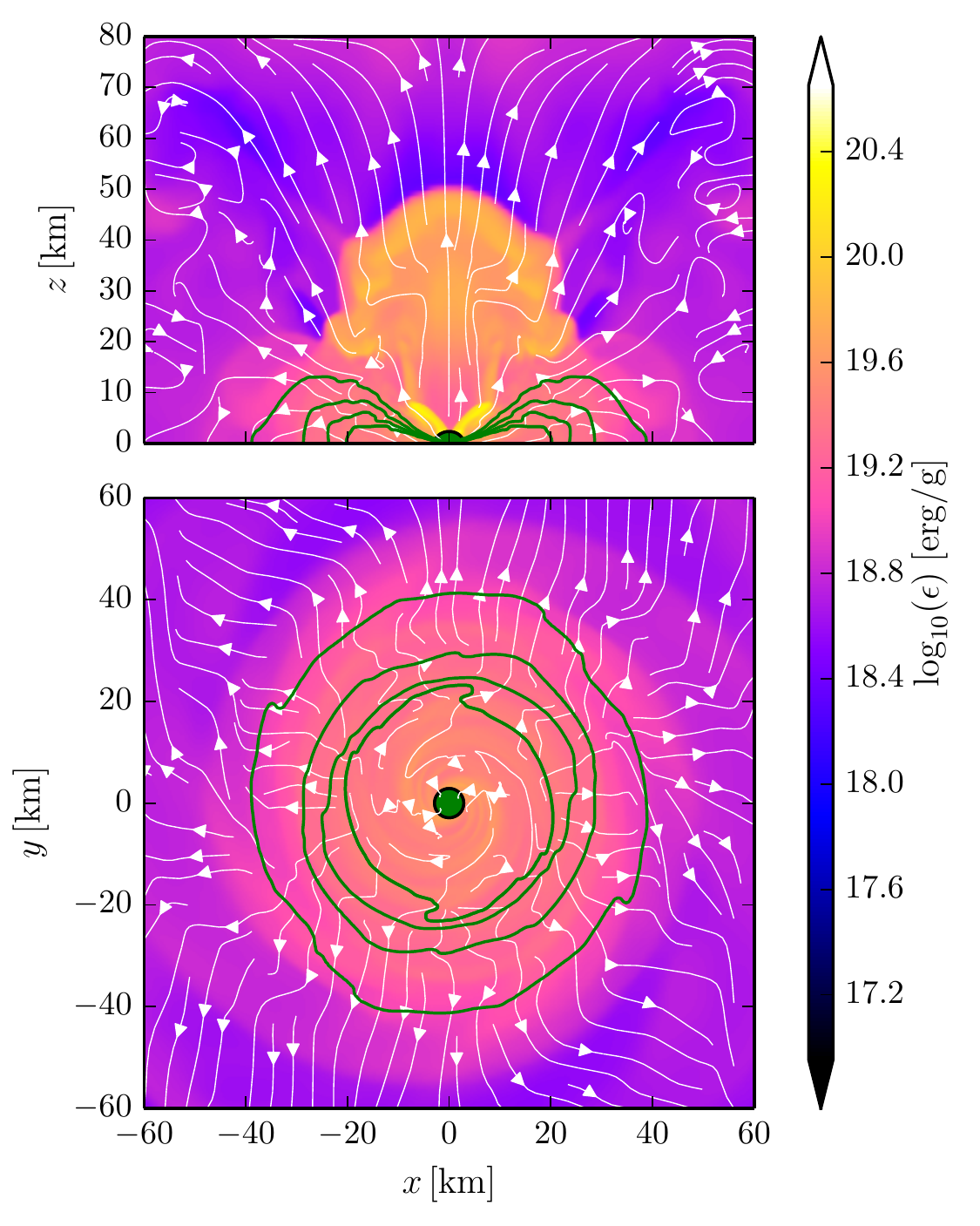}
\includegraphics[width=0.32\textwidth]{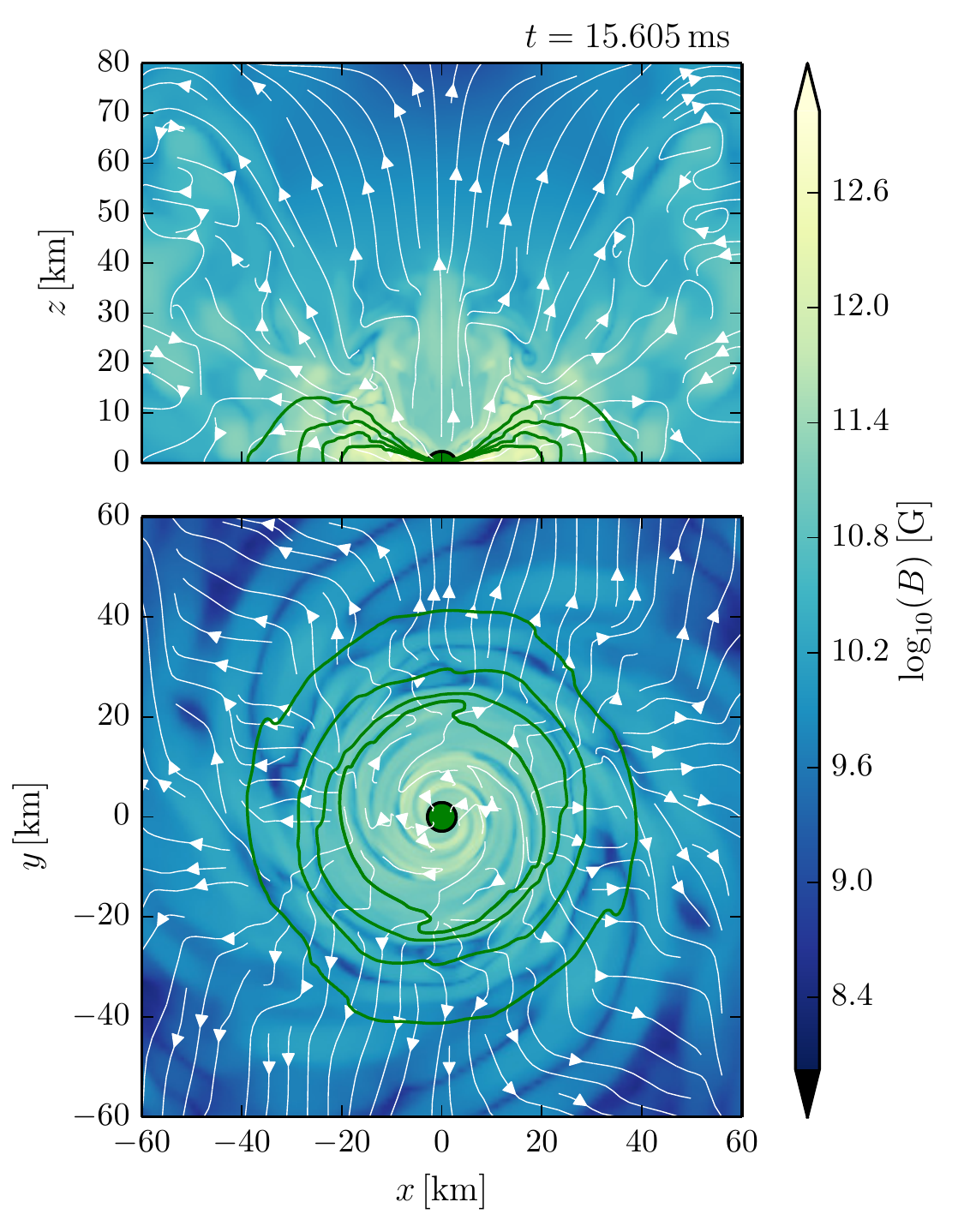}
\includegraphics[width=0.32\textwidth]{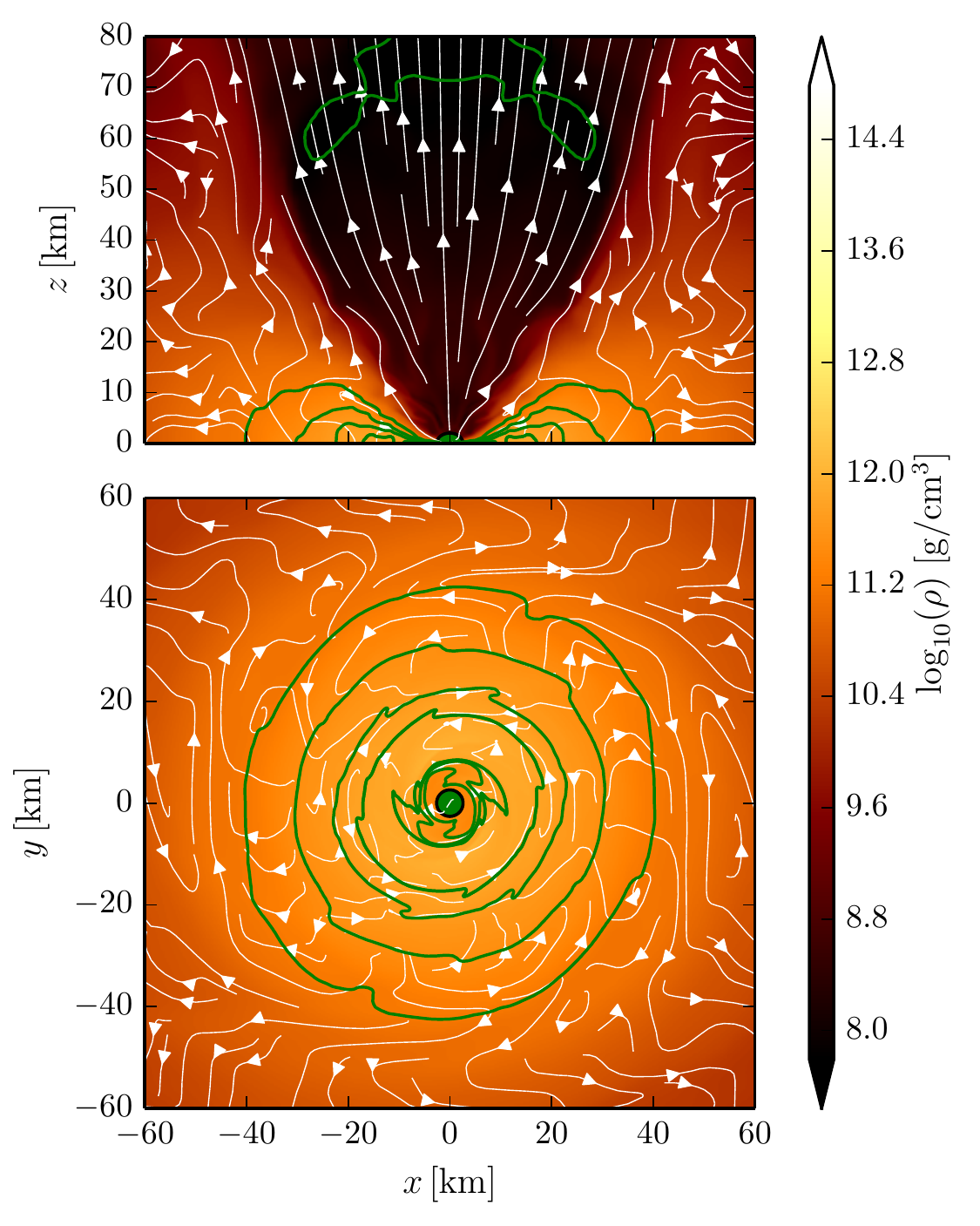}
\includegraphics[width=0.32\textwidth]{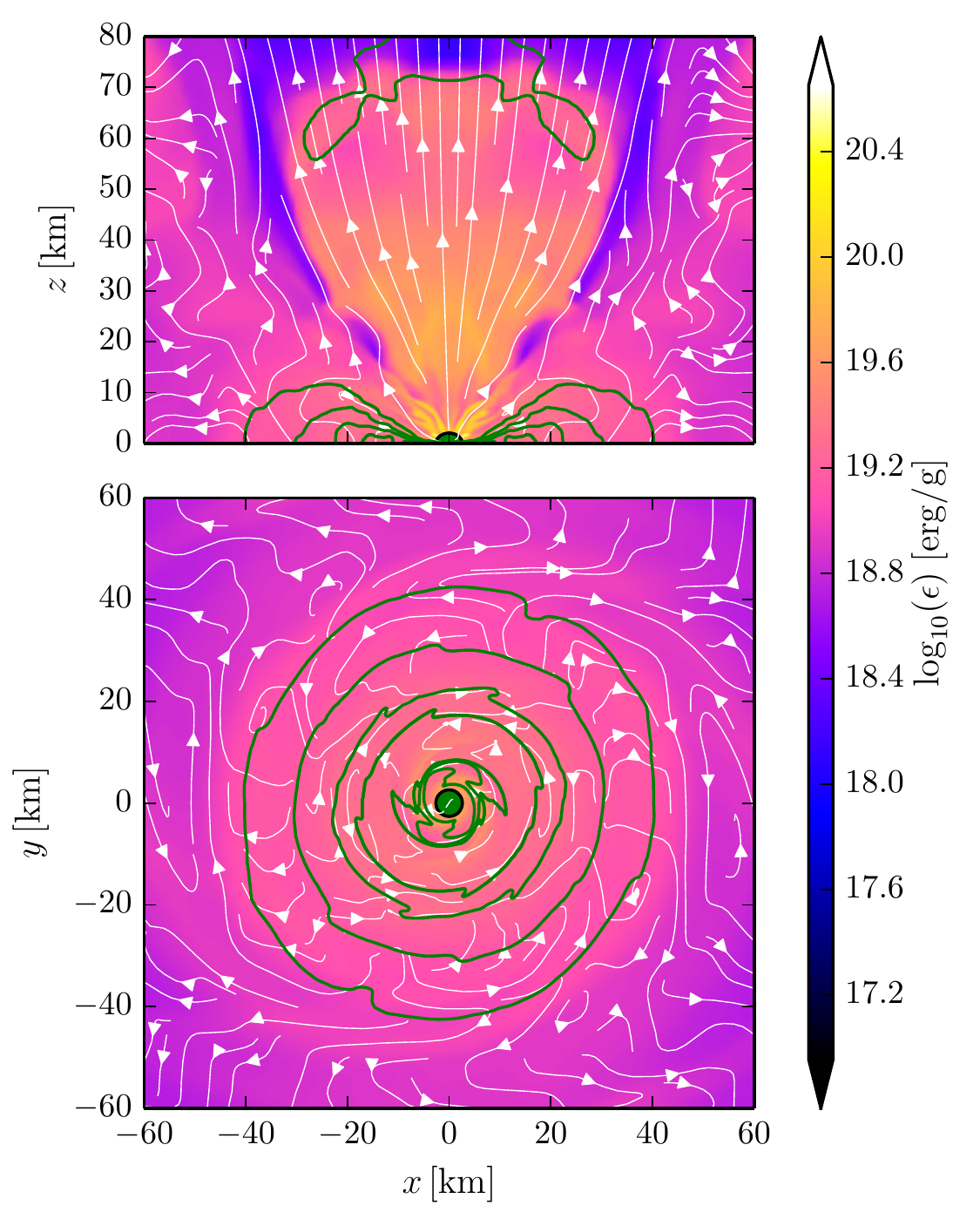}
\includegraphics[width=0.32\textwidth]{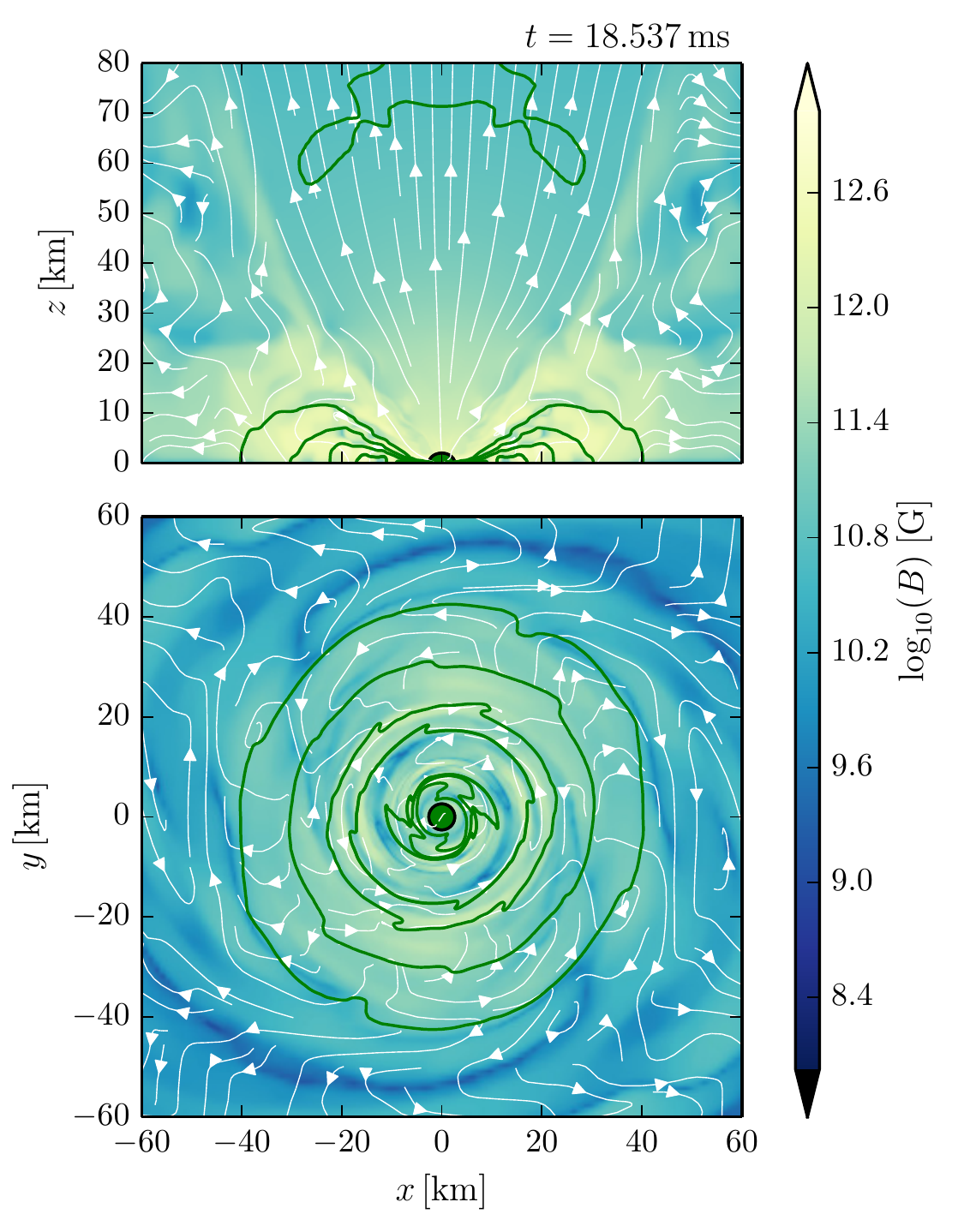}
\caption{The same as in Fig. \ref{fig:bns:rhoB} but for times
  $t=15.61\,\ms$ (top row) and $t=18.54\,\ms$ {(bottom row)}, when a
  black hole has already been formed. Note that, in contrast to the
  dynamics of the HMNS shown in Fig. \ref{fig:bns:rhoB}, the collapse of
  the HMNS leads to the generation of large-scale coherent magnetic
  fields and the emergence of a magnetic-jet structure around the
  black-hole rotation axis. The magnetic field is mostly poloidal in the
  funnel and mostly toroidal in the torus (see also
  Figs. \ref{fig:rho_and_B_xz_large} and \ref{fig:bns:magtorpol}).}
\label{fig:bns:rhoB6}
\end{figure*}

The dynamics of the binary when evolved within the RMHD framework is
summarized in Figs.~\ref{fig:bns:rhoB} and \ref{fig:bns:rhoB6}. The first
one, in particular, reports two-dimensional cuts in the $(x,y)$ and
$(x,z)$ planes of the rest-mass density $\rho$ (left panel), of the
specific internal energy $\epsilon$ (middle panel), and of the modulus of
the magnetic field $|B|:=(B^iB_i)^{1/2}$ (right panel). Marked with white
lines are the projection of the magnetic-field lines on the different
planes, while marked with green solid lines are the isocontours of the
rest-mass density. The snapshots refer to times $t=2.43\,\ms$ {(top
  row)}, $t=3.91\,\ms$ {(middle row)}, and $t=11.60\,\ms$ {(bottom
  row)}. The positions of the two stars are marked with $\times$ and $+$
symbols, and the stars' trajectories are marked with red dashed lines.

Given the small initial orbital separation of $45\,\km$ and the reduced
linear moment, the two stars merge very rapidly. More specifically, at
approximately $t \simeq 0.5\,\ms$, the two stellar surfaces start entering
in contact, although the actual merger takes place at $t \simeq
3.91\,\ms$.\footnote{As is customary, we define the time of merger as the
  time of the first peak of the gravitational-wave amplitude
  \cite{Takami:2014, Bernuzzi2014, Takami2015}.} As the merger takes
place, the two stellar cores become significantly distorted by the large
tidal fields and produce spiral arms. At the leading edges of these
spiral arms, the specific internal energy increases through shock heating
(\cf the middle column of Fig \ref{fig:bns:rhoB}).

The time $t=2.42\,\ms$ in the top row of Fig.~\ref{fig:bns:rhoB}
corresponds to one orbit of the binary, which is sufficient for the
magnetic field to diffuse over the thin transition layer close to the
surface of the stars in an attempt to settle to a new equilibrium
configuration (\cf the discussion in Sec.~IV.C.1 of
Ref.~\cite{Dionysopoulou:2012pp}). Once the magnetic field has diffused out of
the star it continues to propagate also in regions that are dynamically
treated as atmosphere and where the electrical conductivity is set to
zero as if the medium was an electrovacuum. In this way, we achieve a
rather smooth transition between the highly conducting stellar interior
and the electrovacuum exterior. This is shown in
Fig. \ref{fig:bns:sigma}, where we show the electrical conductivity at
three reference times on the $(x,z)$ plane. Note that, the region in black
corresponds to our electrovacuum but does not coincide with the
atmosphere. Indeed, the region in black is filled with tenuous plasma as can be seen in the left columns of
Fig.~\ref{fig:bns:rhoB}. Note also that at this time the magnetic-field
topology is still predominantly poloidal in the exterior of the
star. However, a toroidal component is also being generated as the
highly conducting material in the stellar interior shears the poloidal
magnetic field in the lower-density, high-conductivity spiral arms.

\begin{figure*}[!ht]
\center
\includegraphics[width=0.32\textwidth]{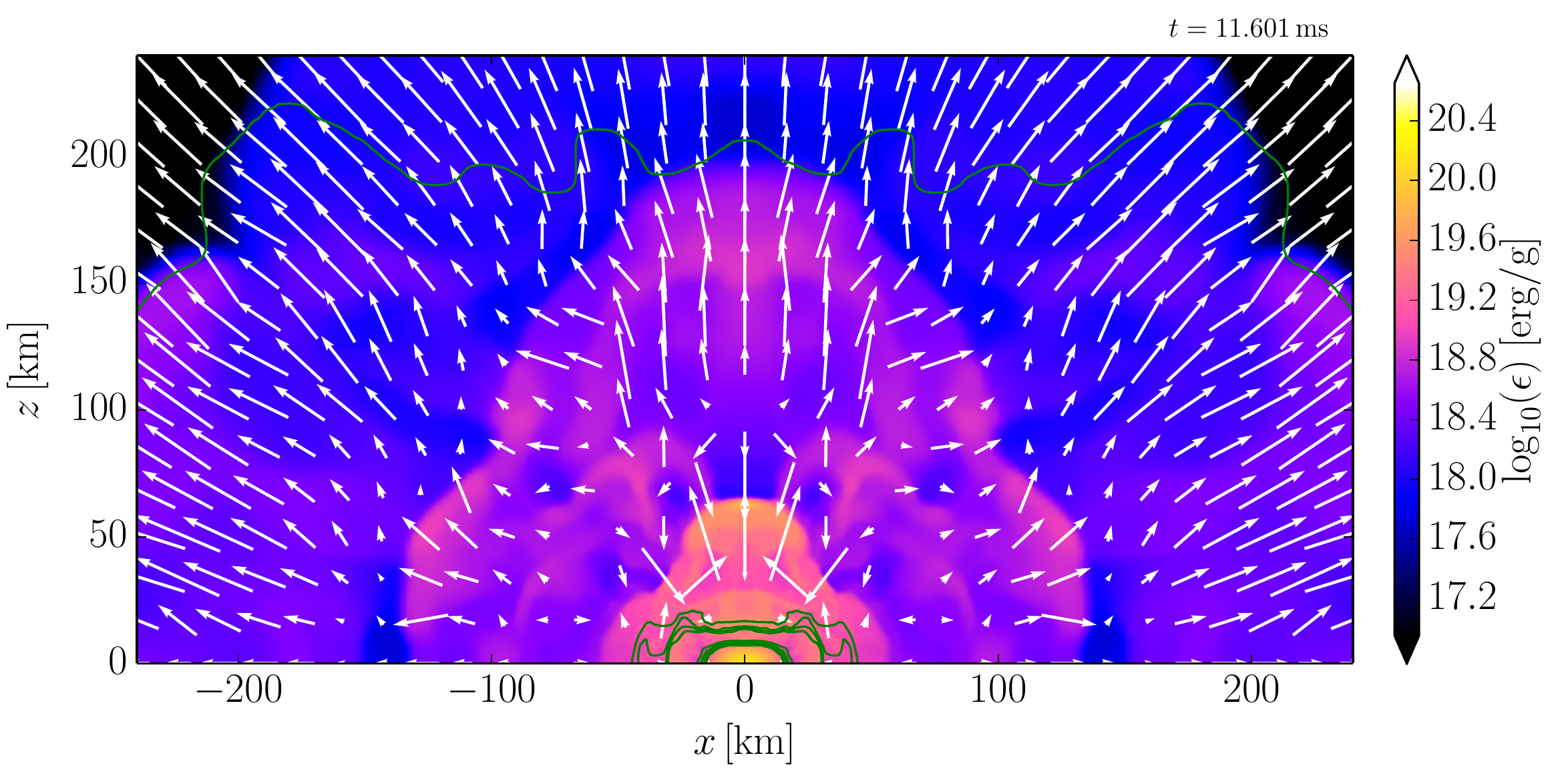}
\hskip 0.2cm
\includegraphics[width=0.32\textwidth]{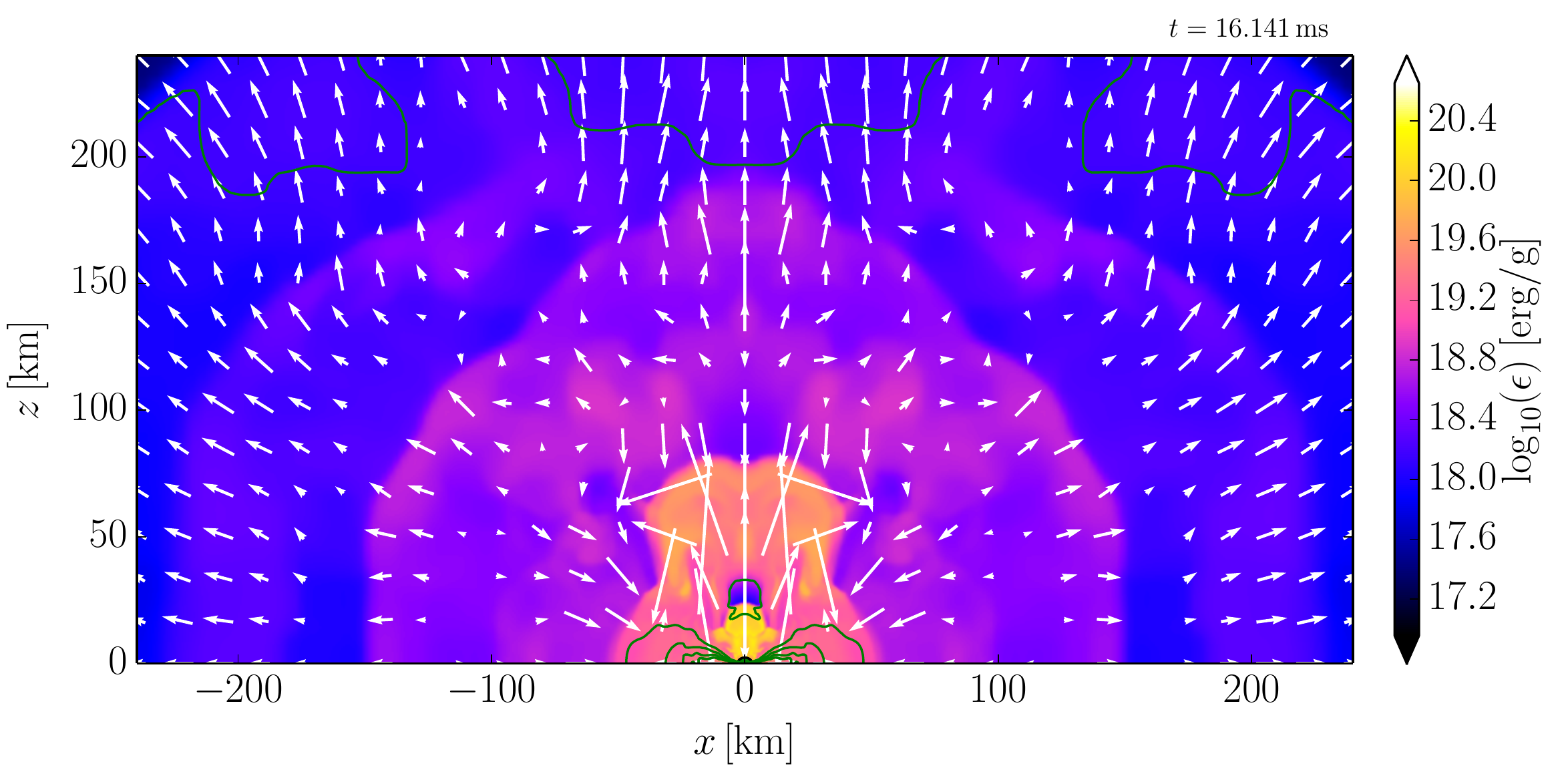}
\hskip 0.2cm
\includegraphics[width=0.32\textwidth]{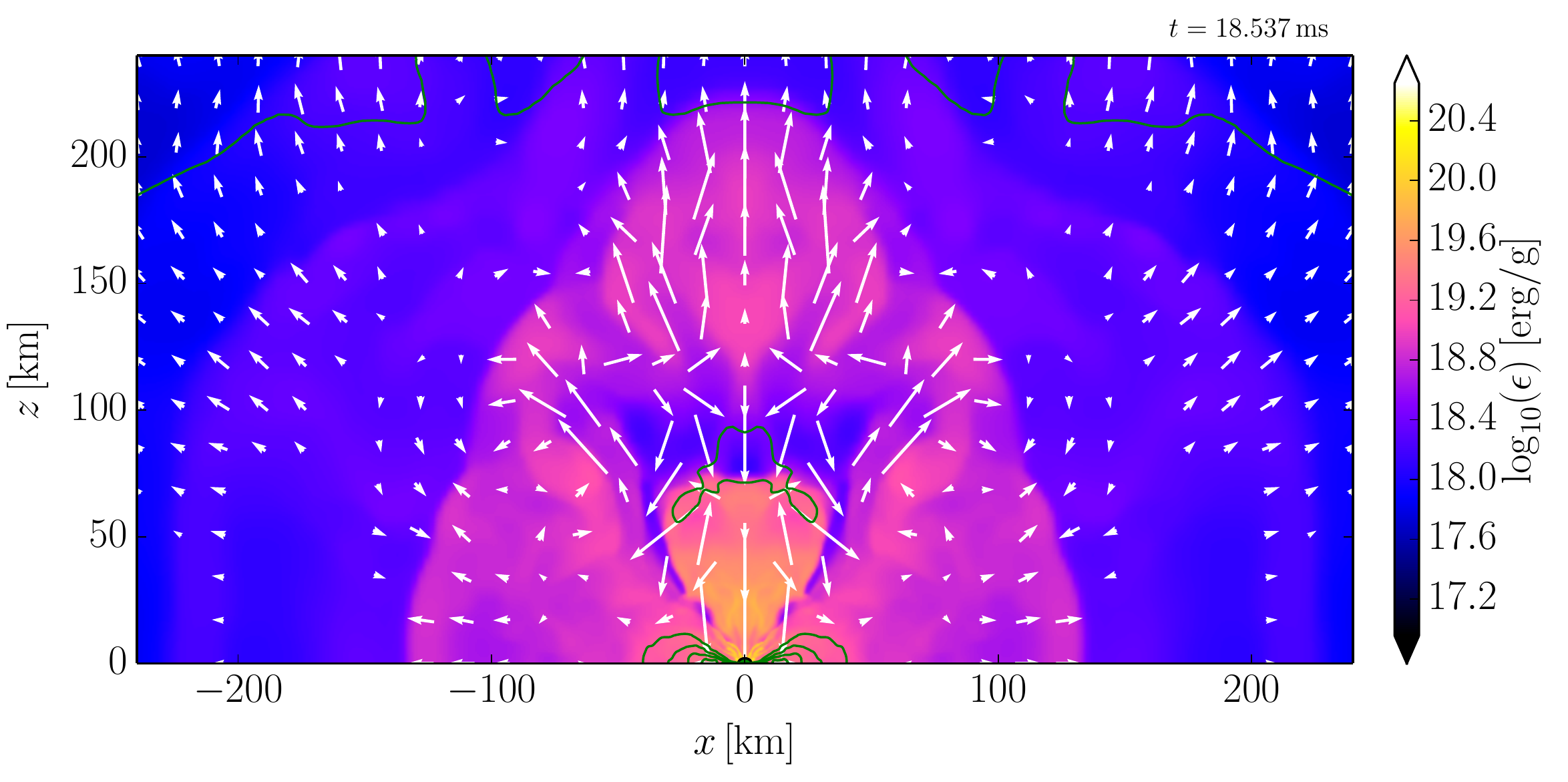}
\caption{Two-dimensional cuts on the $(x,z)$ plane of the specific
  internal energy and of the velocity field in this plane. Note that the
  times selected coincide with those presented in
  Fig. \ref{fig:bns:sigma}, and are useful to gain information on the properties of the flow during
  the HMNS stage and after the collapse to a black hole. The flow is
  essentially given by the magnetically driven wind during the HMNS stage
  (left panel). However, once a funnel is produced, the flow is ingoing
  at heights $z \lesssim 100\,\mathrm{km}$, but becomes outgoing for $z
  \gtrsim 100\,\mathrm{km}$ (middle and right panels).}
\label{fig:bns:epsvel}
\end{figure*}

As mentioned above, at $t=3.91\,\ms$ the merger takes place, at least as
measured from the position of the first peak in the gravitational-wave
amplitude. When this happens, a vortex sheet is created between the two
stars, which could lead to the onset of a Kelvin--Helmholtz instability
and the generation of a large-scale and ultrastrong magnetic field
\cite{Price06} (see the middle row of Fig. \ref{fig:bns:rhoB}). It is
presently a matter of debate whether such large-scale magnetic fields can
be produced with amplifications of several orders of magnitude. Present
direct simulations are not able to reach the resolutions necessary to
resolve the turbulent motion produced by the
instability~\cite{Giacomazzo:2009mp,Giacomazzo:2010}. The results
obtained so far with direct, very high-resolution simulations, either
local~\cite{Obergaulinger10} or global \cite{Kiuchi2014}, indicate that
the amplification of the magnetic field is of a factor 20 at most, most
likely because resistive instabilities disrupt the Kelvin--Helmholtz
unstable vortex~\cite{Obergaulinger10} (but see also
Ref.~\cite{Giacomazzo:2014b} for recent simulations with subgrid modeling
which could lead to much larger amplifications).

In the simulations reported here, we find that the magnetic-field
magnitude increases slightly less than an order of magnitude during this
stage (see Fig. \ref{fig:bns:mag} and the discussion in
Sec.~\ref{sec:bns:comparison}). This moderate growth of the magnetic
field is possibly due to insufficient resolution and our inability to
capture the dynamics of the relevant scales. On the other hand, it could
also indicate that a Kelvin--Helmholtz instability simply does not
develop. A possible reason is that the time scale of the instability might
be longer than the dynamical time scale and the shearing motion could very
rapidly get destroyed as the two stellar cores collide on a time scale of
a fraction of a millisecond. We should also note that resistive effects
might become important at this stage as the reconnection of magnetic-field
lines might lead to the acceleration of matter due to Ohmic heating. We
believe that higher-order schemes, as those presented
in Refs.~\cite{Dumbser2009,Radice2012a,Radice2013b}, could help resolve this
issue.

The bar-deformed HMNS produced after the merger is differentially
rotating, and magnetic braking transfers angular momentum from the inner
core to the outer parts of the star. The spiral arms widen and merge
together generating more shock heating and dissipation. A magnetically
driven wind as a result of differential rotation is launched from the
outer layers of the HMNS \cite{Kiuchi2012b,Siegel2014}. The wind could
play an important role in the modelling of short gamma-ray bursts
(SGRBs), which show an extended x-ray emission
\cite{Rezzolla2014b,Ciolfi2014}. The wind is not constant in time, but
rather characterized by a bursty activity in which high internal energy
plasma blobs (\ie local concentrations of specific internal energy a few
kilometers in size) are launched from near the black-hole horizon and
propagate along the $z$-direction (bottom row of
Fig.~\ref{fig:bns:rhoB}). Interestingly, the bursts observed in the
specific internal energy of the hot rotating halo that forms around the
central object are anticorrelated with the bursts observed in the
modulus of the magnetic field; a more detailed discussion on these bursts
follows in Sec.~\ref{sec:bns:bursts}.

Unfortunately, the resolution used here is insufficient to be able to
track the development of an MRI, which is, however, expected to develop
\cite{Siegel2013, Kiuchi2014} and could significantly amplify the
magnetic field. Blind to this effect, our simulation shows that at
$t=11.60\ms$ (see the bottom row of Fig.~\ref{fig:bns:rhoB6}) magnetic
braking has managed to store enough rotational energy in the winding of
the magnetic-field lines so that the inner core of the star is now less
differentially rotating. The direct consequence of this is that the HMNS
collapses to a black hole of mass $M=~2.88\,M_{\odot}$ and dimensionless
spin $a=J/M^2=0.87$, as measured from the apparent horizon
\cite{Dreyer02a,Thornburg2003:AH-finding}. We postpone the discussion on
how angular momentum is transported outward and how this affects the
lifetime of the HMNS to Sec.~\ref{sec:bns:lifetime}.

At time $t=15.61\,\ms$, the black hole is surrounded by a thick accretion
torus that is responsible for confining and collimating the magnetic
field along the $z$-axis (\cf the top row of Fig.~\ref{fig:bns:rhoB6}). The
properties of the black hole-torus system are shown in
Table~\ref{tab:bns:bhtorus} at approximately $4.74\,\ms$ after the
collapse and show that the mass of the torus is $0.095 M_{\odot}$. The
magnetic-field topology and matter dynamics soon after the collapse are
highly turbulent, but the high degree of symmetry introduced by the black
hole, which is gravitationally dominant over the torus, rapidly
establishes some order in this system. After about one orbital period, in
fact, the torus becomes essentially axisymmetric and the matter in the
polar region is rapidly accreted onto the black hole, giving rise to a
funnel where the rest-mass density reaches values close to
that of the atmosphere (\cf the bottom row of Fig.~\ref{fig:bns:rhoB6}).

\begin{figure*}
\centering
\includegraphics[width=0.48\textwidth]{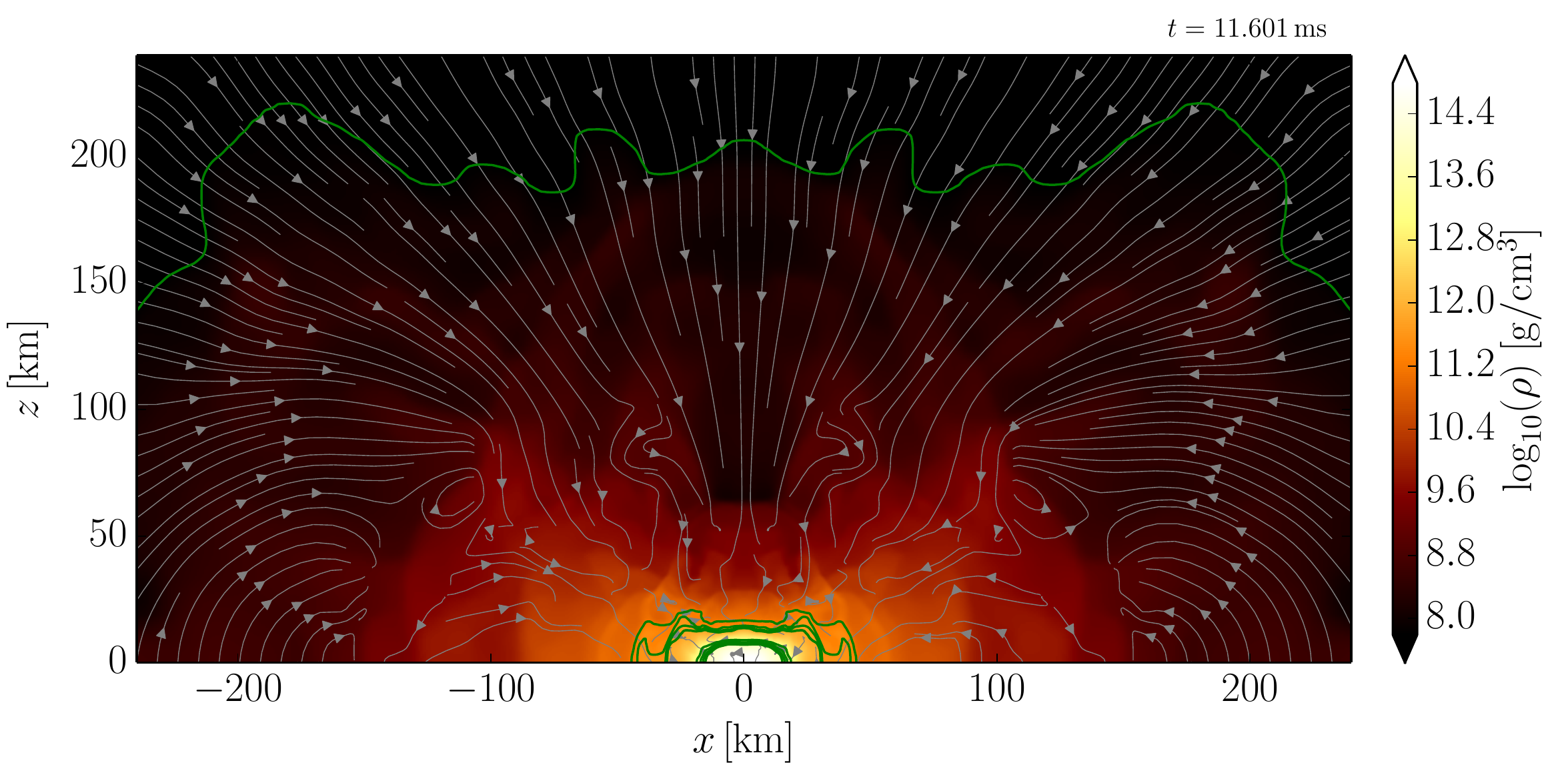}
\hskip 0.5cm
\includegraphics[width=0.48\textwidth]{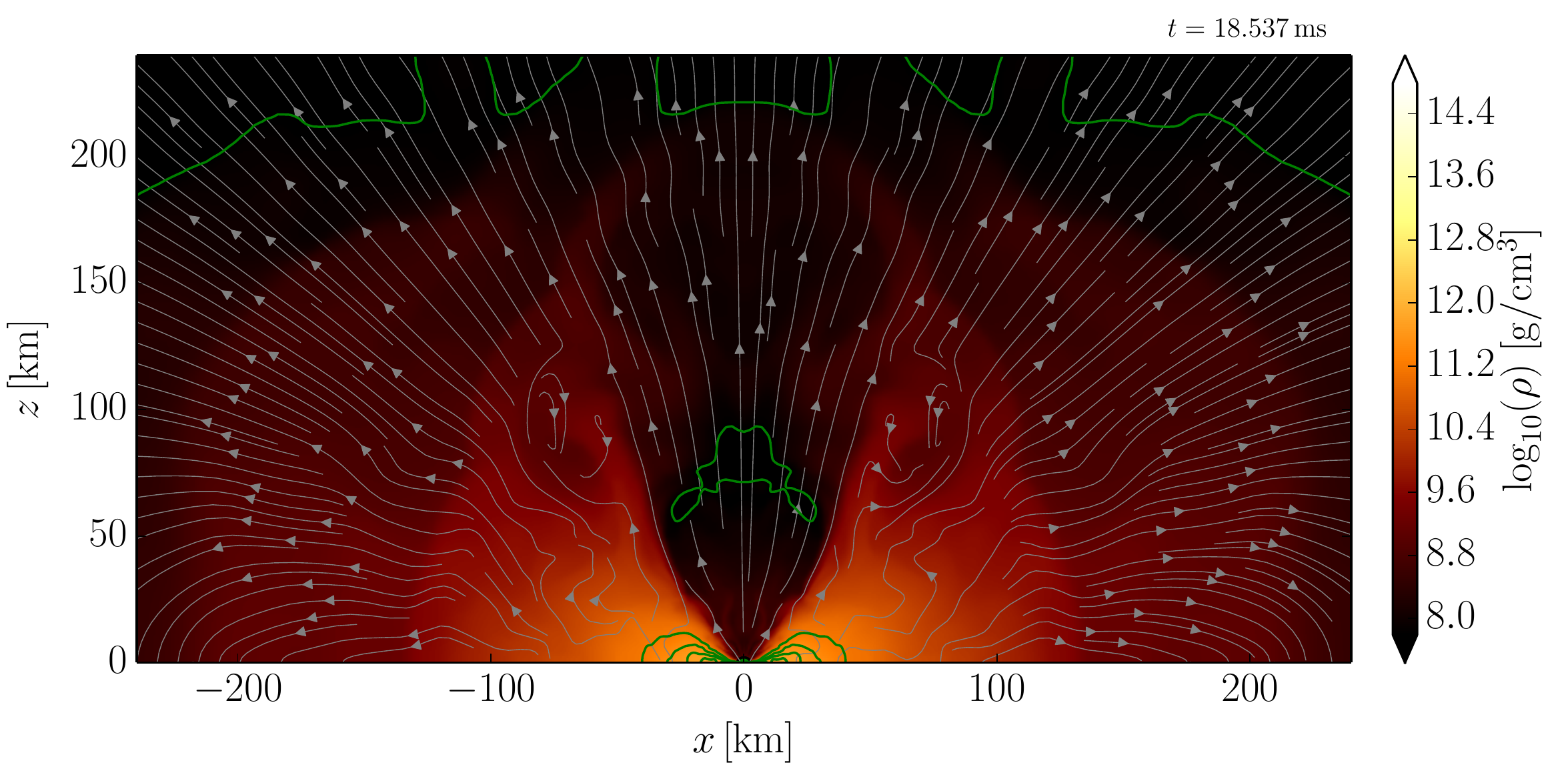}
\hskip 0.5cm
\includegraphics[width=0.48\textwidth]{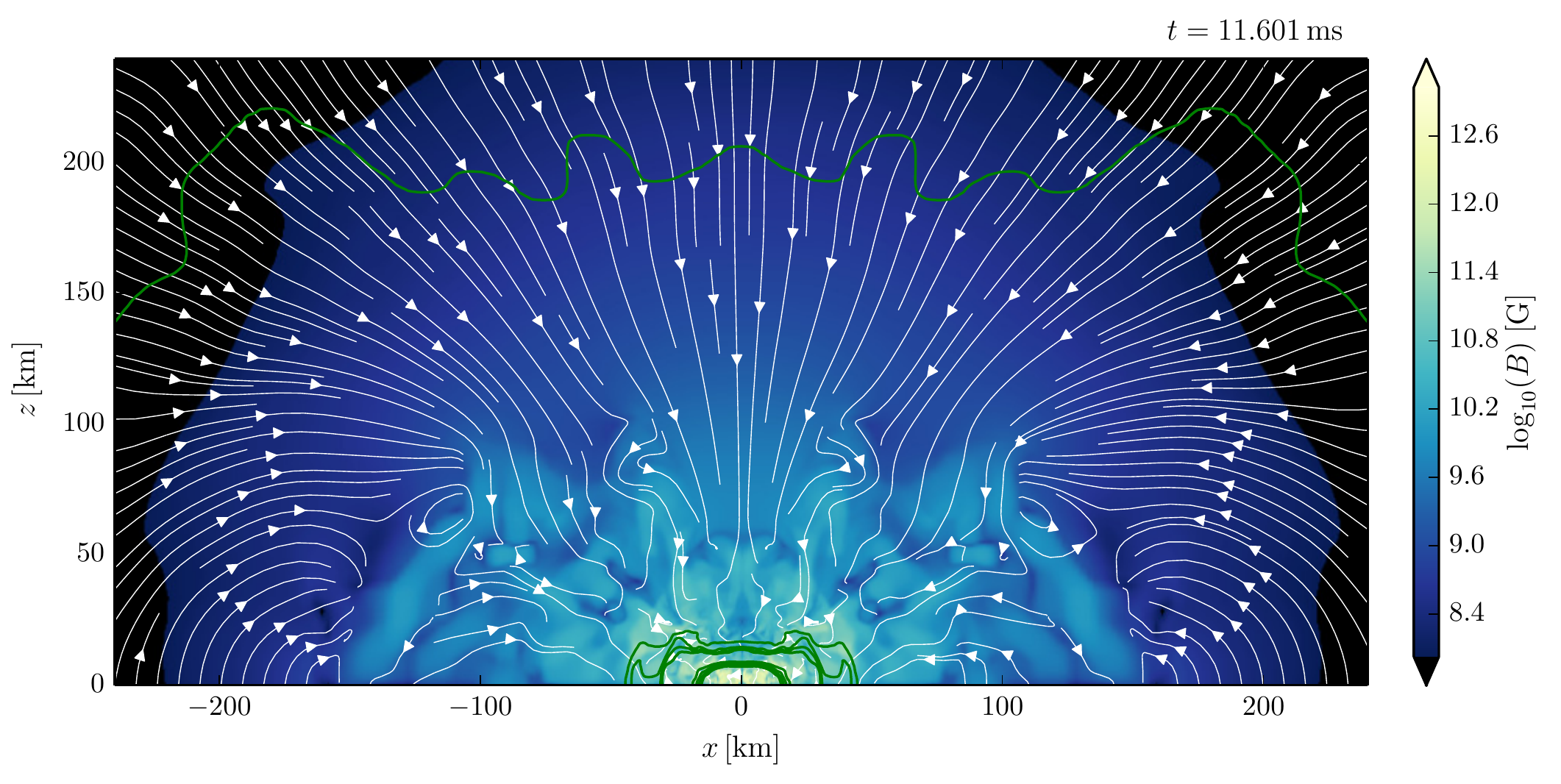}
\hskip 0.5cm
\includegraphics[width=0.48\textwidth]{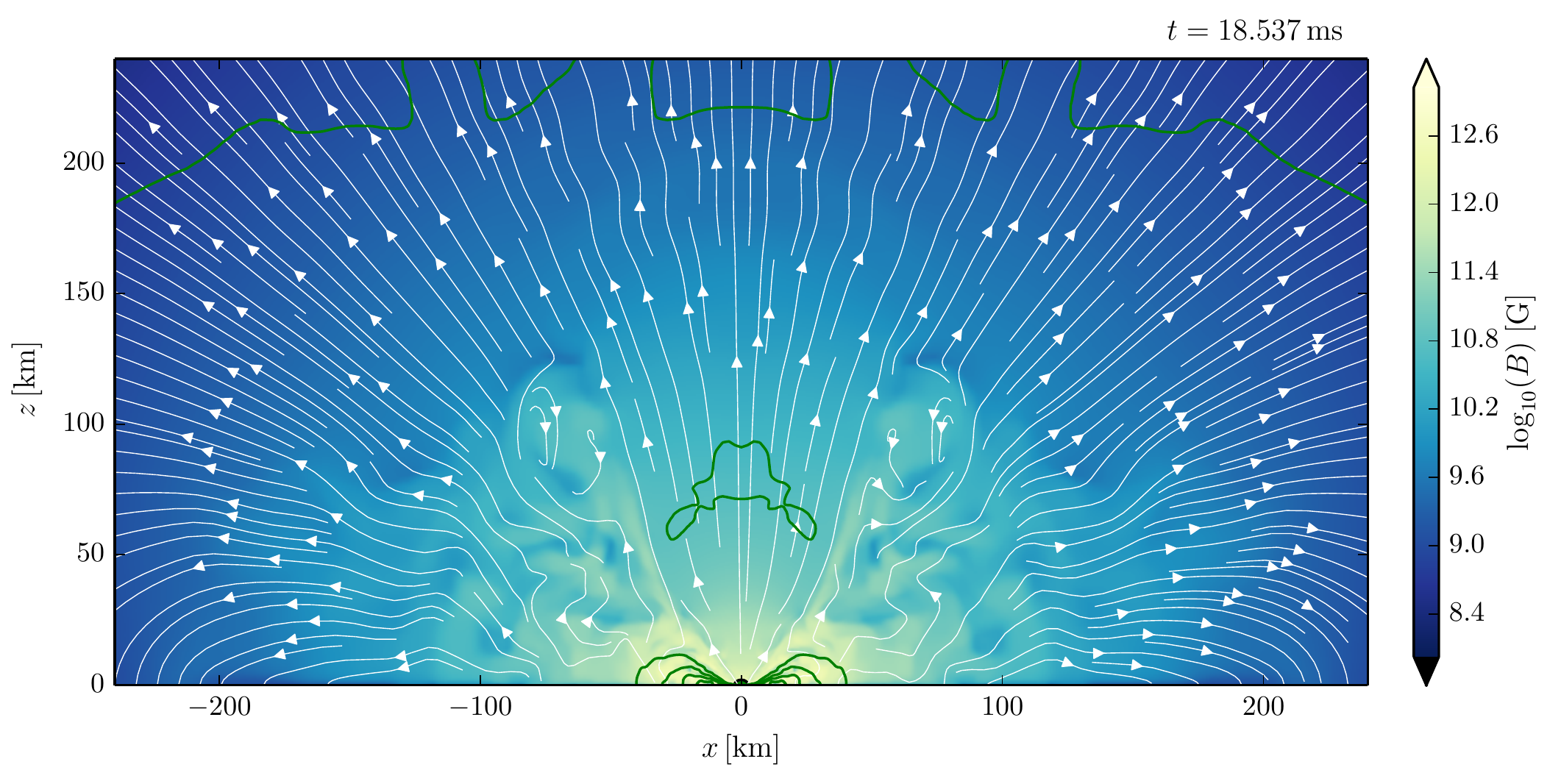}
\caption{Large-scale two-dimensional snapshots on the $(x,z)$ planes of
  the rest-mass density (top row) and of the magnetic field (bottom
  row). The two columns refer to $t=11.60\,\ms$ (left column), when the
  HMNS has not yet collapsed and to $t=18.54\,\ms$ {(right column)}, when
  a black hole has already been formed. Note again the formation of a
  magnetic-jet structure around the black-hole rotation axis, which
  extends on scales that are much larger than those of the accreting
  torus (\ie $\sim \pm 60\,\km$) and of the black hole (\ie $\sim \pm
  5\,\km$) (see also Figs. \ref{fig:bns:rhoB6}, \ref{fig:bns:magtorpol},
  \ref{fig:rho_and_B_xz_large.imhd}, and \ref{fig:rho_and_B_xz.imhd}).}
  \label{fig:rho_and_B_xz_large}
\end{figure*}

After black-hole formation, the plasma dynamics in the funnel is far from
being stationary and continues with repeated bursts having a period of
about $2.4-3.7\,\ms$. While a more detailed discussion of these bursts is
postponed to Sec.~\ref{sec:bns:bursts}, it is useful to remark here that
the ejected material does not have sufficient energy to reach large
distances away from the black hole. This is probably due to the fact that,
although the magnetic field is comparatively strong, the material in the
funnel is still matter dominated, with $\beta_{_{\rm P}} \sim
10^{4}-10^{6}$. Such large values are not particularly surprising since
the initial magnetic field in the stars is rather small, \ie $\sim
10^{12}\,\G$, and is not amplified significantly. At the same time, the
torus angular velocity profiles have become nearly Keplerian, and the MRI
could develop (\cf Fig. \ref{fig:bns:omega}). However, as for the HMNS,
also the spatial resolution of the grid covering the torus is too small
to capture the fastest growing modes of the instability in the torus (see
discussion in Sec. \ref{sec:bns:torus}).

For completeness, we report in Fig. \ref{fig:bns:epsvel} three different
two-dimensional cuts on the $(x,z)$ plane of the specific internal energy
and of the velocity field in this plane. The times selected are the same
as those presented in Fig. \ref{fig:bns:sigma}, and are useful to gain information on the properties
of the flow during the HMNS stage and after the collapse to a black
hole. Note that during the HMNS stage the flow consists mostly of the
intense magnetically driven outgoing wind discussed above (left
panel). However, once a funnel is produced, the flow is ingoing at
heights $z \lesssim 100\,\mathrm{km}$ but becomes outgoing for $z
\gtrsim 100\,\mathrm{km}$ (middle and right panels). The location of this
stagnation point varies with time and follows the expansion of hot fluid
which is produced by the accretion process and that can be followed via
the increases in the specific internal energy. In all cases considered,
the vertical flow very close to the black hole is \emph{ingoing}.

\subsection{Magnetic-field topology and magnetic-jet structure}
\label{sec:bns:magjet}
 
As mentioned above, the high degree of symmetry near the rapidly rotating
black hole induces a quick rearrangement of the matter and of the
magnetic fields. As a result, the magnetic-field topology at time
$t=18.54\,\ms$ changes in the funnel and develops a dominant poloidal
component, giving rise to a well-defined magnetic-jet structure, which is
almost axisymmetric. This result is similar to what was already found in the
simulations of Ref.~\cite{Rezzolla:2011}, with the important difference
that this configuration has been reached with a higher spatial resolution
and a consistent treatment of the resistivity.

\begin{figure*}
\centering
\hskip -0.55cm
\includegraphics[width=0.365\textwidth]{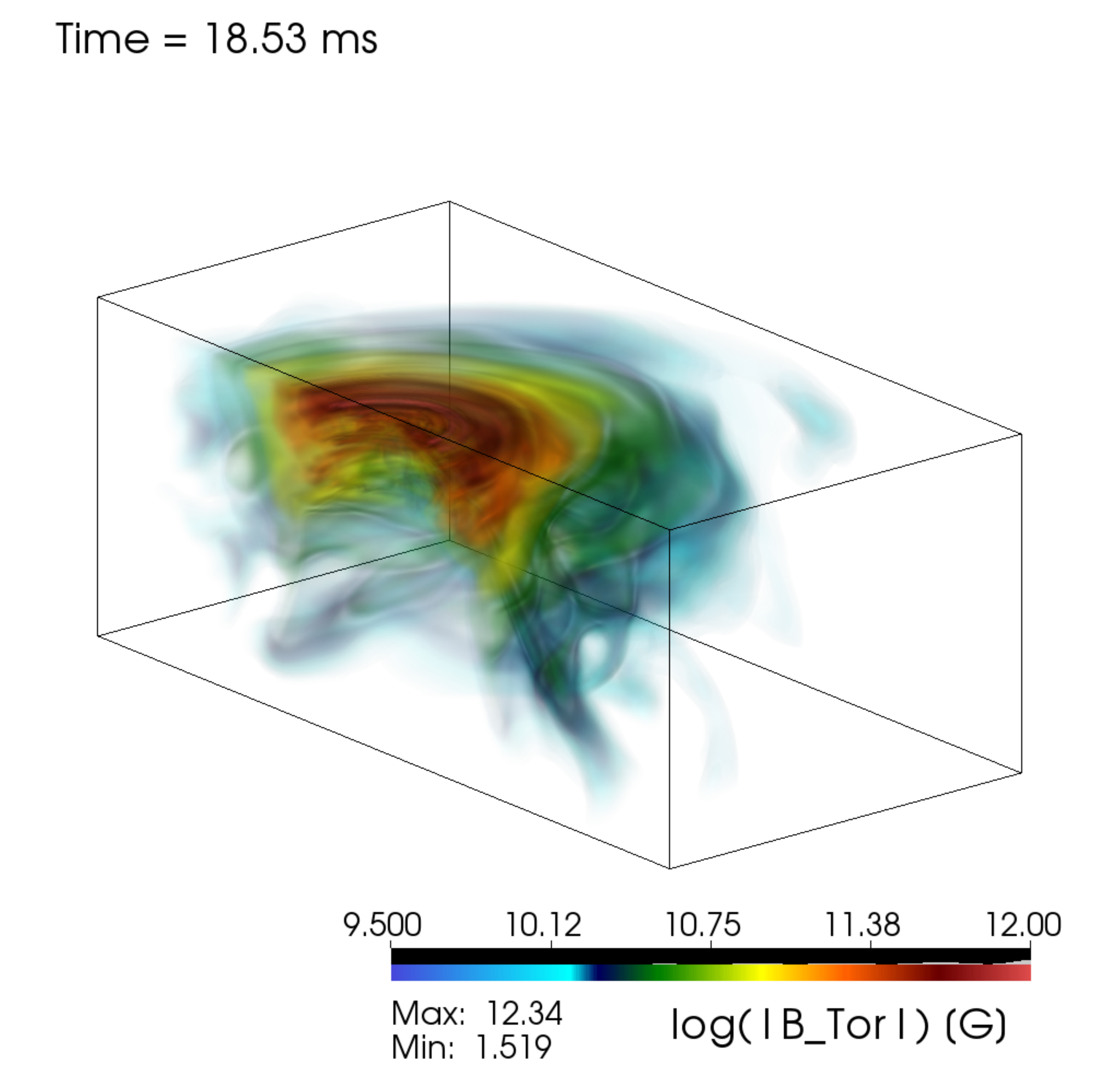}
\hskip -0.40cm
\includegraphics[width=0.365\textwidth]{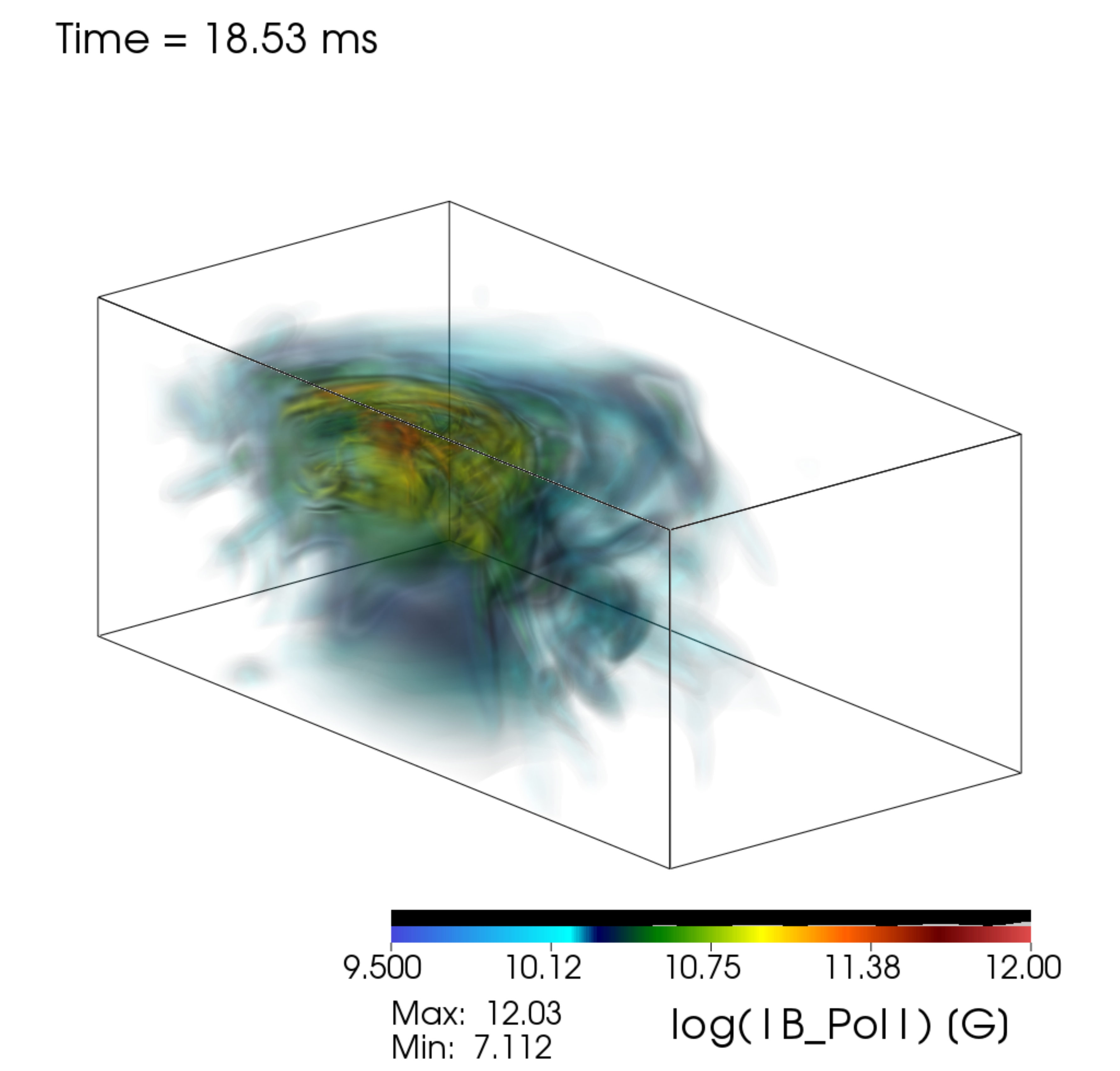}
\hskip -0.40cm
\includegraphics[width=0.335\textwidth]{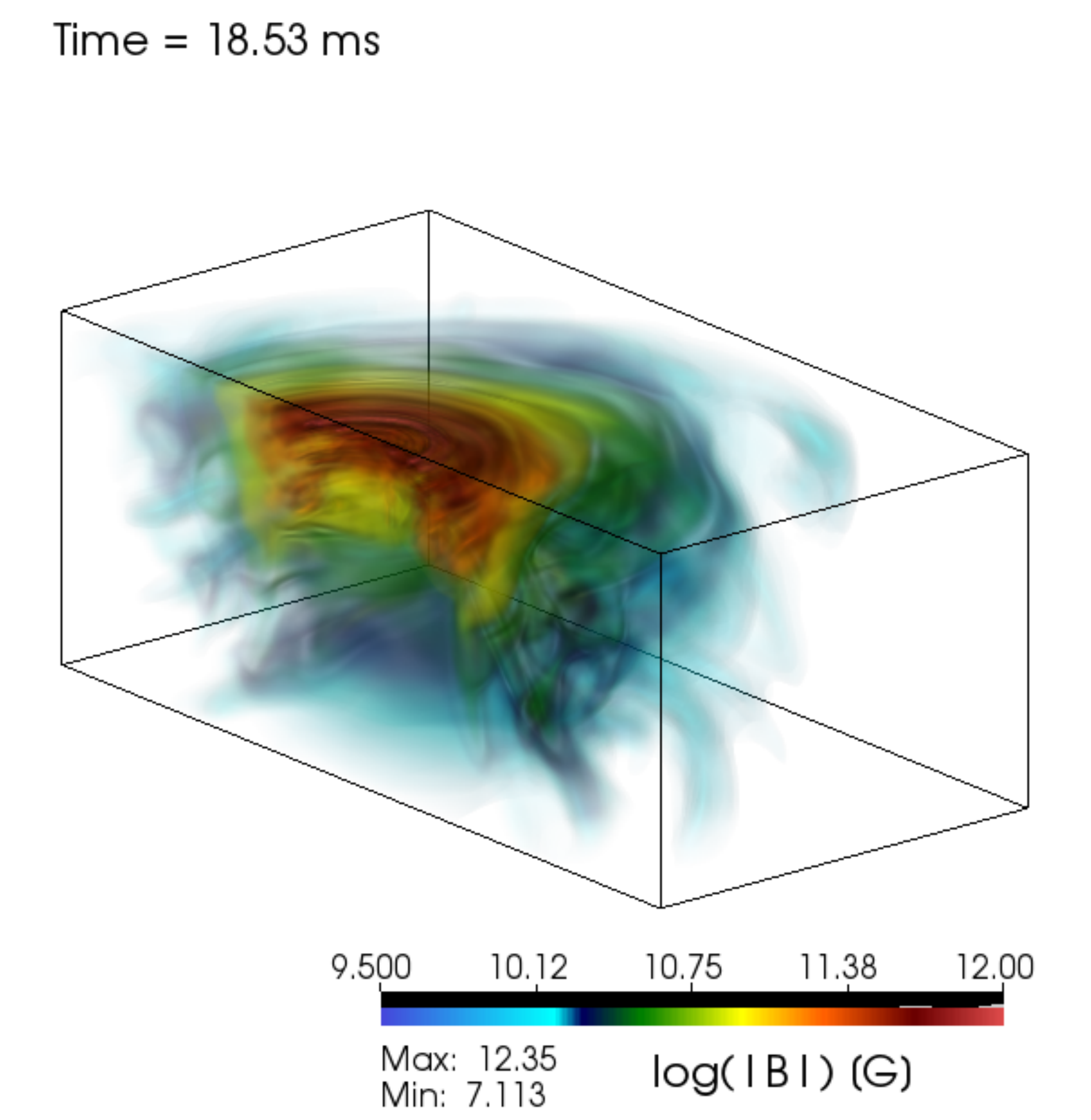}
\caption{Three-dimensional snapshots of the norm of the toroidal magnetic
  field $|B_{\rm tor}|$ (left panel), of the poloidal one $|B_{\rm pol}|$
  (middle panel), and of the total magnetic field $|B|$ (right panel),
  all at time $t=18.3\,\ms$. Additionally, we plot the modulus of the
  magnetic field at the same time to illustrate where the toroidal and
  poloidal contributions become dominant. Note that the magnetic field at
  the edges of the funnel starts developing a toroidal component that
  exhibits signs of twisting, while the magnetic field in the evacuated
  region is predominantly poloidal. The figures show a cut through the
  torus, with the $z$-axis facing down, in order to be able to see the
  evacuated region formed along the $z$-axis and the funnel-wall
  structure that develops at the interface with the interstellar
  medium. The domain plotted corresponds to a rectangular grid with
  dimensions
  $[0\,\km,115.8\,\km]\times[-115.8\,\km,115.8\,\km]\times[0\,\km,92.16\,\km]$}
  \label{fig:bns:magtorpol}
\end{figure*}

\begin{figure*}
\centering
\hskip -0.55cm
\includegraphics[width=0.355\textwidth]{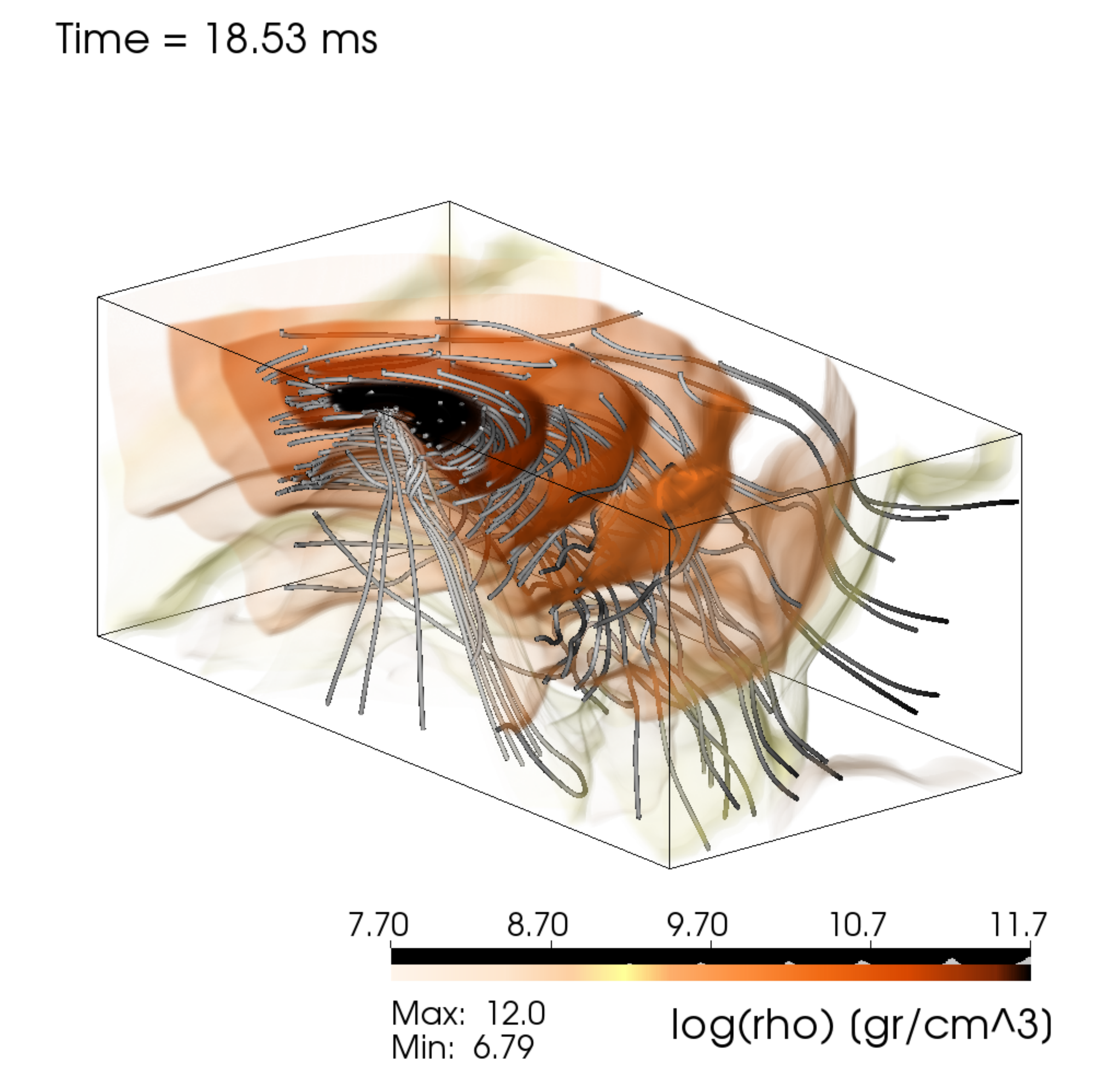}
\hskip -0.40cm
\includegraphics[width=0.355\textwidth]{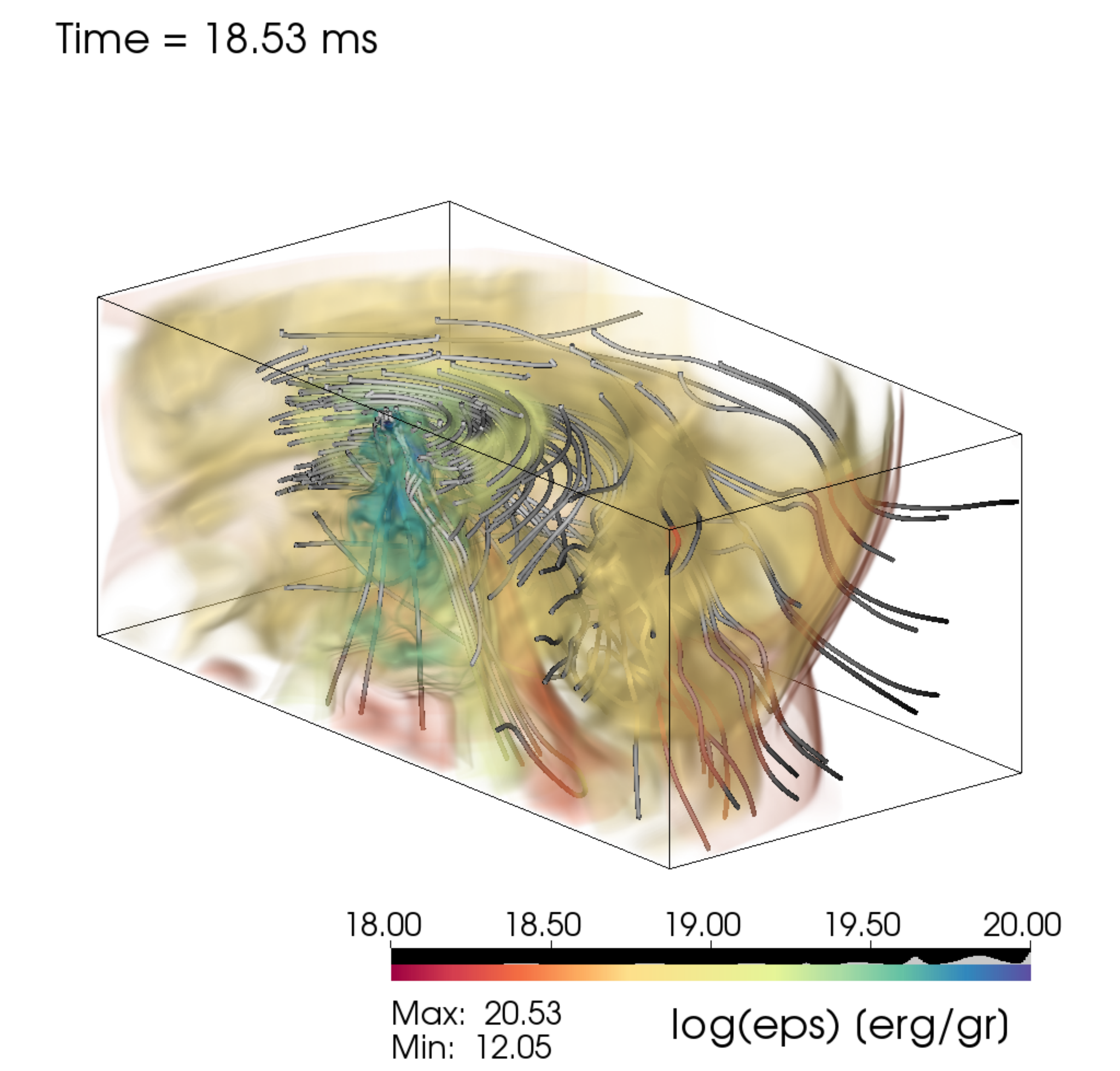}
\hskip -0.40cm
\includegraphics[width=0.355\textwidth]{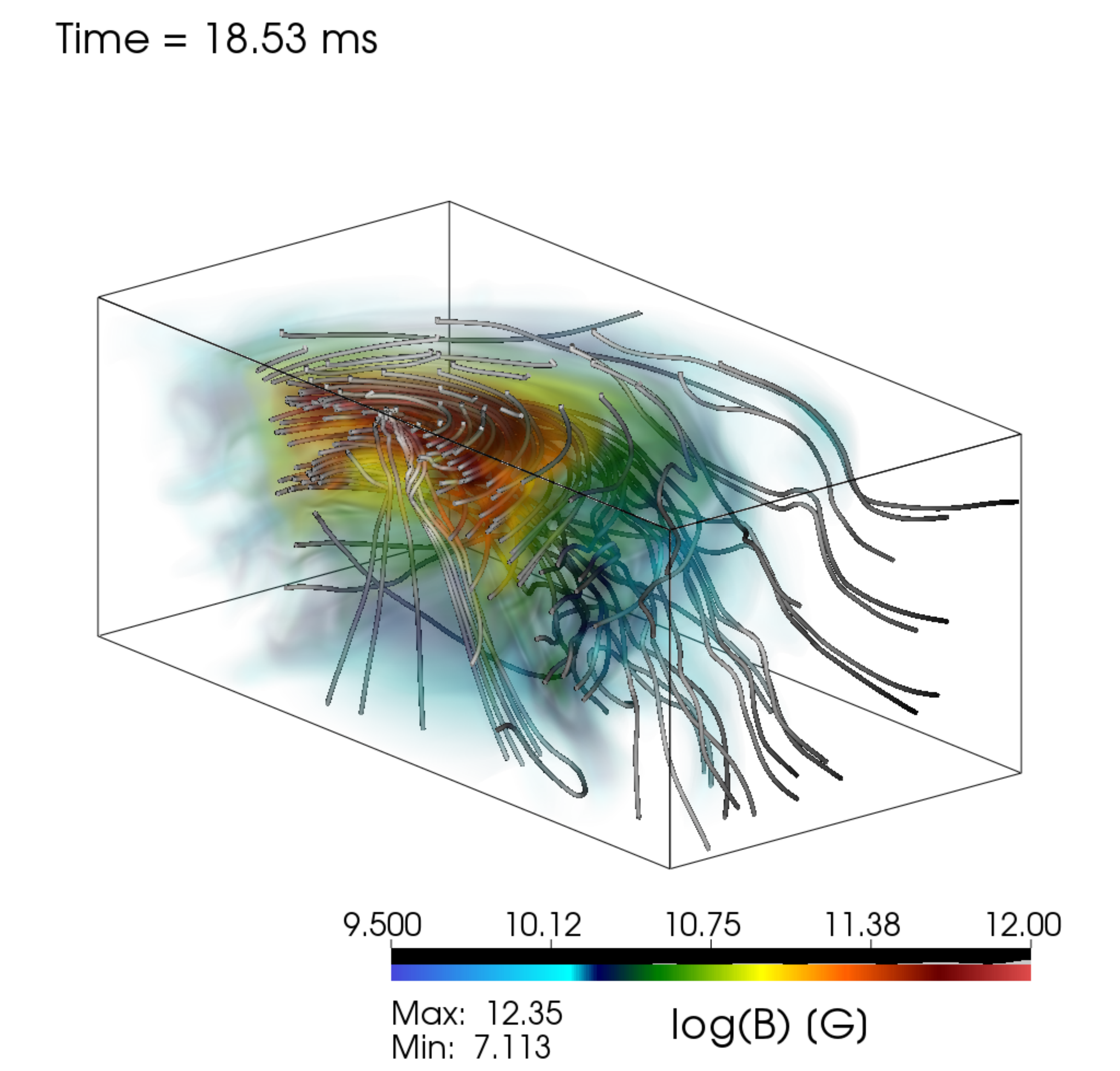}
  \caption{Three-dimensional snapshots of the rest-mass density $\rho$
    (top left panel), specific internal energy $\epsilon$ (top right
    panel), and modulus of the magnetic field $|B|$ (bottom panel) at
    $t=18.3\,\ms$. Additionally, we plot the magnetic-field lines on top
    of these quantities to illustrate the topology of the magnetic
    field. Because of the symmetries applied in our simulation we only
    show a quadrant of the black hole--torus system. The figures show a
    cut through the torus, with the $z$-axis facing down, in order to be
    able to see the evacuated region formed along the $z$-axis and the
    funnel-wall structure that develops at the interface with the
    interstellar medium. The domain plotted corresponds to a rectangular
    grid with dimensions $[0\, \km,115.8\, \km]\times[-115.8\,
      \km,115.8\, \km]\times[0\, \km,92.16\, \km]$.}
  \label{fig:bns:mag:fieldlines}
\end{figure*}

The rest-mass density of the plasma in the funnel is close to that of the
atmosphere, but also slightly larger. Hence, given our choice of the
conductivity profile in Eq. \eqref{eq:bns:conductivityprofile}, the
conductivity is essentially zero everywhere in the funnel (see the middle and
right panels of Fig.~\ref{fig:bns:sigma}). This has two important
consequences. First, the dynamics of the electromagnetic fields in this
region is not that prescribed by the IMHD equations but rather that of
electromagnetic waves in vacuum. At the same time, because of its
(comparatively) small rest-mass density and pressure, the matter in the
funnel tends to move along the field lines. We should clarify that this
behavior is not achieved because the test-particle limit of the RMHD
equations is reached (the rest-mass density and pressure are in fact
nonzero) but rather because the matter in the funnel can only move in
the vertical direction, either accreting onto the black hole or moving
outward (the matter in the funnel has low or zero specific angular
momentum).

Modelling the dynamics of the matter in the funnel is among the most
  challenging aspects of these calculations. Although the matter there has
  the largest magnetic-pressure support (\ie $\beta_{_{\rm P}} \gtrsim
  10^{4}$), this is still about 4 orders of magnitude away from being
  magnetically dominated. The reason for this behavior is most likely
  due to the comparatively weak magnetic fields that we are able to build
  in the funnel at these resolutions and the short evolution times.
Second, the zero-conductivity plasma in the funnel is just adjacent to
the high-conductivity plasma of the torus; this large jump in the
conductivity helps in preserving the magnetic-jet structure and in
providing a natural agent for the collimation of the flow at low
latitudes.

It is useful to remark that in close analogy with what found in
Ref.~\cite{Rezzolla:2011} the magnetic-jet structure produced here is
\emph{not} a relativistic outflow. Instead, it can just be viewed as an
almost quasistationary magnetic structure confining the tenuous plasma
in the funnel and confining it away from the dynamics of the ultradense
plasma in the torus. In fact, despite the resistive losses, the plasma in
the funnel does not have yet sufficient internal energy to be able to
launch a relativistic outflow. It is possible that the strong magnetic
fields in the vicinity of the black hole could provide the conditions for
electromagnetic extraction of the black hole's rotational energy through
the Blandford--Znajek mechanism \cite{Blandford1977} or through a
generalized Penrose process~\cite{Lasota2014}. Alternatively, the energy
required for launching a relativistic outflow could also be efficiently
deposited along the baryon-poor funnel by reconnection processes not
fully modelled here or by neutrino pair annihilation \cite{Ruffert96b,
  Aloy:2005}. Clearly, additional work is needed to assess the robustness
of our modelling of the funnel region and to assess whether and how
energy can be deposited in the magnetic-jet structure.

A closer look at the magnetic-jet structure is offered in
Fig. \ref{fig:rho_and_B_xz_large}, which shows two-dimensional snapshots
on the $(x,z)$ planes of the rest-mass density (top row) and of the
magnetic field. The two columns refer to $t=11.60\,\ms$ (left column), when the HMNS is just about to collapse, and to $t=18.54\,\ms$ {(right
  column)}, when the black hole--torus system is toward reaching a
quasistationary equilibrium. The different panels in
Fig. \ref{fig:rho_and_B_xz_large} should be compared with the
corresponding ones in Figs. \ref{fig:bns:rhoB6} and
\ref{fig:bns:magtorpol} but are represented here on much larger spatial
scales (see also Appendix \ref{sec:appendix:imhd} for a closer comparison
of the magnetic-field structure in the IMHD simulations). Interestingly, the
magnetic-jet structure extends well beyond the scale of direct influence
of the black hole and shows a coherent structure on scales of $\gtrsim
250\,\km$. The scale of the magnetic structure is much larger than that
of the accreting torus (\ie $\sim \pm 60\,\km$) and of the black hole
(\ie $\sim \pm 5\,\km$).

Additional information on the magnetic-field topology can be appreciated
by considering three-dimensional views of the magnetic-field strength and
field lines. This can be seen in Figs.~\ref{fig:bns:magtorpol}, where we
show a three-dimensional snapshot of the toroidal (left panel), poloidal
(middle panel), and total magnetic field (right panel). It is important to
remark that a well-defined magnetic-jet structure has been recently
reported also in Ref. \cite{Paschalidis2014} from IMHD simulations of the
merger of a black hole--neutron-star binary. In addition to a jet
structure not very different from the one reported here, the authors in
Ref.~\cite{Paschalidis2014} are also able to produce a sustained outflow from
the accretion torus. At the same time, the IMHD simulations reported in
Ref.~\cite{Kiuchi2014}, where a very high spatial resolution was used, do not
reveal the formation of such a magnetic-jet structure. It is difficult to
assess at the moment the origin of these differences, partly because of
the limited amount of information provided on the simulations in
Ref.~\cite{Kiuchi2014}. A more extended discussion of the properties of the
magnetic-field topology and dynamics, made along the lines suggested in
this paper (see also Sec. \ref{sec:spd}) would be useful to clarify if
there are really differences and their origin.

\begin{figure*}
\centering
\includegraphics[width=0.4\textwidth]{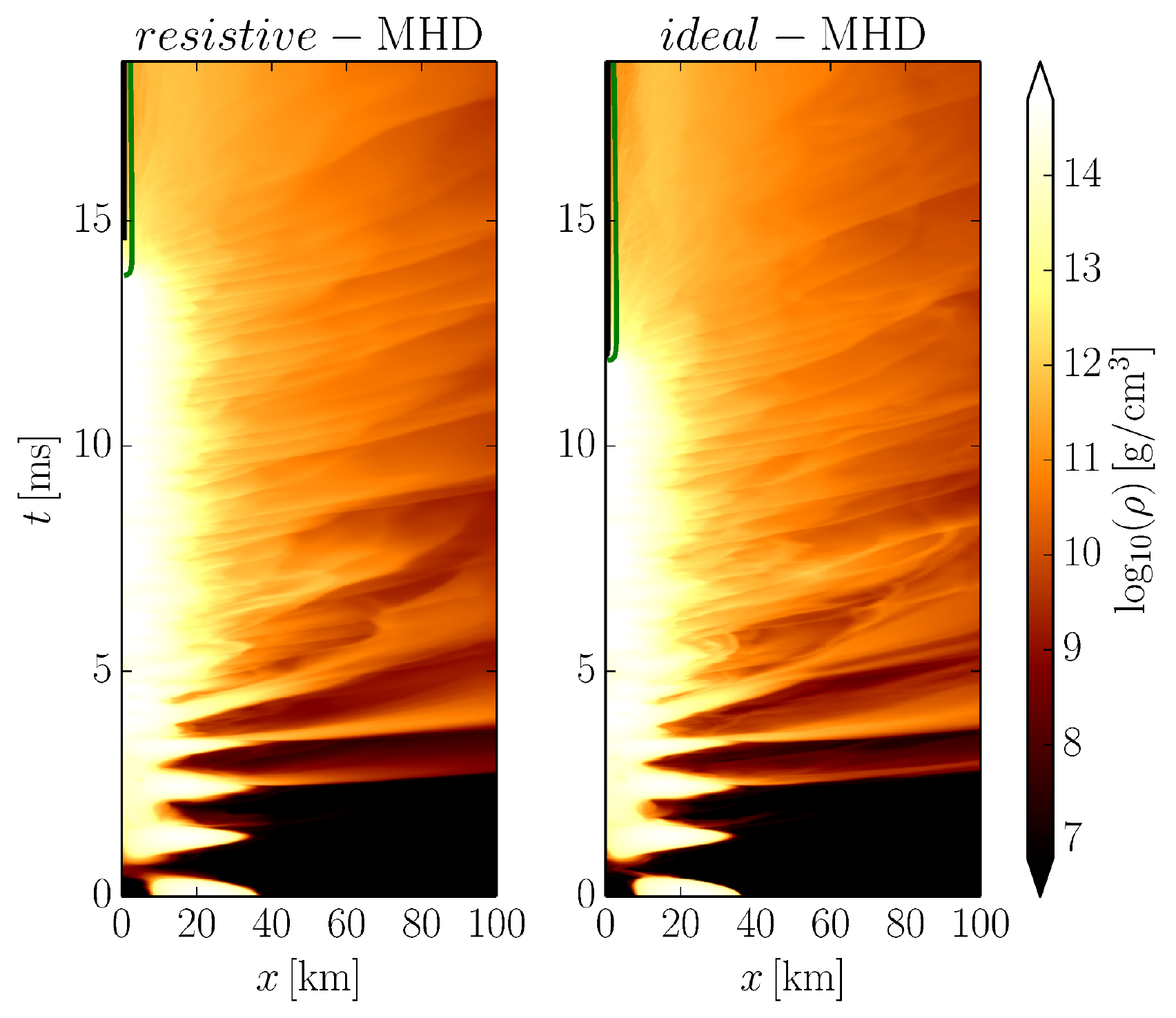}
\includegraphics[width=0.4\textwidth]{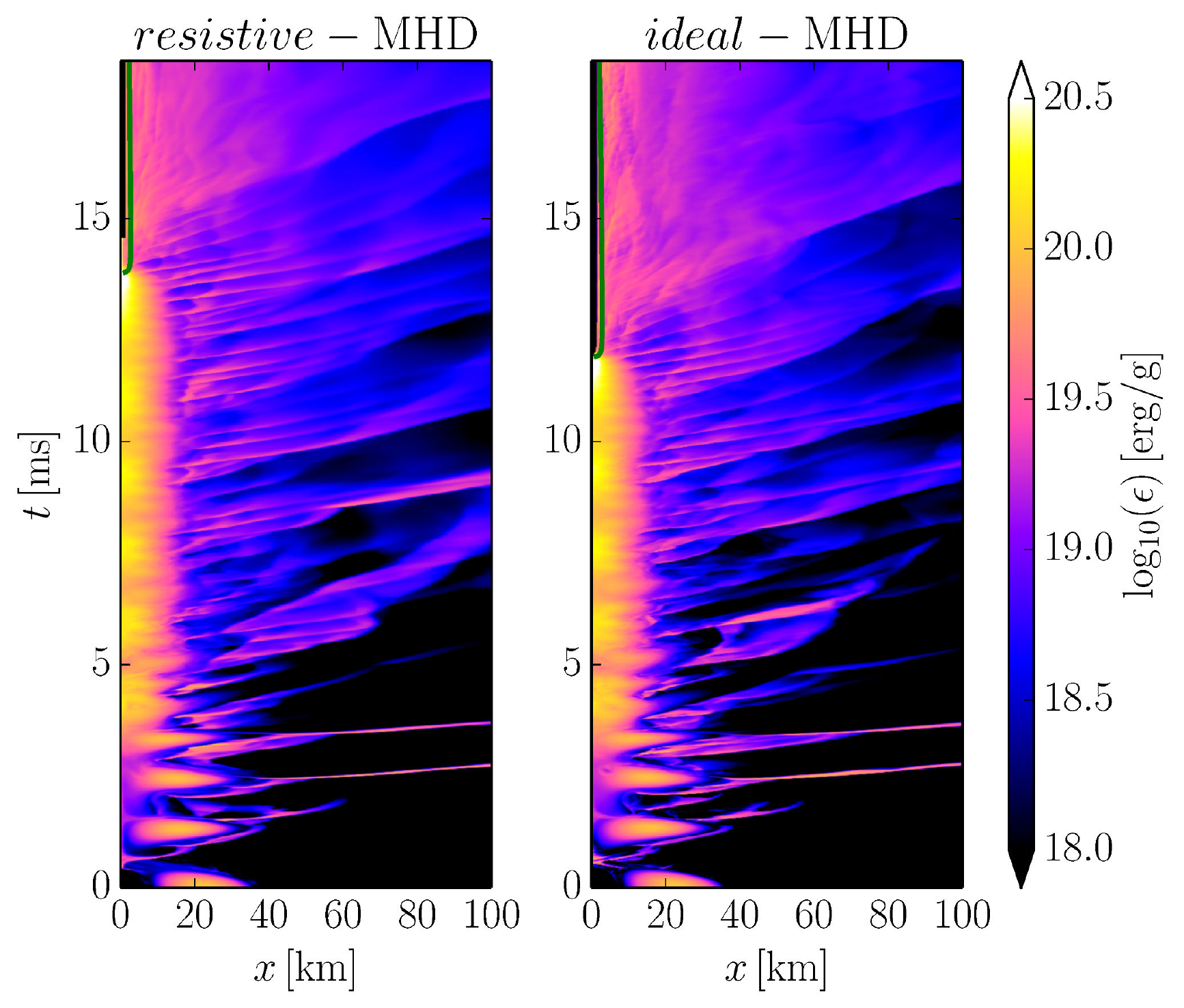}
\includegraphics[width=0.4\textwidth]{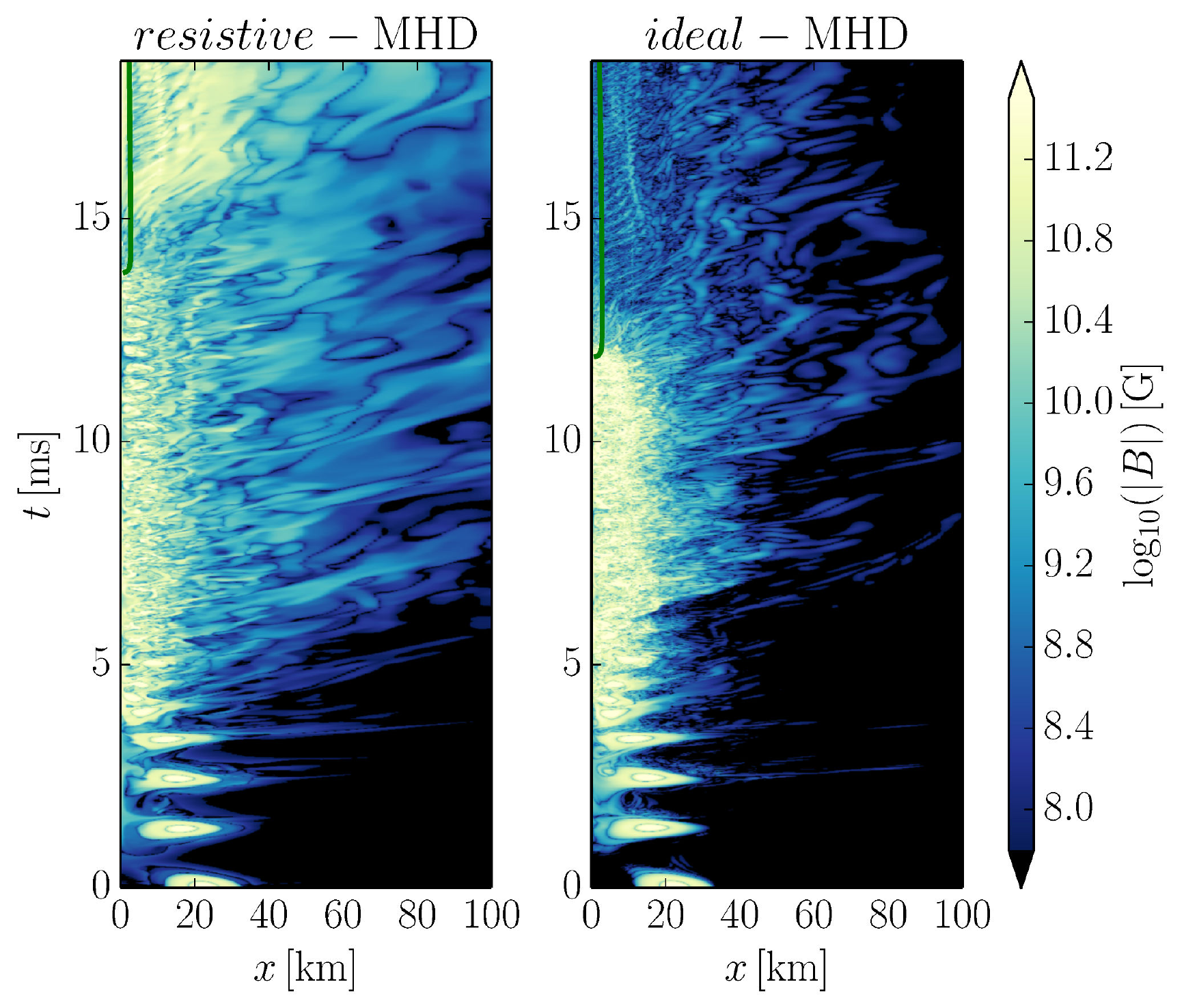}
\caption{Two-dimensional spacetime representation of the evolution of the
  rest-mass density $\rho$ (left panel), of the specific internal energy
  $\epsilon$ (middle panel), and of the magnetic field norm $|B|$ (right
  panel). All quantities are measured along the $x$-axis and are therefore
  representative of motions on the equatorial plane. Indicated with a
  solid green line is the evolution of the apparent horizon; note that
  the values in the color bars are saturated and do not correspond to the
  minimum and maximum values of the corresponding fields.}
\label{fig:bns:spt_far1}
\end{figure*}

When studying the magnetic-field topology, \ie whether it is mostly
poloidal or toroidal, it is unavoidable to discuss where certain
measurements are made, as the magnetic field can be at the same time
mostly poloidal and mostly toroidal but in two different regions. A quick
inspection of the three panels suggests that the magnetic field in the
low-density funnel is predominantly poloidal (clearly shown in the middle
panel of Fig.~\ref{fig:bns:magtorpol}). It is quite natural to expect
that the magnetic field will be essentially poloidal near the rotation
axis, where matter has a specific angular momentum that is intrinsically
small. Equally natural is to expect that a toroidal component will start
to develop away from the axis and as one approaches the regions filled by
the torus. This behavior can be explained by the fact that some of the
magnetic-field lines in the funnel are anchored to the highly conducting
material at the edges of the torus, which is rotating at nearly Keplerian
velocities. Indeed, the left panel of Fig.~\ref{fig:bns:magtorpol} shows
that the magnetic field acquires a toroidal component near the edges of
the funnel, and then becomes essentially toroidal in the torus. This is
also shown in Fig.~\ref{fig:bns:mag:fieldlines}, which offers
three-dimensional snapshots of the rest-mass density $\rho$ (left panel),
of the specific internal energy $\epsilon$ (middle panel), and of the
modulus of the magnetic field $|B|$ (right panel) at $t=18.3\,\ms$. Also
reported are the magnetic-field lines, of which the three-dimensional
representation confirms that the magnetic field is mostly poloidal in the
magnetic-jet structure, acquiring a twist and a kink when it reaches the
edges of the funnel, and becoming essentially toroidal inside the torus.

Finally, we note that a vigorous outflow develops at the interface
between the magnetic-jet structure and the torus. The resulting shearing
boundary layer could be the site for the development of a
Kelvin--Helmholtz instability, which unfortunately we cannot investigate
at the present resolutions and without a more sophisticated treatment of
the conductivity in the transition between large and small values (the
instability does potentially develop in a particularly difficult
region). It is clear, however, that the dynamics along the torus walls
has all the potential of yielding interesting observational features and
should be investigated in the future, possibly using advanced numerical
techniques such as those presented in Ref. \cite{Duffell2013}.

A word of caution should be spent before concluding this section. While
the behavior described above appears reasonable and is possibly the
expected one on the basis of rather simple considerations, this behaviour
is ultimately the result of our choice for the conductivity profile in
Eq. \eqref{eq:bns:conductivityprofile} and of our choice of rest-mass
density in the atmosphere. The large computational costs associated with
these simulations prevent us from presenting at this point a systematic
investigation of how sensitive the results are on the choice for $\sigma$
and $\rho_{\rm atm}$. We are aware that this represents a limitation of
our investigation, which we plan to resolve with future simulations.

\subsection{Comparison with IMHD simulations}
\label{sec:bns:comparison}

\begin{figure*}
\centering
\includegraphics[width=0.4\textwidth]{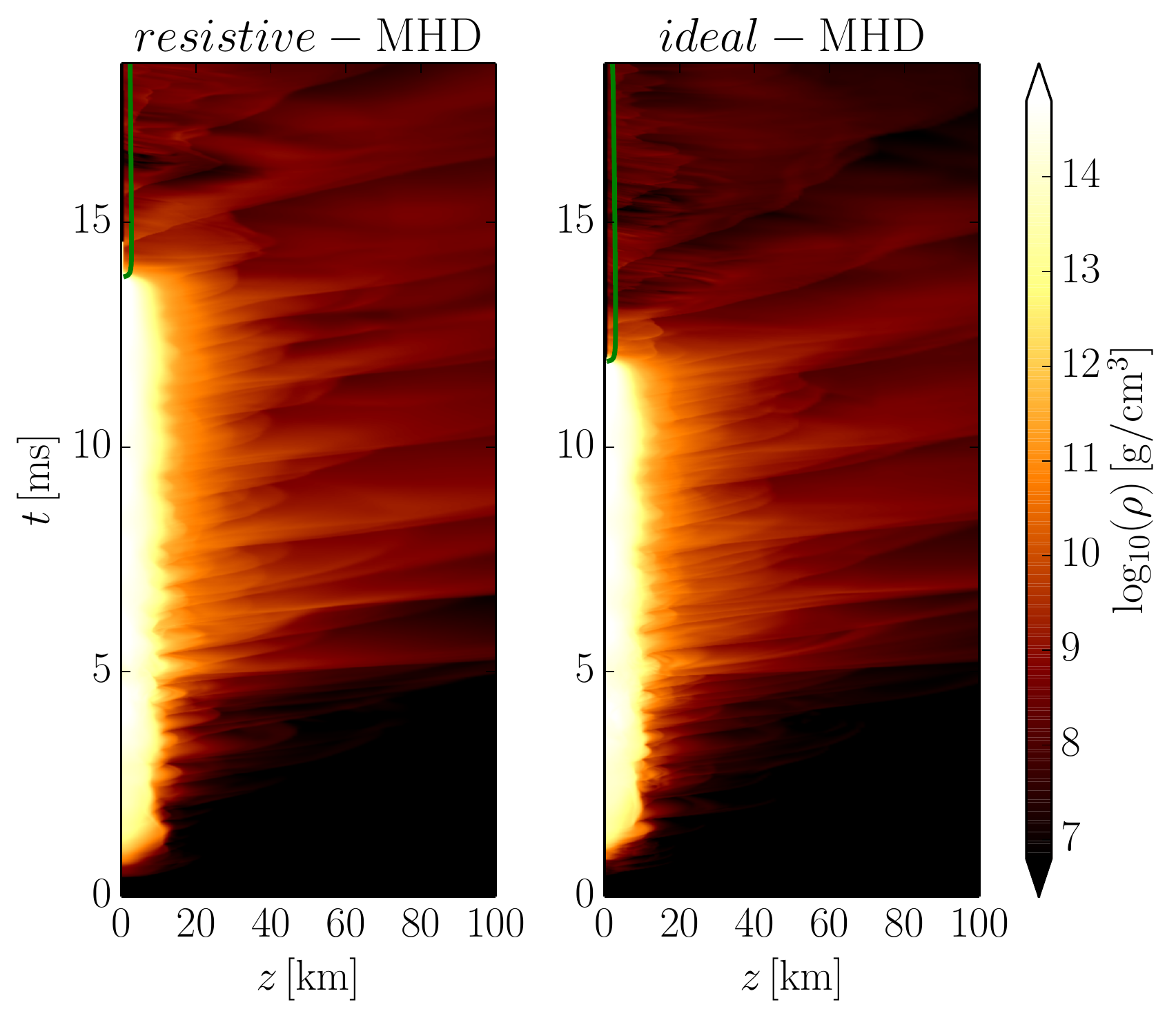}
\includegraphics[width=0.4\textwidth]{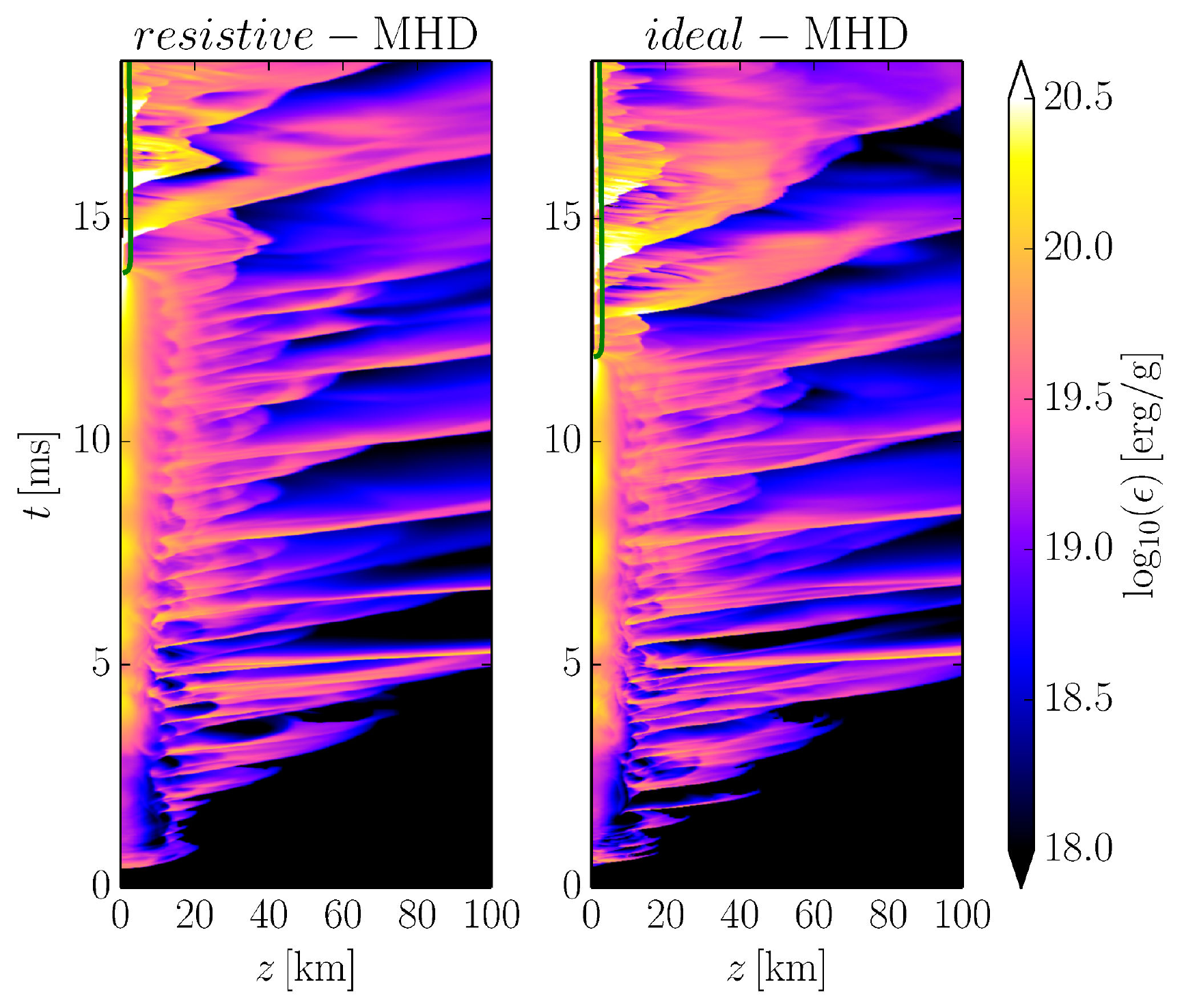}
\includegraphics[width=0.4\textwidth]{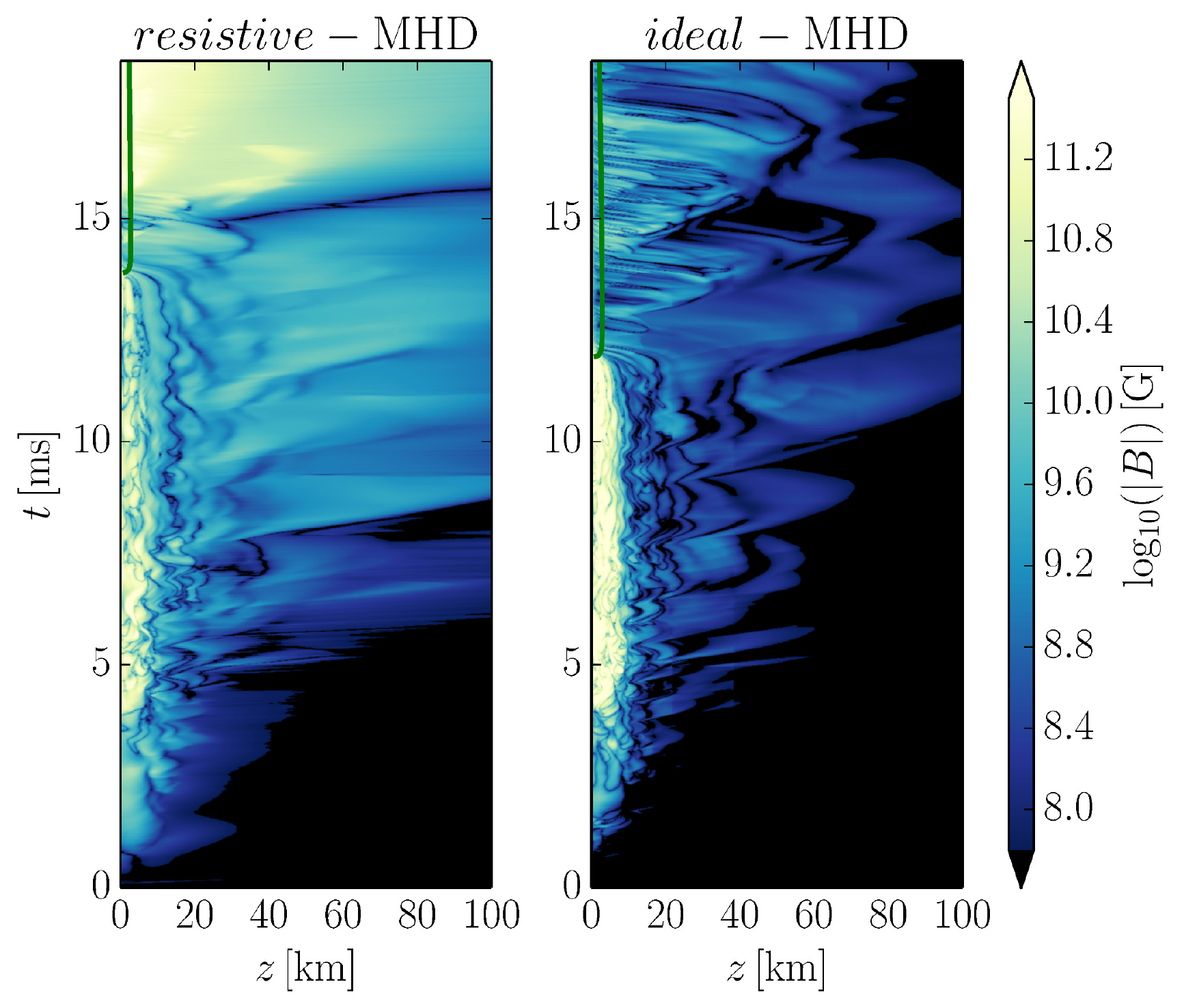}
\caption{The same as in Fig. \ref{fig:bns:spt_far1} but for quantities
  measured along the $z$-axis and therefore representative of motion in
  the polar regions.}
\label{fig:bns:spt_far2}
\end{figure*}

As described in the previous section, the dynamics of the RMHD
simulations is rather similar, at least qualitatively, to the one
observed in IMHD simulations of the same binary presented in
Ref.~\cite{Giacomazzo:2010}. Yet, there are some important differences, and
these can be best appreciated if we perform a careful comparison of the
two evolutions. To this scope, we have performed additional simulations
of the same binary discussed in the previous section when, however, the set
of equations solved are those of general-relativistic IMHD. We note that
this was necessary because the simulations in Ref.~\cite{Giacomazzo:2010} used
a different grid structure and resolution but also investigated
quasicircular initial data in contrast to the reduced linear momenta we
have considered here.

In the following we present the results of this side-by-side comparison,
first using two-dimensional spacetime diagrams and then moving to
standard one-dimensional snapshots.

\subsubsection{Spacetime diagrams}
\label{sec:spd}

We have found that a very efficient way of carrying out a comparison
between two MHD evolutions which are qualitatively similar is to use
two-dimensional spacetime diagrams. This technique, which was first
introduced in Ref.~\cite{Rezzolla:2010}, provides a color-coded evolution of
various scalar quantities as measured along principal axes (\eg the $x$-
and $z$-directions) and has the advantage of summarizing simply even
rather complex dynamics.

As representative examples, we show in Figs.~\ref{fig:bns:spt_far1} and
\ref{fig:bns:spt_far2} the differences between the RMHD and IMHD
implementations in the evolution of the rest-mass density $\rho$ (top left
panel), of the specific internal energy $\epsilon$ (top right panel) and of
the magnetic field norm $|B|$ (bottom panel). Note that each panel is
split into two diagrams, with the left one referring to the RMHD solution
and the right one to the IMHD solution. Furthermore, while
Fig.~\ref{fig:bns:spt_far1} refers to quantities measured along the
$x$-axis and hence is representative of motions on the equatorial plane,
Fig.~\ref{fig:bns:spt_far2} refers to the $z$-axis and is therefore
representative of motions in the polar region. In all panels we indicate
with a solid green line the evolution of the apparent horizon (see
the discussion in Ref. \cite{Baiotti04} on how to interpret such a
line). Note that the values in the color bars are saturated and do not
correspond to the minimum and maximum values of the corresponding fields.

It is clear from both figures that the evolution of all quantities is
very similar during the inspiral, when indeed the IMHD and RMHD
evolutions should be mathematically identical given that our fields are
contained in the stars. As the binary reaches the merger, matter is
expelled from the stars mainly in the equatorial plane, as is evident in
Fig.~\ref{fig:bns:spt_far1}, which also shows that the ejected matter
moving through the low-density medium generates shocks that heat up the
plasma and appear as thick lines. The subsequent bursts of matter are
mainly associated with the fundamental mode ($f$-mode) of oscillation of
the HMNS~\cite{Stergioulas2011b,Takami2015}. The equatorial ejections are
also accompanied by four or five subsequent ``bursts'' along the
$z$-axis, starting in both simulations at approximately $5\,\ms$, and can
be seen in Fig.~\ref{fig:bns:spt_far2}. Clearly, the violent oscillations
experienced by the HMNS launch matter essentially isotropically.

We have already mentioned that the differentially rotating HMNS does
collapse promptly to a black hole by first rearranging its angular
velocity profile and mass distribution through magnetic braking. During
this phase, angular momentum is transported outward, with the outer
fluid elements moving further away from the star and the inner ones
moving toward the center as a result of the angular-momentum
losses. Clearly, this redistribution of angular momentum will be
different in the RMHD and IMHD evolution, with the latter being more
efficient in transporting angular momentum outward (in IMHD the fluid
can only move along magnetic-field lines, while it can also partially
cross them in RMHD). As a result, the collapse can take place slightly
earlier, occurring at $11.9\,\ms$ in the IMHD evolution and at
$13.8\,\ms$ in the RMHD simulation (\cf
Figs.~\ref{fig:bns:spt_far1} and \ref{fig:bns:spt_far2}).

We have already commented that one of the major differences between
the IMHD and RMHD simulations is that our implementation of the latter
allows for a description of propagating electromagnetic waves in
vacuum. By contrast, the atmosphere treatment of the IMHD
implementation is such that it does not evolve the magnetic field as
the fluid velocities are reset to zero there. As a result, already
after the first $0.5\,\ms$ the magnetic field manages to diffuse out
of the stars, heating up the plasma in the outer layers and forcing it
to expand. In this way, magnetic energy is converted into internal
energy, and therefore the magnetic field at the center of the stars is
(slightly) lower than in the corresponding IMHD simulation. In
contrast, the magnetic field along the $z$-axis in the RMHD simulation
is higher than in the IMHD simulation in the first few
milliseconds. Indeed, the differences between the IMHD and RMHD
simulations become particularly evident after black-hole formation,
when the funnel is evacuated and the magnetic-jet structure is built
(see the bottom panels of Figs.~\ref{fig:bns:spt_far1} and
\ref{fig:bns:spt_far2}). The magnetic diffusivity in the RMHD
simulation acts so rapidly that in a bit more than one crossing time
the whole computational domain in the RMHD run is filled with
electromagnetic fields despite the fact that the magnetic field was
initially constrained to the stellar interior.\footnote{We prefer to be
  repetitive rather than confusing: the magnetic field diffuses out of
  the stellar matter because of the finite resistivity at the surface
  of the neutron stars, of the HMNS, or of the torus. Once in the
  atmosphere, however, the magnetic fields propagate as
  electromagnetic waves in vacuum.}

Note that the modulus of the magnetic field along the $z$-axis is about
2 orders of magnitude larger in the RMHD simulation (\cf $\sim
10^{10}\,\G$ in IMHD vs $\sim 10^{12}\,\G$ in RMHD). This is due
(mainly) to the intense currents produced by the rapidly rotating torus
and (partly) to the magnetic-field diffusion of the strong magnetic field
in the torus, which diffuses across the walls of the funnel. This
behavior of the magnetic field is in fact similar to the one observed in
a stable magnetized star with extended magnetic fields which was studied
in Ref.~\cite{Dionysopoulou:2012pp}, where after an initial transient, the
system relaxed to a solution consisting of a large-scale, nearly
electrovacuum, dipolar magnetic field configuration in the exterior,
which was anchored to the highly conducting neutron star. The larger
values of the magnetic field in the funnel are particularly encouraging
as strong magnetic fields are necessary to produce a large acceleration
along the $z$-direction. The maximum Lorentz factors achieved after the
collapse in both simulations is $W \approx 2.0-2.7$, with the highest
values occurring at the end of the simulations.

The magnetic field in the torus is also stronger in the RMHD simulation
than the corresponding field in the IMHD run, although the differences in
this case are only of 1 order of magnitude. The reason for this has to
be found in the fact that in the RMHD simulation the torus is more
massive and therefore able to sustain larger amounts of magnetic fields
(as for isolated stars, also a self-gravitating torus in MHD equilibrium
will be able to sustain stronger magnetic fields for increasing masses;
\cf Table \ref{tab:bns:bhtorus}).

\subsubsection{Angular-momentum transfer and HMNS lifetime}
\label{sec:bns:lifetime}

\begin{figure}
\centering 
\includegraphics[width=0.98\columnwidth]{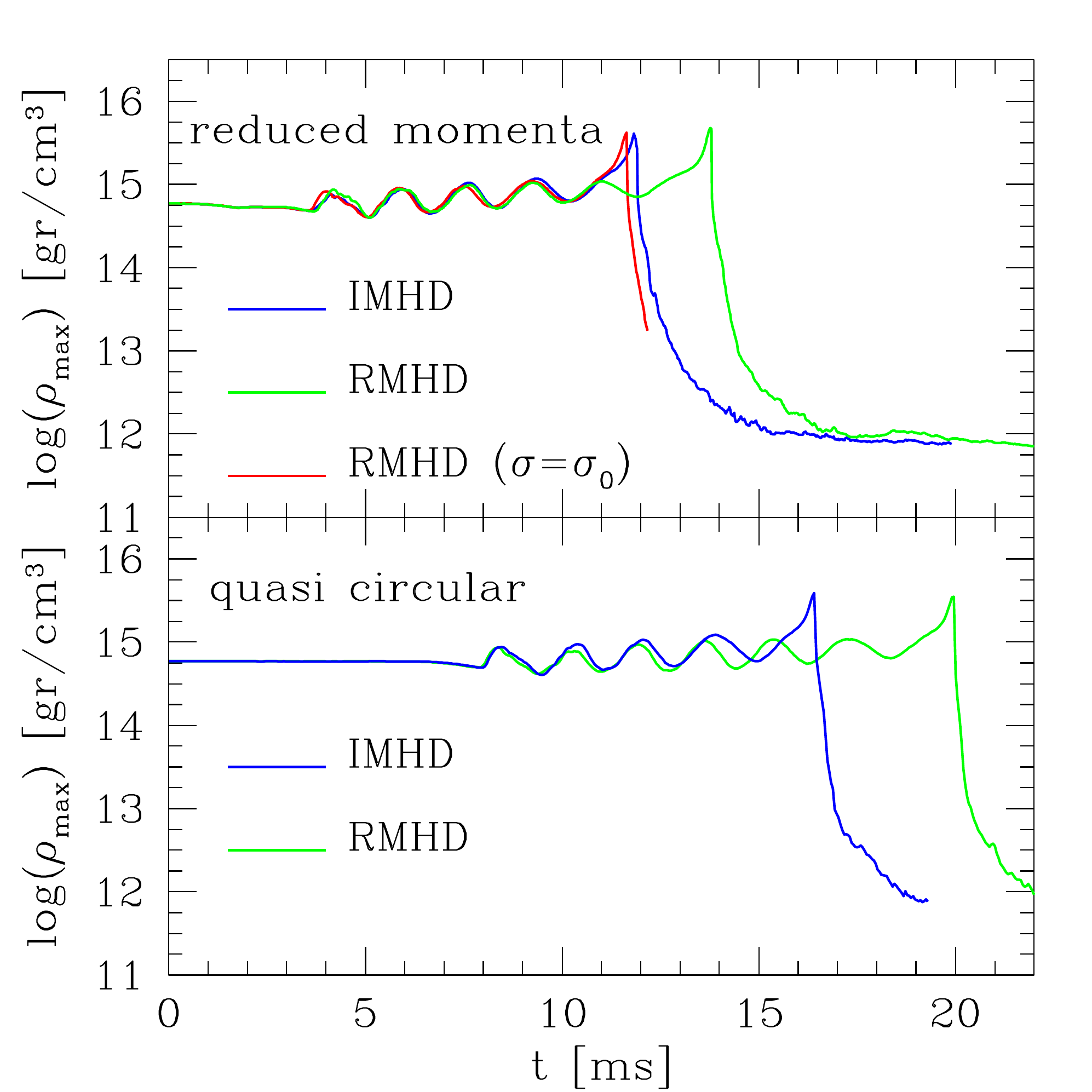}
\caption{Evolution of the maximum of the rest-mass density for both the
  IMHD (blue solid line) and the RMHD simulation (green solid line). The
  top panel refers to the initial data with reduced linear momenta, while
  the bottom one refers to data in quasicircular orbits. Also shown in the top
  panel is the evolution of the RMHD set of equations with a large and
  uniform conductivity (red solid line); in this case the evolution
  should mimic the IMHD one, as indeed it does.}
\label{fig:bns:rho}
\end{figure}

\begin{figure*}
\centering
\includegraphics[width=0.85\textwidth]{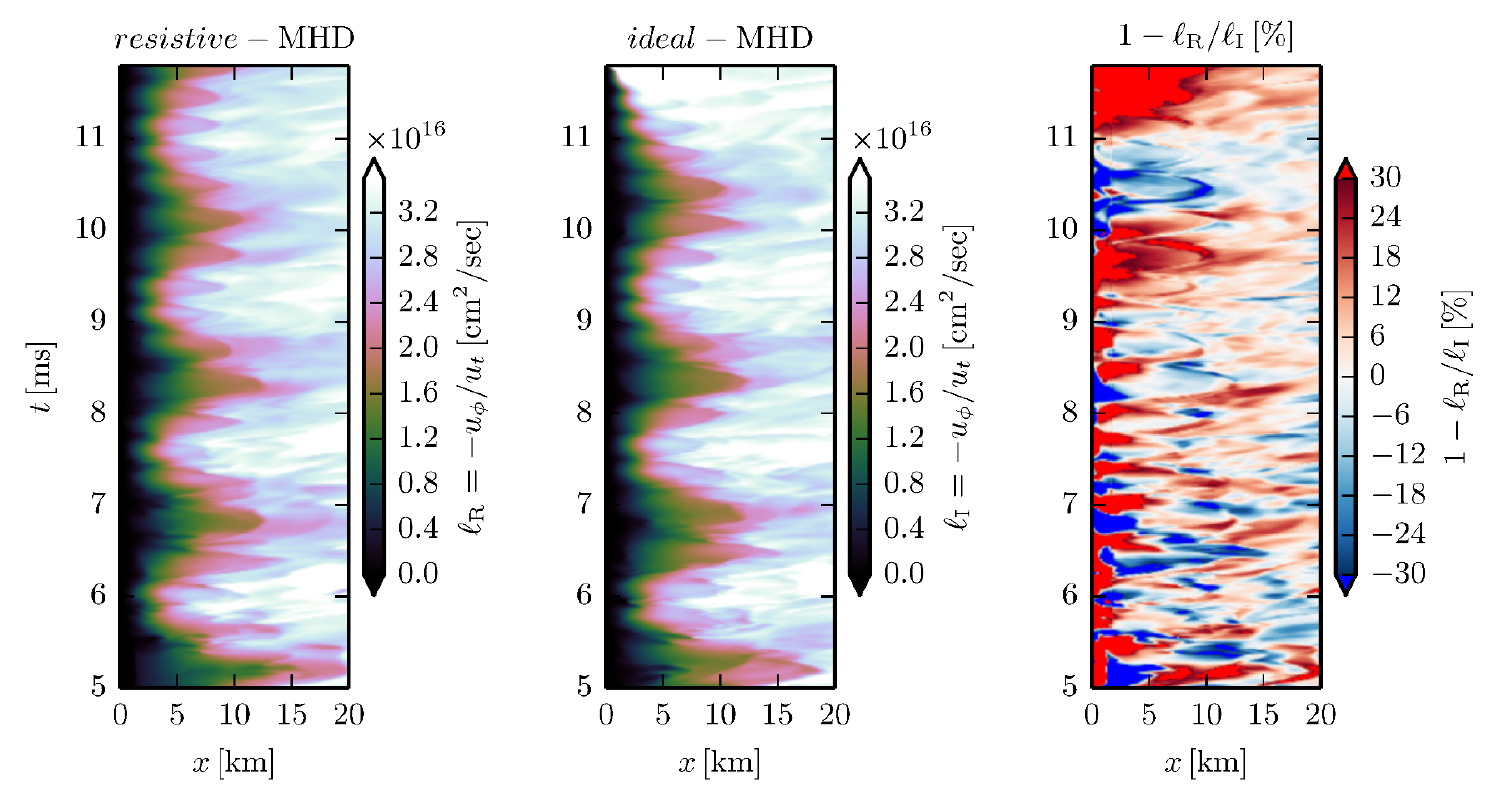}
  \caption{Spacetime diagrams of the specific angular momentum
    ${\ell}:=-u_\phi/u_t$ for both the RMHD (left panel) and IMHD (middle
    panel) simulations. In addition, we show the relative difference in
    the specific angular momenta between the RMHD and IMHD
    implementations (right panel).}
  \label{fig:bns:angularmomentum}
\end{figure*}

The top panel of Fig.~\ref{fig:bns:rho} reports the evolution of the
maximum of the rest-mass density $\rho_{\rm max}$ for both the IMHD (blue
solid line) and the RMHD simulation (green solid line). After the merger
takes place at $t\simeq 3.5\,\ms$, the maximum rest-mass density
oscillates at the $f$-mode frequency and experiences a sudden drop when
an apparent horizon is found since matter inside the apparent horizon is
excluded from the calculation of $\rho_{\rm max}$. It is quite obvious
that the behavior of $\rho_{\rm max}$ is different in the two
simulations, with the HMNS in the RMHD surviving for a longer time. Note,
however, that the first increase in the central rest-mass density in the
top panel of Fig.~\ref{fig:bns:rho} is slightly larger in the RMHD case,
most likely because of the resistive increase of the internal energy at
the merger.

Also shown in the top panel of Fig.~\ref{fig:bns:rho} is
the evolution of a simulation in which the set of RMHD equations is used
together with a large and uniform conductivity $\sigma=\sigma_0=10^6$
(red solid line). In this case, despite the very different set of
equations solved and the different approach for enforcing the
divergence-free condition of the magnetic field, the RMHD evolution
should mimic the IMHD one. This is indeed the case, and it provides
considerable confidence on the robustness of our RMHD approach. It
suggests that the delayed collapse is an effect associated with the
choice of physical resistivity and not due to numerical artifacts.

Despite the complex dynamics, the differences between the IMHD and RMHD
runs are not difficult to explain. As mentioned in the previous section,
in fact, an important difference between the IMHD and RMHD simulations is
that in the latter the magnetic field cannot be perfectly locked with the
plasma. As a result, the IMHD evolution is more efficient in
redistributing the angular momentum in the system and, in particular, in
transporting it outward. This magnetic-braking process deprives the
HMNS core of the angular-momentum support, and this leads to an earlier
collapse. Clearly, since the conductivity in the HMNS interior is very
high also in the RMHD simulation (although not infinite), magnetic flux
freezing is very efficient here as well and the differences in the
dynamics of the IMHD and RMHD simulations can only be small. This
explains why overall the time of collapse varies by only $\sim
1.9\,\ms$. In addition, the important differences in this dynamics are
also expected to take place in the outer layers of the HMNS, where the
conductivity decreases as a response to the conductivity
profile~\eqref{eq:bns:conductivityprofile}. We should note that the
difference in the lifetime of the HMNS is not related to the use of
initial data with modified momenta, but is present also for a binary
of which the initial data is on a quasicircular orbit. This is shown in the
bottom panel of Fig.~\ref{fig:bns:rho}, where the corresponding
quantities are shown, and where it is clear that also in this case the
HMNS collapses earlier to a black hole in the IMHD simulation. We also
recall that for the binary in quasicircular orbit no additional
refinement level is added after the merger. Hence, when comparing in
Fig. \ref{fig:bns:rho} the RMHD evolution of the binaries with reduced
momenta and on quasicircular orbits, one is also effectively comparing
the evolution of the HMNS at two different resolutions, \ie $148$ and
$296~\m$, respectively. The fact that they both yield a longer lifetime
of the HMNS provides an indirect validation of the numerical consistency
of the RMHD solution.

Because the differences in the magnetic braking between the RMHD and IMHD
implementations are intrinsically small, it is not easy to show that it
is exactly these differences that are responsible for the earlier
collapse of the HMNS in the IMHD simulations. However, such evidence is
offered in Fig.~\ref{fig:bns:angularmomentum}, which reports the
spacetime diagrams along the $x$-direction of the specific angular
momentum ${\ell} : = -u_\phi/u_t$ \cite{Rezzolla_book:2013}.\footnote{A
  very similar behaviour is observed if the specific angular momentum is
  shown in terms of ${\ell} : = h u_\phi$ (see \cite{Rezzolla_book:2013}
  for a discussion in the differences in the two definitions).} The left
panel refers to $\ell_R$, the specific angular momentum of the RMHD run,
while the middle panel refers to the corresponding quantity for the IMHD
simulation, $\ell_I$, and the right panel refers to the relative difference: $1
- \ell_R/\ell_I$.

A rapid inspection of the left and middle panel reveals that in both the
RMHD and IMHD simulations the specific angular momentum increases
outward (as it should for a rotating fluid satisfying the Rayleigh
stability criterion), but also that the profiles are not constant in time
and show instead periodic variations that are in phase in the two
simulations. These variations reflect the large oscillations of the HMNS,
and, indeed, the oscillations in $\ell_R, \ell_I$ take place at the same
frequency as those in the rest-mass density and shown in the other
spacetime diagrams in Fig.~\ref{fig:bns:spt_far1} (note the different
scale in the $x$-axis). However, there are important small differences in
the dynamics of $\ell_R$ and $\ell_I$, which are apparent in the right
panel of Fig.~\ref{fig:bns:angularmomentum}, where regions in red
indicate that the specific angular momentum in these regions is higher
(of $\sim 20-30\%$) in the IMHD simulation than in the RMHD simulation,
while blue regions are exactly the opposite. It is apparent that the
differences are larger in the central parts of the HMNS, while the
specific angular momenta are very similar in the outer layers, \ie for
$x\gtrsim 5\,\km$. We recall that an excess of specific angular momentum
at a given position on the $x$-axis reflects fluid elements that are
rotating at a larger frequency ($\ell = \Omega x^2$ in the Newtonian limit)
and this is indeed what one would expect if the magnetic fields and the
fluid are tightly coupled. Hence, the red regions in the right panel
Fig.~\ref{fig:bns:angularmomentum} can be taken to signal a more
efficient transfer of angular momentum from the inner regions of the
HMNS. A direct consequence of this transfer of angular momentum is the
appearance of blue regions adjacent to the red ones and signalling
therefore fluid elements that have slowed down.

Although the differences in $\ell_R$ and $\ell_I$ are small, the transfer
continues steadily and with an increased rate up to $t \approx 11\,\ms$,
when a much larger transfer of angular momentum takes place. This signals
the onset of the instability to gravitational collapse in the IMHD
simulation, which effectively takes place soon after, \ie at $t \approx
12\,\ms$. 

Three remarks should be made before concluding this section. First, the
fact that an RMHD simulation with a uniform conductivity yields the same
collapse time as an IMHD simulation gives us great confidence about the
correctness of the RMHD evolution with nonuniform conductivity. Second,
although the difference in the survival time between the RMHD and IMHD
evolution is here rather small, it can be much larger if smaller values
of the resistivity are chosen for the stellar interior and for less
massive HMNSs; unfortunately present astronomical observations do not set
any stringent constraint on the values of the conductivity at these
temperatures, rest-mass densities, and magnetic fields. More importantly,
however, a longer survival time is a useful new result in the modelling
of binary neutron stars, as it points out that the HMNS can survive on
comparatively longer time scales than those computed so far in pure
hydrodynamics or in IMHD simulations. This is not a minor detail as the
most recent modelling of SGRBs with an extended x-ray emission invokes
the existence of a magnetized HMNS that is able to survive on time scales
of the order of $10^3-10^4\,\sec$ before collapsing to a black hole
\cite{Zhang2001, Metzger2008, Metzger:2011, Bucciantini2012,
  Rezzolla2014b, Ciolfi2014}. Finally, computational constraints have
prevented us from extending much past black-hole formation the evolution
of the RMHD/IMHD simulations with initial data on quasicircular
orbits. Nevertheless, the fact that already the dynamics of the HMNS is
unaffected by the initial reduction in linear momenta and that the HMNS
collapses to a black hole earlier in both cases, provides us with
confidence that the magnetic field dynamics discussed in
Sec. \ref{sec:bns:magjet} will be very similar also for binaries on
quasicircular orbits. This will also be the focus of our future work.

\begin{figure*}
\centering
\includegraphics[width=0.32\textwidth]{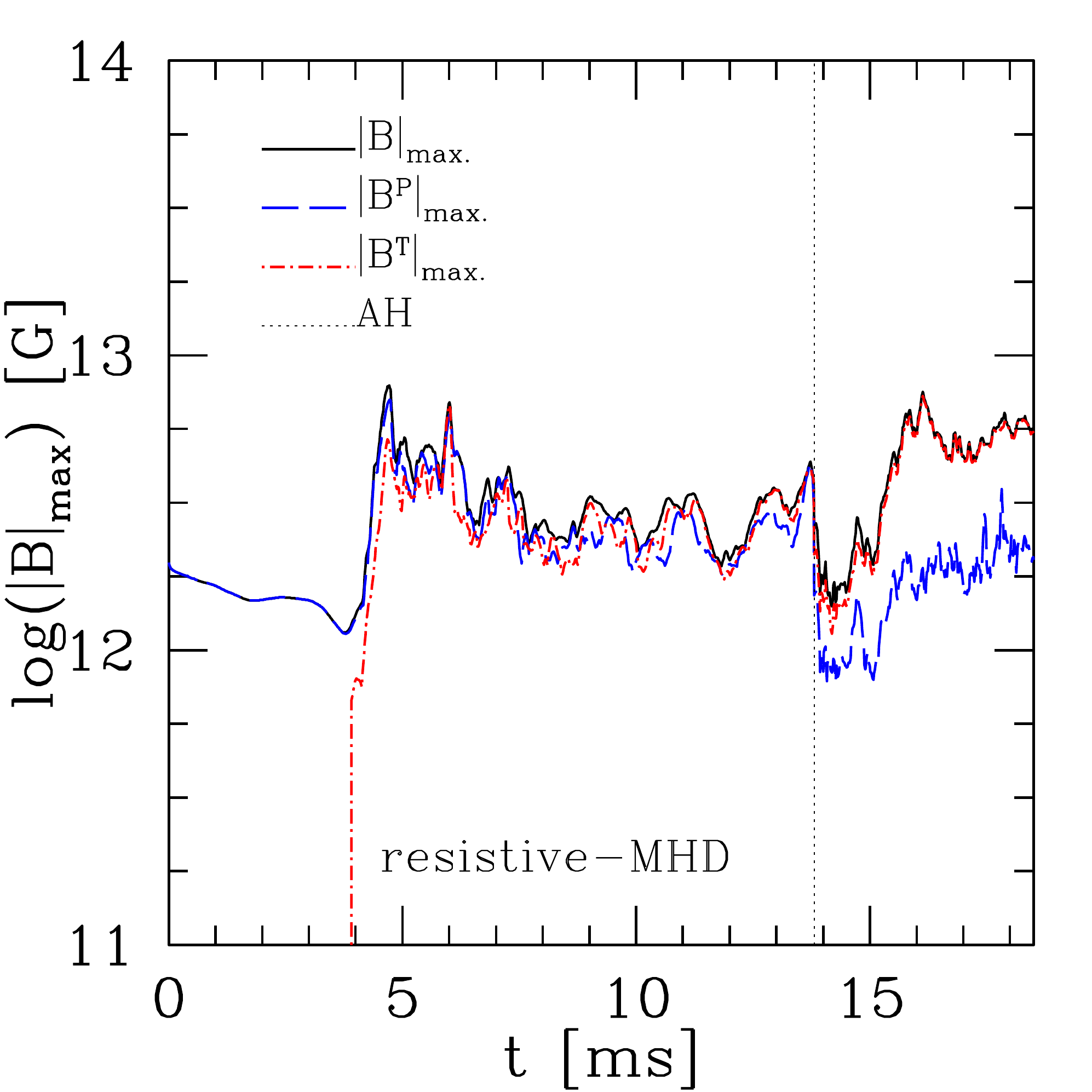} 
\includegraphics[width=0.32\textwidth]{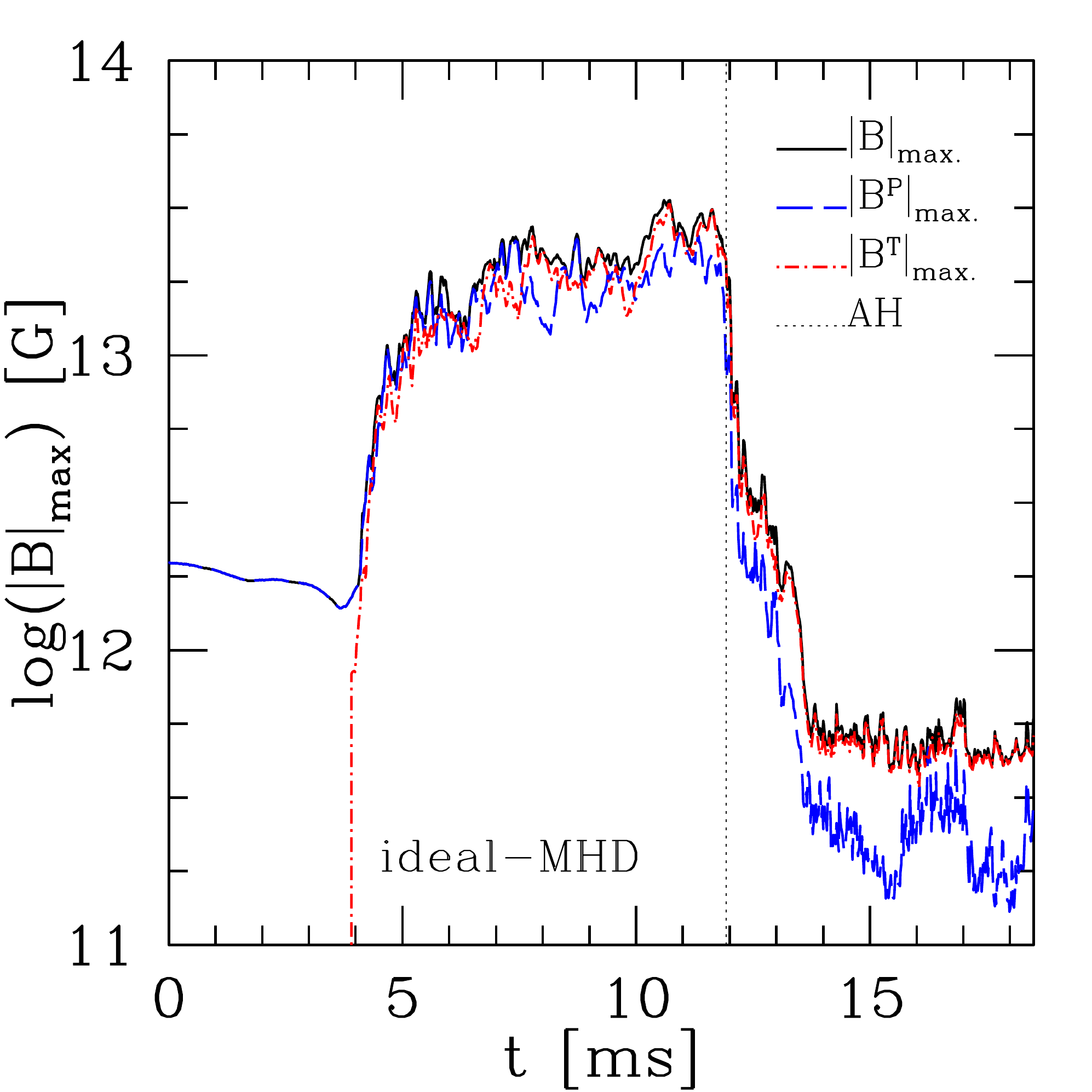}
\includegraphics[width=0.32\textwidth]{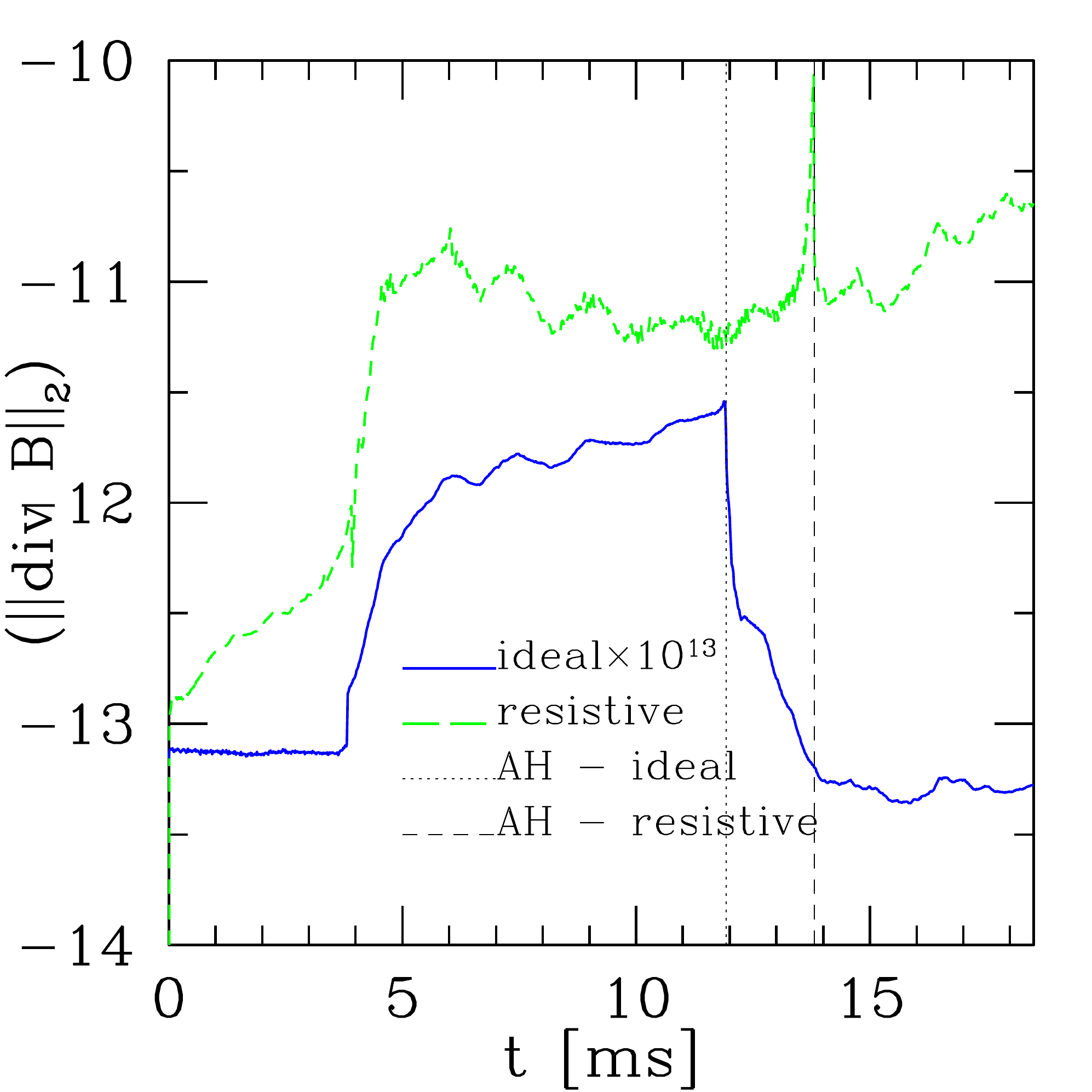}
\caption{Evolution of the maximum of the magnetic-field strength
  $|B|_{\rm max}$ for both the RMHD (left panel) and the IMHD simulation
  (middle panel). The black solid lines correspond to the time series of
  the maximum magnetic-field modulus; the blue dashed lines correspond to the
  time series of the maximum poloidal magnetic-field norm, $|B|_{\rm
    max}$; and the red dotted-dashed lines correspond to the maximum toroidal
  magnetic field, $|B^T|_{\rm max}$. Additionally, we show the
  $L_2$-norm of the divergence of the magnetic field, for the RMHD
  simulation (green dashed line) and IMHD one (blue solid line). The
  collapse times are depicted with black dotted or black dashed lines.}
  \label{fig:bns:mag}
\end{figure*}

\subsubsection{magnetic field growth}
\label{sec:mfd}

In Secs. \ref{sec:bns:magjet} and \ref{sec:spd}, we have already discussed
the properties of the evolution of the magnetic fields, but have not
quantified in detail how the magnitude of the magnetic field changes with
time and how this evolution varies in the RMHD and IMHD simulations. This
is done now in Fig.~\ref{fig:bns:mag}, where we report the evolution of
the maximum of the modulus of the magnetic field (black lines) but also
of its toroidal (red lines) and poloidal (blue lines) components, either in
the RMHD simulation (left panel) or in the IMHD simulations (middle
panel).

Note that the evolution of the magnetic field in the two implementations
is very similar during the inspiral, but also that this changes
considerably after the merger. While in both cases the toroidal magnetic
field grows exponentially, the growth is of about 1 order of magnitude
in the IMHD simulation but of a factor 2 smaller for the RMHD
simulation. Furthermore, the magnetic field reaches values of $3\times
10^{13}\,\G$ just before the collapse in the IMHD simulation and
$8.3\times 10^{12}\,\G$ only just after the merger in the RMHD
simulation. This different behavior is not difficult to explain and is
simply due to the fact that the shearing of magnetic-field lines is less
efficient in RMHD because of the finite conductivity of the
matter. Because the growth at the merger is mostly due to the shear layer
between the two impacting stars, it is quite natural that a resistive
calculation will lead to a smaller magnetic field, quite independently of
how well the instability is resolved.

As for the rest-mass density (\cf Fig. \ref{fig:bns:rho}), the collapse
of the HMNS to a black hole leads to a rapid decrease of the maximum
value of the magnetic field, as shown in Fig.~\ref{fig:bns:mag} where the
vertical lines signal the first appearance of the apparent horizon. Also
in this case, the strongest magnetic fields are hidden inside the horizon,
and the maximum values reported are those relative to the magnetic field
in the torus. As remarked already when commenting on the spacetime diagrams
in Sec. \ref{sec:spd}, the larger values of the magnetic field in the
RMHD simulation are the result of a more massive torus produced in this
case (\cf Table \ref{tab:bns:bhtorus}).

In the right panel of Fig.~\ref{fig:bns:mag} we complement the evolution
of the magnetic fields with the evolution of the $L_2$-norm of the
divergence of the magnetic field for the RMHD simulation (green dashed
line) and for the IMHD one (blue solid line) after it is multiplied by
$10^{13}$. We recall that the IMHD simulation makes use of a constrained
transport scheme \cite{Toth2000}, and hence it is able to maintain the
violations of this constraint down to machine precision. The RMHD
simulation, on the other hand, makes use of a divergence-cleaning scheme
\cite{Dedner:2002}, which is widely known to be less efficient in
suppressing the violations. Yet, the purpose of making this comparison is
mostly that of highlighting that the divergence-cleaning approach used
here may not be as efficient as the constrained transport, but it yields
nevertheless very small violations. Indeed, the evolution of the
  $L_2$-norm of the ratio between the divergence and the magnetic field
  strength, \ie $||\nabla_i B^i/ \sqrt{B_i B^i}||_2$ and not shown in
  Fig.~\ref{fig:bns:mag}, is of the order of $\sim 10^{-2}$, thus similar
  to the values reported by similar works in the
  literature~\cite{Palenzuela2013}. Note that the late-time moderate
growth of the divergence in the RMHD simulation is due to the
amplification of the magnetic fields in the torus, and it would be
interesting to investigate if a dynamically adapted dissipation parameter
for the divergence cleaning method could help reduce such a growth.

\subsubsection{Magnetically driven wind and bursting activity}
\label{sec:bns:bursts}

\begin{figure}
\centering
\includegraphics[width=0.8\columnwidth]{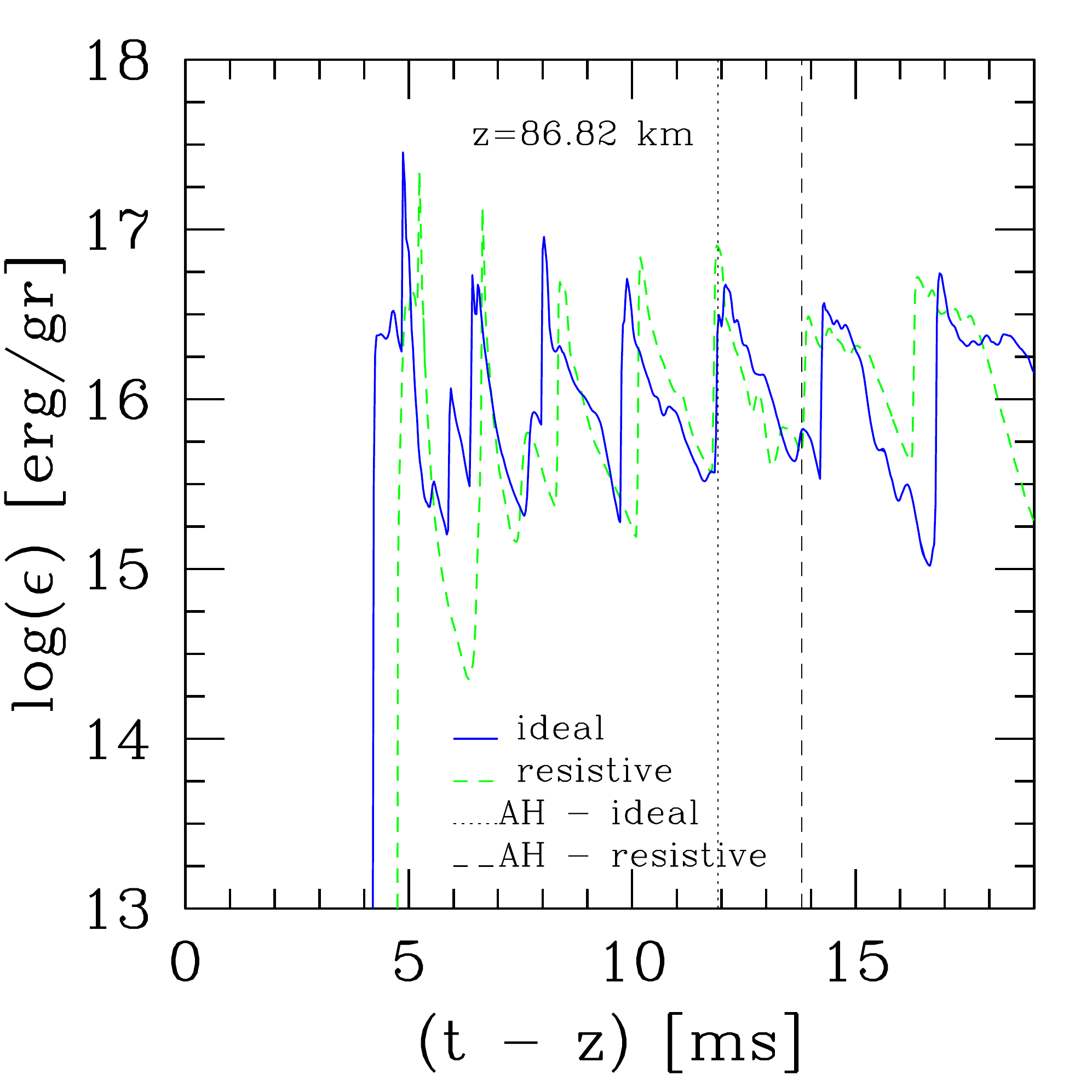}
\caption{The time series of the specific internal energy measured at
  $(x,\,y,\,z)=(0,\,0,\,86.82)\,\km$ are shown for the RMHD simulation
  with a green dashed line and for the IMHD simulation with a solid blue
  line. The time of collapse for the RMHD case is depicted with a black
  dashed line, while it is depicted with a black dotted line for the IMHD
  one.}
  \label{fig:bns:bursts}
\end{figure}

In Secs. \ref{sec:bns:features} and \ref{sec:spd}, we have already
anticipated that a wind is produced after the merger, either as a result
of shock heating at the merger or because of magnetic winding and
consequent pressure imbalance in the outer layers of the HMNS or because
of neutrino losses~\citep{Metzger2014,Perego2014}. In addition to an
almost quasistationary and quasi-isotropic wind, both the RMHD and the
IMHD simulation show the existence of mildly anisotropic and
quasiperiodic launching of low rest-mass density, high internal energy
blobs of matter that we will refer to as bursts. Overall, seven
bursts are launched during the total time of the simulation, with five
bursts relative to the HMNS stage and two being produced after black-hole
formation.

In particular, the first two bursts eject material that is moving through
the low-density atmosphere with an average speed of $\approx~
0.4-0.6$ but that decelerate as they move away from the HMNS, reaching a
final outward velocity of $\sim 0.16-0.18$. This behavior is in part due
to the natural conversion of kinetic energy to binding energy but also
to the interaction of the bursting material with the slow isotropic
wind. This interaction, which is obviously accompanied by shocks,
provides a damping mechanism on the propagation of the ejected
material. However, the slow down of the baryon rich material could be
revived later on if the slow wind is impacted by a faster, baryon poor
wind, as suggested in the ``two-winds'' model of
Ref.~\cite{Rezzolla2014b}.

We have investigated the properties of the bursts by studying the
evolution of the specific internal energy as measured by an observer on
the $z$-axis. This is reported in Fig.~\ref{fig:bns:bursts} for an
observer at $(x,\,y,\,z)=(0,\,0,\,86.82)\,\km$, for both the RMHD
simulation (green dashed line) and the IMHD one (blue solid
line). Clearly, after the merger the specific internal energy exhibits a
quasiperiodic behavior in both simulations, with a very rapid
increase. The increase in the specific internal energy is of slightly
less than 2 orders of magnitude and is followed by a slower decay.

As already mentioned in Sec. \ref{sec:spd}, this behavior can be
associated with the oscillations of the HMNS and is observed also in the
evolution of the rest-mass density (\cf the five peaks in
Fig~\ref{fig:bns:rho} and in Fig.~\ref{fig:bns:spt_far2}, top left
panel). The characteristic frequency of these peaks is related to the
$f$-mode frequencies of the bar-deformed HMNS~\cite{Stergioulas2011b,
  Takami2015} and thus these bursts occur every $\sim
1.7-2.0\,\ms$. Similar peaks (although less marked) can be observed also
in the magnetic field along the $z$-axis (\cf
Fig.~\ref{fig:bns:spt_far2}, bottom panel). The difference is that they
occur slightly later than the specific internal energy ones, possibly
indicating a conversion of kinetic energy into magnetic energy and
vice versa. The rest-mass density in the blobs of low-density material
ejected in the bursts depends on height, but, at a distance of
$z=86.82\,\km$ along the $z$-axis, it is $6\times 10^{6}\,\gr\,\cm^{-3} -
6\times 10^8\,\gr\,\cm^{-3}$, while the magnetic field, at the same
location, has a strength of $\sim 10^{9}-5 \times 10^{10}\,\G$.

The bursting activity continues also after the collapse, but with
somewhat different properties. First, the frequency is now set by the
radial epicyclic frequencies of the oscillating torus as deduced, for
instance, when analyzing the time series of the specific internal energy
at a $\sim 60\,\km$ on the $x$-axis (see
Refs. \cite{Zanotti03,Rezzolla:2010} for an introduction to these
frequencies in rotating tori). Second, the rest-mass density of the blobs
ejected is smaller and of the order of $5\times10^7\,\gr\,\cm^{-3}$,
while the magnetic field oscillates between $10^8-10^9\,\G$ and is
stronger probably as a result of the magnetic field increase in the
funnel.

With only two bursts observed after black-hole formation, the time span
is too short to reach a firmer conclusion on the origin of the bursts
after the collapse. However, all present evidence seems to suggest that
the postmerger bursts are triggered by an increased mass-accretion rate
as the torus approaches the black hole during the inward phase of its
epicyclic oscillation. Of course also resistive reconnection processes
could lead to the conversion of magnetic energy into internal energy and
thus may be invoked to explain this phenomenology. We find this not a
likely explanation mostly because of the ordered magnetic field structure
that builds up in the funnel and which seems rather stationary. Clearly,
a more detailed study of long-term evolutions with different
prescriptions for the conductivity profiles is necessary before reaching
more robust conclusions.

We finally note that the outflows produced either by shock-heating,
magnetically driven winds or by the periodic bursts, eject a substantial
amount of matter. More specifically, the total rest-mass flux across a
spherical surface located at $r=295.4\,\km$ is measured to be
$0.5-2.0\,M_{\odot}\,\sec^{-1}$, which amounts to a total of $\sim
0.01\,M_{\odot}$ ejected from the beginning of the simulation and over
the survival time of the HMNS (\ie $13.8\,\ms$).  We also note that the
mass-ejection rates reported here are larger than those obtained
in Ref.~\cite{Siegel2014}, where mass fluxes of $\sim
10^{-3}-10^{-2}\,M_{\odot}\,\sec^{-1}$ were reported. However, this is not particularly
surprising and for a number of reasons. First, the HMNS considered here
is the self-consistent result of a binary merger, while the one studied
in Ref.~\cite{Siegel2014} was built using an axisymmetric differentially
rotating equilibrium with a standard (but somewhat arbitrary) law of
differential rotation. Second, as mentioned in Sec. \ref{sec:id}, the
linear momenta in our initial data are artificially reduced to accelerate
the inspiral. This also leads to a more violent merger and to larger mass
losses at least till black-hole formation. After the HMNS collapses, in
fact, the mass flux saturates at $0.2\,M_{\odot}\,\sec^{-1}$. Finally,
the initial magnetic field in Ref.~\cite{Siegel2014} is about 2 orders of
magnitude larger, and this facilitates substantially the loss of MHD
equilibrium at the surface of the HMNS and thus the mass loss.

\subsubsection{Black hole--torus properties}
\label{sec:bns:torus}

\begin{figure}
\centering
\includegraphics[width=0.8\columnwidth]{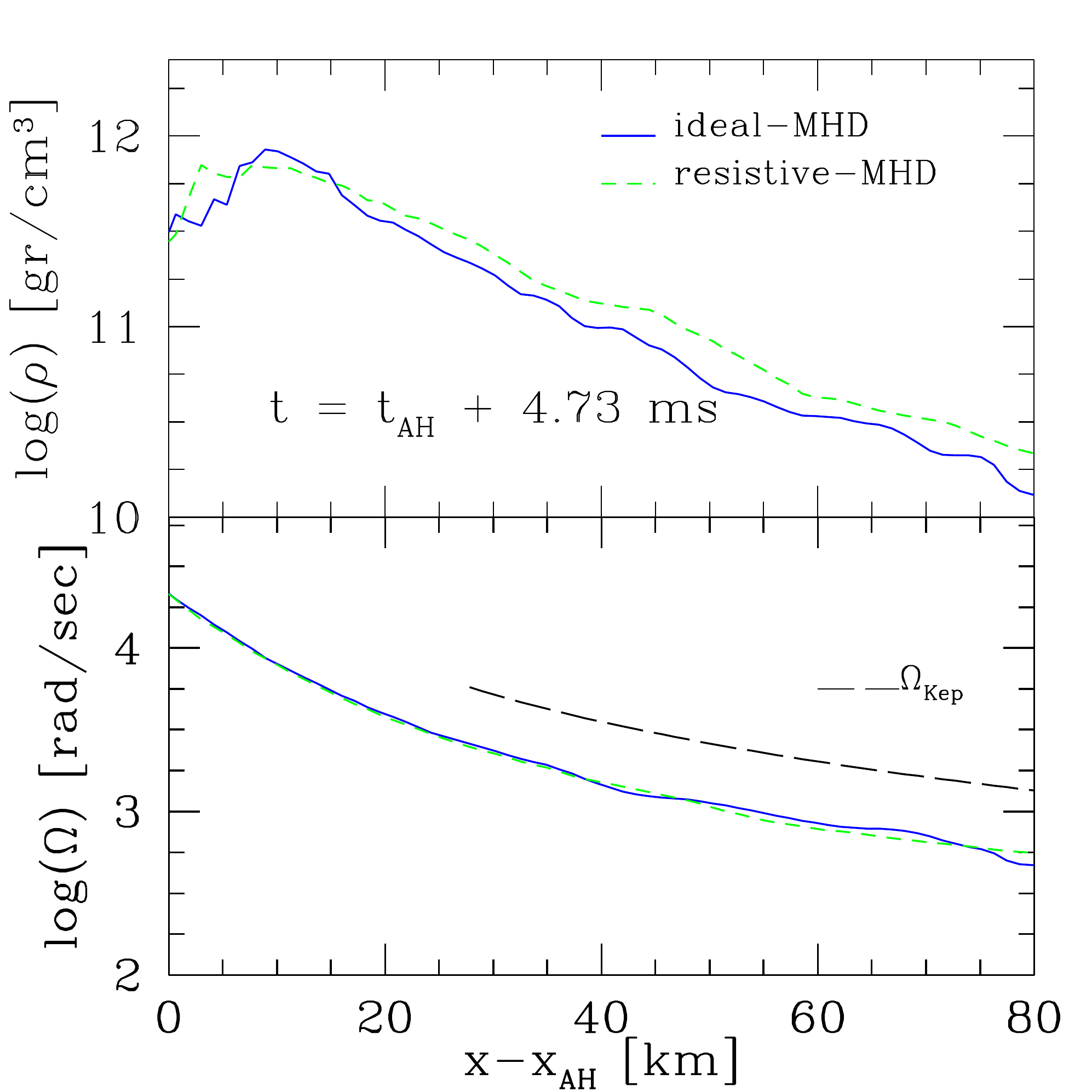}
\caption{\emph{Upper panel}: rest-mass density profile along the $x$-axis
  for the RMHD (green dashed line) and IMHD (blue solid line) simulations
  at time $t =4.74\,\ms$ after the apparent-horizon
  formation. \emph{Bottom panel}: the same as above but for the angular
  velocity; also shown as a reference is a Keplerian profile (black
  dashed line).}
  \label{fig:bns:omega}
\end{figure}

Because the magnetic field is not strong enough to alter the torus
dynamics significantly~\cite{Giacomazzo:2010}, it is natural to expect
the two solutions (IMHD and RMHD) to be very similar in terms of
dynamical properties once a nearly quasistationary state is established
after the collapse to a black hole. This expectation is confirmed in the
upper panel of Fig.~\ref{fig:bns:omega}, which shows the rest-mass
density distribution along the $x$-direction at about $4.74\,\ms$ after
the formation of the apparent horizon. Clearly, the two distributions are
very similar but not identical. The RMHD simulation, in particular,
yields larger rest-mass densities in the outer portions of the torus,
which, in turn, are responsible for larger rest masses (\cf Table
\ref{tab:bns:bhtorus}) and stronger magnetic fields (see the discussion in
Sec. \ref{sec:spd}). Furthermore, the RMHD simulation also yields a
larger torus as measured from the average position on the $x$-axis where
the rest-mass density falls below $\rho=10^{10}{\rm g\,cm}^{-3}$.

\begin{table}
\centering
\vspace{0.2in}
\begin{tabular}{lcccc}
\hline
& $M~(M_{\odot})$ & $J/M^2$ & $M_{\rm tor.}~(M_{\odot})$ &  $r_{\rm tor.}~(\km)$\\
\hline
RMHD & 2.88 & 0.873 & 0.095 & 105.9 \\
IMHD & 2.91 & 0.884 & 0.075 &  88.9 \\
\hline
\end{tabular}
\caption{Properties of the black hole--torus system at $t=4.74\,\ms$
  after the appearance of the apparent horizon for both the resistive and
  IMHD simulations. Shown are the mass and dimensionless spin of the
  black hole, as well as the rest mass and size of the torus as estimated
  with a cutoff on the rest-mass density at $\rho=10^{10}{\rm
    g\,cm}^{-3}$.}
\label{tab:bns:bhtorus}
\end{table}

Additionally, the bottom panel of Fig.~\ref{fig:bns:omega} illustrates
the angular velocity profile along the $x$ direction $\Omega
:=u^\phi/u^t$ for the RMHD implementations (green dashed line) and for
the IMHD one (blue solid line), at the same time as the upper
panel. Also shown as a black dashed line is a reference Keplerian
profile, \ie scaling like $x^{-3/2}$ and which is well matched by both
distributions. As remarked in Ref. \cite{Rezzolla:2010}, the fact that
the outer parts of the torus have a quasi-Keplerian behavior has two
important implications. First, it suggests that the tori will be
stable and not subject to dynamical instabilities that would lead to
their rapid destruction (see, \eg Refs. \cite{Zanotti05, Kiuchi2011b,
  Korobkin2013}). Second, a quasi-Keplerian profile also provides
optimal conditions for the development of an MRI in the torus, thus
opening the possibility of further amplification of the magnetic fields that
are present after the collapse of the HMNS.

\begin{figure*}
\centering
\includegraphics[width=0.9\columnwidth]{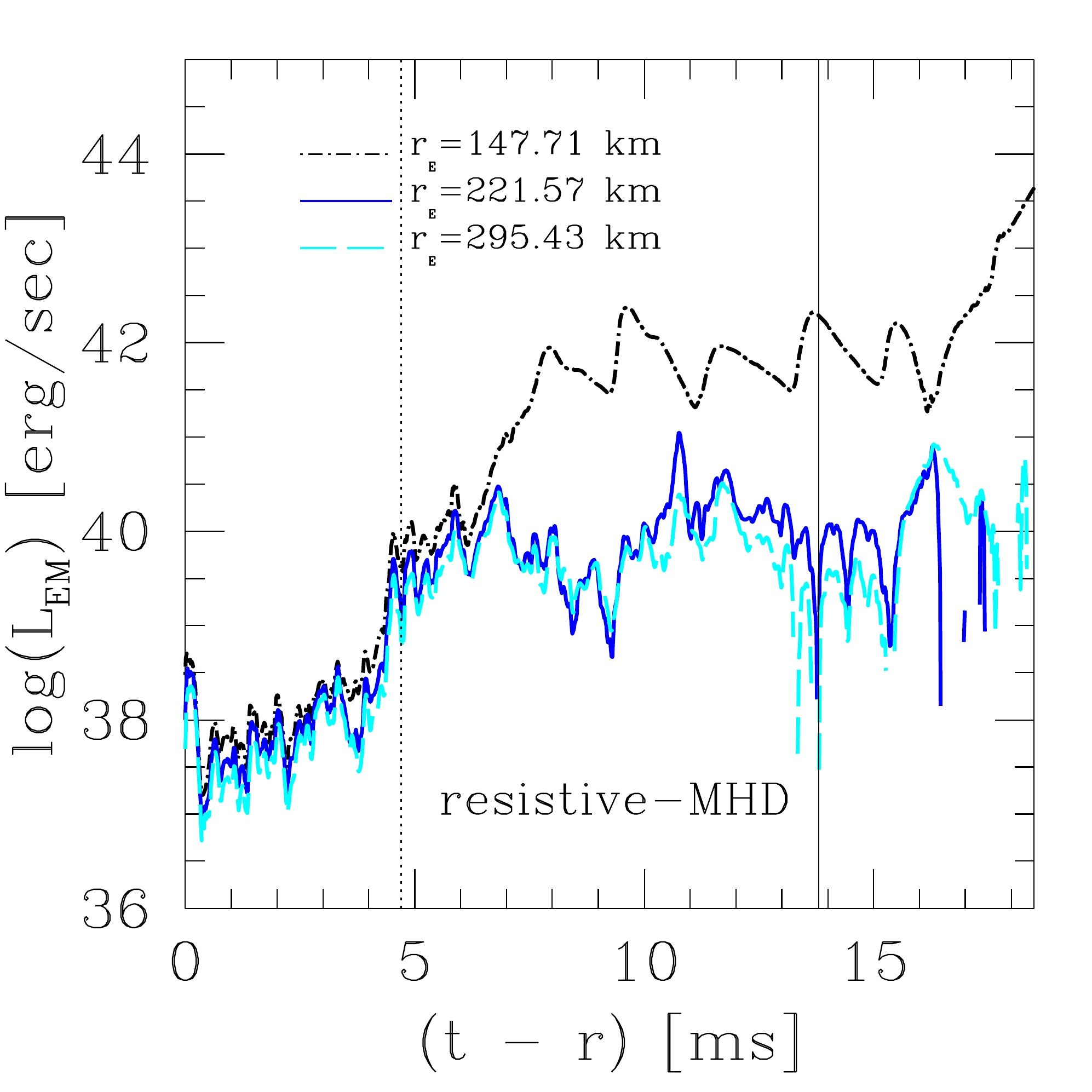}
\hskip 1.5cm
\includegraphics[width=0.9\columnwidth]{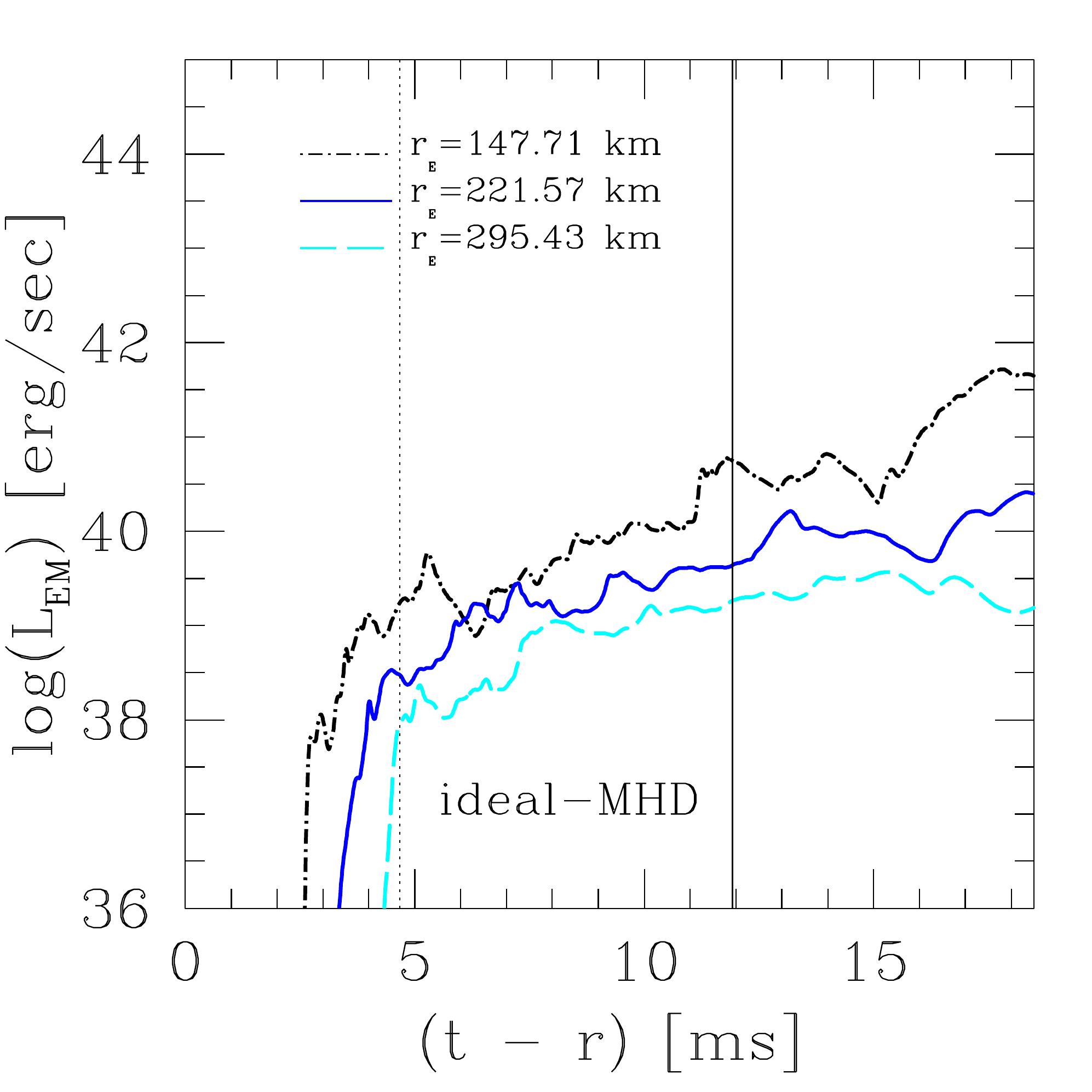}
\caption{Time series of the Poynting flux computed at different extraction
  radii $r_{_E}=\{147.71,\,221.57,\,295.43\}\,\km$ are shown here for the
  RMHD implementation (left panel) and the IMHD one (right panel). The
  dotted line represents the time of the merger, and the black solid line
  represents the time of horizon formation.}
  \label{fig:bns:poynting}
\end{figure*}

\subsubsection{Electromagnetic luminosities}

Despite the exploratory nature of the simulations carried out here, we
have computed for both the RMHD and IMHD implementations the total
electromagnetic luminosity $L_{_{\rm EM}}$ emitted. This has been
estimated as a surface integral of the Poynting flux over spherical
surfaces placed at representative coordinate radii $r_{_E}$, where
$r_{_E}$ has been varied to guarantee that the measurement is an
asymptotic one and is not affected by the local plasma dynamics. We
recall, in fact, that, because in the IMHD approximation the magnetic
fields are locked with the plasma, the electromagnetic luminosity
estimates can be heavily influenced by the presence of matter in the
outer regions and thus not correspond to a genuine amount of
electromagnetic energy flux leaving the system. Unfortunately, there is
no simple way within the IMHD approximation of determining whether the
integral of the Poynting flux computed on the numerical grid is genuinely
asymptotic. However, it is certainly reassuring if the values of
$L_{_{\rm EM}}$ are independent from the extraction radius.

The evolution of the electromagnetic luminosity is illustrated in
Fig.~\ref{fig:bns:poynting}, where we report it as computed at three
different extraction radii, $r_{_E}=\{147.71,\, 221.57,\, 295.43
\}\,\km$. Note that the luminosity during the inspiral phase of the RMHD
simulation is much larger than the corresponding one in the IMHD
simulation because of the diffusion of the magnetic field across the
stars' surfaces, that fills the entire domain with vacuum electromagnetic
fields. As remarked already in Sec. \ref{sec:rmhdatmo} and \ref{sec:spd},
these magnetic fields are in areas which are treated as atmosphere from a
hydrodynamical point of view, but where the Maxwell equations are solved
with zero conductivity, so that the electromagnetic fields can propagate
freely. 

We note that a nominal value of $\sigma=0$ does not imply that the
electromagnetic luminosity will be zero, since the motion of the compact
stars will introduce perturbations in the external electric and magnetic
fields, and thus a net Poynting flux (see Ref.~\cite{Moesta:2009} for the
electromagnetic emission of inspiralling binary black holes in
electrovacuum). Indeed, we have found that the electrovacuum luminosity
before the merger is $L_{_{\rm EM}} \sim 10^{38}\,\erg\,\sec^{-1}$, which
is smaller than the one reported in ~Ref.\cite{Ponce2014} (\ie $L_{_{\rm EM}}
\sim 10^{41}\,{\rm erg\,\sec}^{-1}$), where the stellar exteriors were
modelled in the force-free approximation. Although it has already been
found that the electrovacuum luminosity is slightly smaller than the
force-free one for the same system (see Refs.~\cite{Palenzuela:2010b,
  Moesta2011, Alic:2012}), the differences found here are larger than
expected and this may be due to the rather different way in which the
exterior regions of stars are treated. By contrast, the electromagnetic
luminosity before the merger in the IMHD simulation, where the magnetic
fields are always contained inside the stars, is essentially zero.

After the merger, the electromagnetic luminosity grows rapidly of about
2 orders of magnitude, essentially as a result of the growth of the
magnetic field already discussed in the left panel of
Fig. \ref{fig:bns:mag} (we recall that the electromagnetic luminosity
should scale quadratically with the magnetic field). During the
postmerger phases, the luminosity ranges from $\sim 10^{39}$ to
$10^{41}~\erg\,\sec^{-1}$, to reach values up to
$10^{42}~\erg\,\sec^{-1}$ after the collapse of the HMNS to a black hole.

In the left panel of Fig.~\ref{fig:bns:poynting}, the luminosity computed
on a surface of radius $r_{_E}=147.71\,\km$ (black dot-dashed line) does
not overlap with those computed on larger radii (dark-blue solid line and
light-blue dashed solid lines), signalling that this radius is too close
to the central object and contaminated by the presence of
matter. Fortunately, however, the luminosities at $r_{_E}=221.57\,\km$
and $r_{_E}=295.43\,\km$ are very close to each other, confirming the
robustness of these measurements. By contrast, the three luminosities in
the IMHD simulation reported in the right panel of
Fig.~\ref{fig:bns:poynting} provide three different values for the
luminosity, indicating that at least two of them (\ie those at the
smaller extraction radii) are probably contaminated by the presence of
bound matter and hence not reasonable.

\section{Conclusions}
\label{sec:binaries:conclusions}

We have presented general-relativistic simulations of the inspiral and
merger of binary neutron stars when evolved solving the coupled set of
the Einstein equations and those of RMHD. Our main interest here has been
to assess the impact that resistive effects have on the dynamics of these
binaries, and which are usually investigated in the more idealized
framework of IMHD.

Because the differences with an IMHD description could be rather small in
certain stages of the process (\eg during the inspiral), we have carried
out a close comparison between two simulations evolving the same binary,
either in the context of RMHD or in that of IMHD. More specifically, we
have studied the dynamics of an equal-mass binary system of neutron stars
with a total ADM mass $M_{_{\rm ADM}} = 3.25\,M_{\odot}$ and an initial
orbital separation of $45\,\km$. The stars are initially irrotational and
with zero magnetic field. A dipolar magnetic field is therefore added
before the evolution is started, which is entirely contained inside
the stars, at least initially. Furthermore, to reduce computational costs
and ``accelerate'' the inspiral, we have slightly reduced the linear
momenta of the initial data as done in Ref.~\cite{Kastaun2013}, so that the
merger occurs after approximately one orbit.

A crucial goal of our RMHD approach has been that of attaining a smooth
resistive description from the highly conducting, high-density stellar
interior, out to regions of very low-density plasma, where the
electromagnetic fields decouple from the fluid. Falling between these two
regimes is the large amount of high-density, small-velocity material that
is ejected during and after the merger by the HMNS and, once the latter
collapses to a black hole, by the accreting torus. This material,
occupies a large portion of the computational domain and is produced
either by the spiral arms launched at the merger, or by the magnetic
winding and consequent pressure imbalance in the outer layers of the
HMNS. Neutrino losses can also be a source of a wind, but we do not model
this here.

While there are several ways of potentially reaching a smooth transition
between the IMHD limit in the stellar interior and an electrovacuum
behavior, we have here adopted the same approach we have extensively
investigated with isolated neutron stars in
Ref.~\cite{Dionysopoulou:2012pp}.  In essence, this matching is achieved
through a carefully chosen conductivity profile, where the conductivity
is directly related to the conserved rest-mass density and is set to zero
once the latter reaches a value close to the atmospheric floor. This
prescription has at least two free parameters. First, they ensure that
the transition region covers only a thin layer close to the surface of
the star. Second, they guarantee that this layer remains ``thin'' even
in the first steps of the evolution, when the outer layers of the star
expand due to a nonzero pressure in the atmosphere. While we have set
these two parameters to sensible values, their influence on the results
still needs to be fully explored.

Overall, we have found that there are many similarities between the RMHD
and IMHD evolutions, but also one important difference, namely, that the
survival time of the hypermassive neutron star, which increases in a RMHD
simulation. The increased lifetime of the HMNS appears to be due to a
less efficient magnetic-braking mechanism in the resistive regime, in
which matter can move across magnetic-field lines, so that the outward
transport of angular momentum is reduced. This interpretation is
supported by the analysis of the evolution of the specific angular
momentum, and it shows that the transport is more efficient in the IMHD
simulation. An extended lifetime of the HMNS could have intriguing
astrophysical consequences, since a longer-lived magnetized hypermassive
neutron star brings support to the recent modelling of SGRBs in terms of
long-lived magnetarlike objects produced by the merger~\cite{Zhang2001,
  Metzger2008, Bucciantini2012, Rezzolla2014b}.

Another important result of these simulations is the confirmation that a
magnetic-jet structure is formed in the low-density funnel produced by
the black hole--torus system. We note that these simulations have been
carried out at higher resolutions and with a different grid structure
than those in Ref. \cite{Rezzolla:2011}. In the RMHD simulations the
magnetic-jet structure is far more regular, essentially axisymmetric, and
extending out to the largest scale in our system. This is most likely the
result of the effective decoupling established between the dynamics of
the plasma and that of the electromagnetic fields. In the IMHD
simulations, a magnetic-jet structure is still present, but on the scale
of the torus. This difference is due to the fact that a decoupling of the
electromagnetic fields from the plasma is not possible in the IMHD
approximation, and the magnetic field follows tightly the turbulent
dynamics of the matter. In this case, the magnetic-field lines are almost
parallel to the $z$-axis (in analogy with what was shown in
Ref. \cite{Rezzolla:2011}) and the topology becomes more turbulent on
large scales. In both regimes, the magnetic field is predominantly
toroidal in the highly conducting torus and predominantly poloidal in the
nearly evacuated funnel, although in the IMHD simulation, this happens
near the rotation axis. The matter in the funnel does not have an
internal energy sufficiently large to launch a relativistic
outflow. However, it is reasonable to expect that reconnection processes
or neutrino annihilation occurring in the funnel, none of which we model
here, could potentially increase the internal energy in the funnel.

The final comment of this work is in fact a caveat. While the dynamics of
the magnetic-field results presented here appears reasonable, matching
the expectations for this type of system as well as previous simulations,
we should remark that our results are ultimately dependent on the choice
made for Ohm's law and for the conductivity profile. Again, while our
choice is a very conservative and a plausible one, it represents a choice
nevertheless. The large computational costs associated with these
simulations have prevented us from presenting a systematic investigation
of how sensitive the results are on the choices for Ohm's law, for the
conductivity profile, or on the treatment of the atmosphere. All of these
issues deserve further investigation and will be the focus of our future
work.

\begin{acknowledgments}
  We thank W. Kastaun for the visualization library used for the
  two-dimensional plots and are grateful to B. Mundim and K. Takami
  for their valuable comments on the manuscript. We appreciate useful
  discussions with N. Andersson, I. Hawke and C. Palenzuela regarding
  ideas and further applications of resistive-MHD in astrophysical
  scenarios. Finally, we thank B. Giacomazzo for help with the initial
  data and the ideal-MHD code. Partial support comes from the Science
  and Technology Facilities Council STFC, Grant No. ST/H002359/1, the DFG Grant
  No. SFB/Transregio~7, the ``NewCompStar'' COST Action MP1304, the
  LOEWE-Program in HIC for FAIR. The simulations were performed on
  SuperMUC at LRZ-Munich, on Datura at AEI-Potsdam, on the DiRAC
  BlueGene/Q cluster at University of Edinburgh-ACSP33, and on the
  LOEWE at CSC-Frankfurt.
\end{acknowledgments}

\appendix 

\section{Magnetic-jet structure in the IMHD simulations}
\label{sec:appendix:imhd}

\begin{figure*}
\centering
\includegraphics[width=0.48\textwidth]{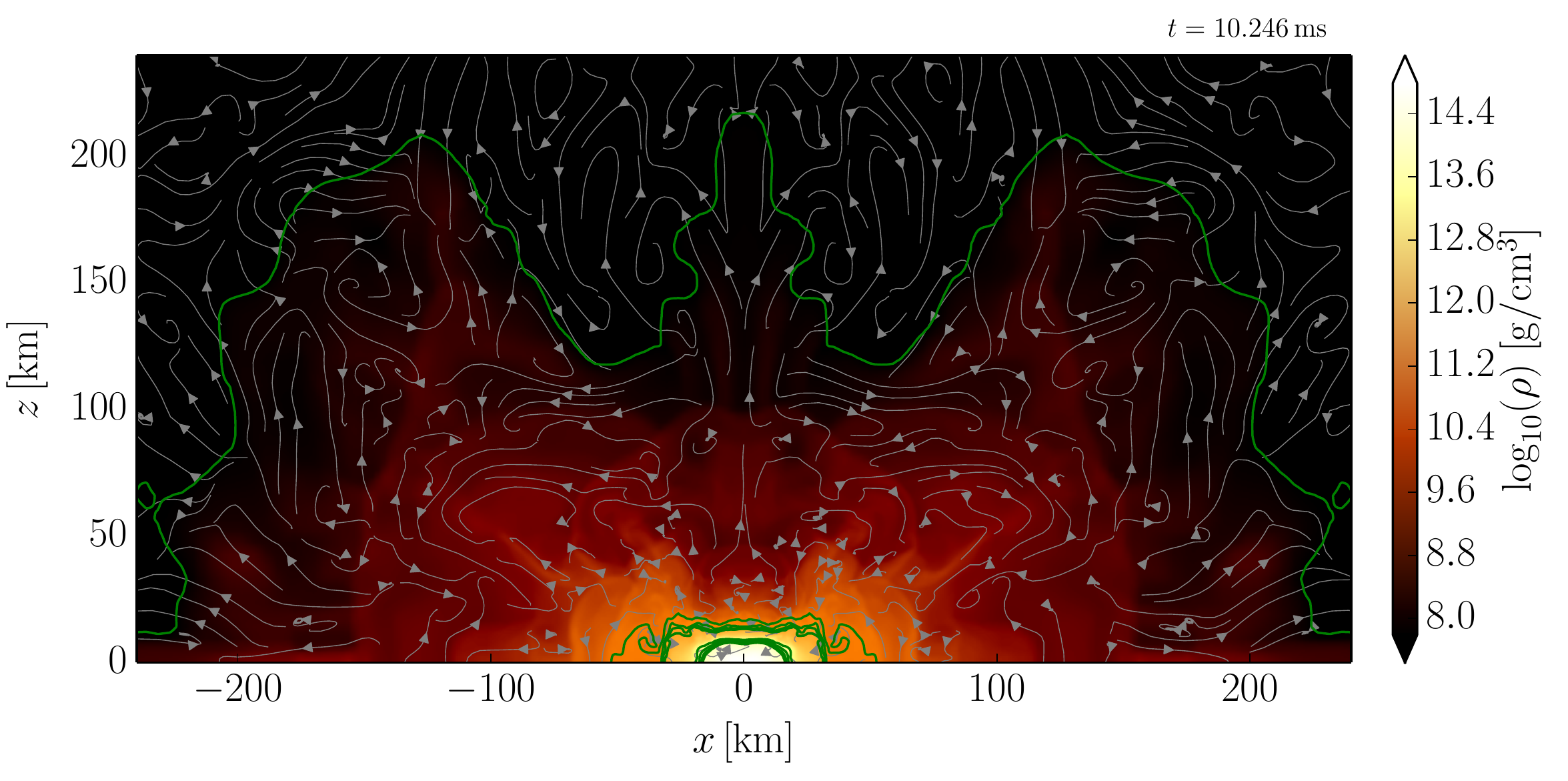}
\hskip 0.5cm
\includegraphics[width=0.48\textwidth]{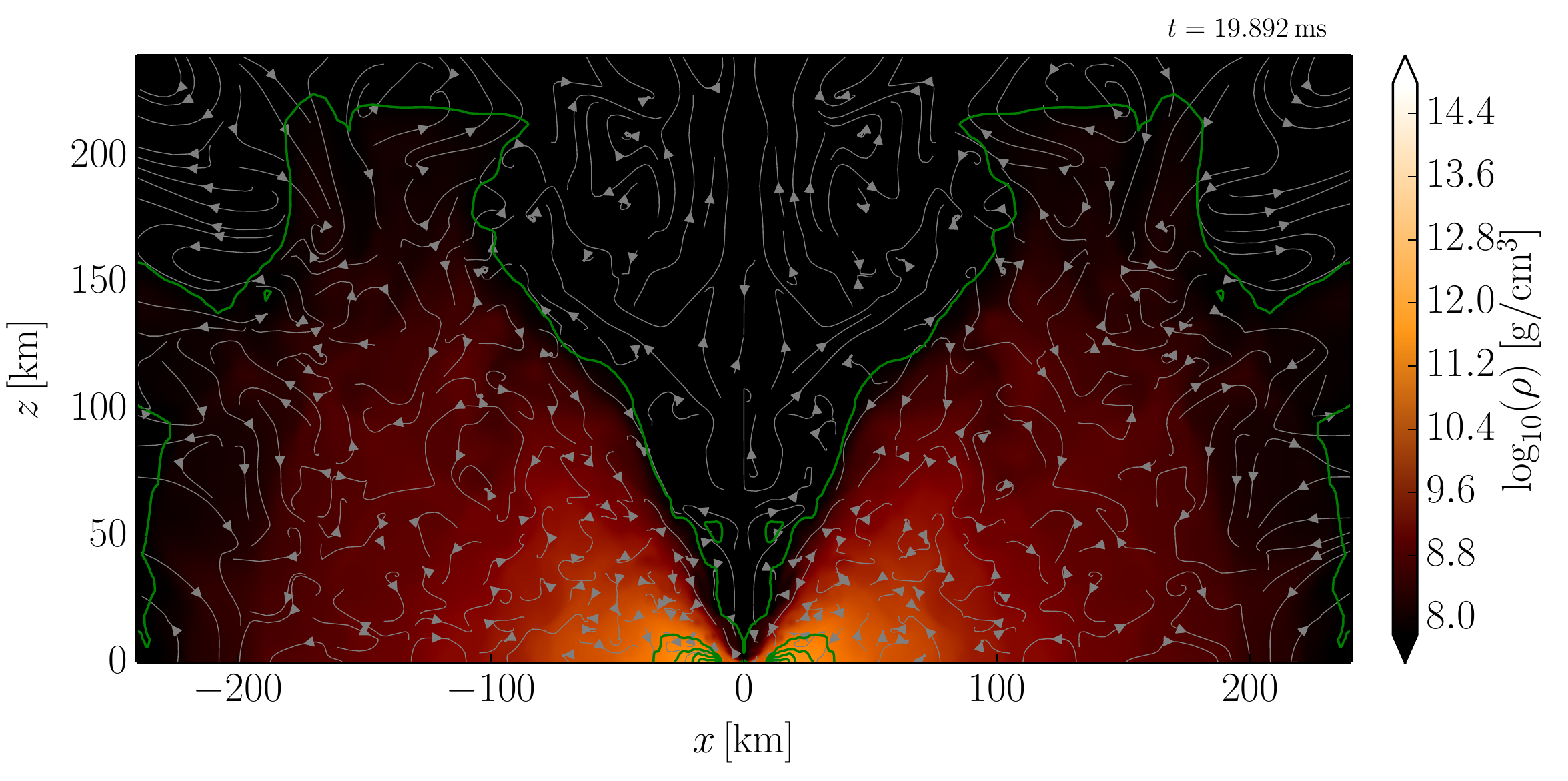}
\hskip 0.5cm
\includegraphics[width=0.48\textwidth]{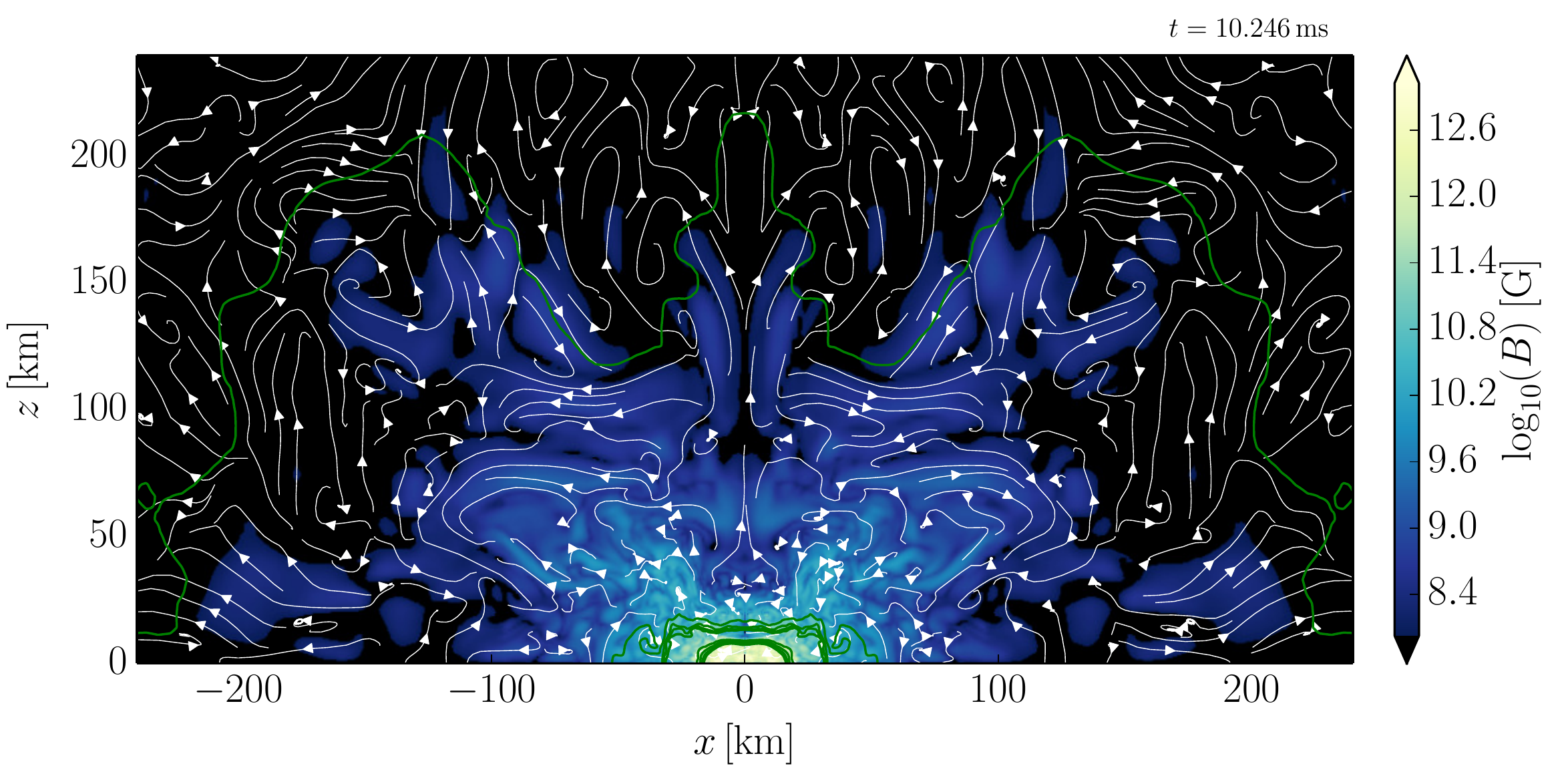}
\hskip 0.5cm
\includegraphics[width=0.48\textwidth]{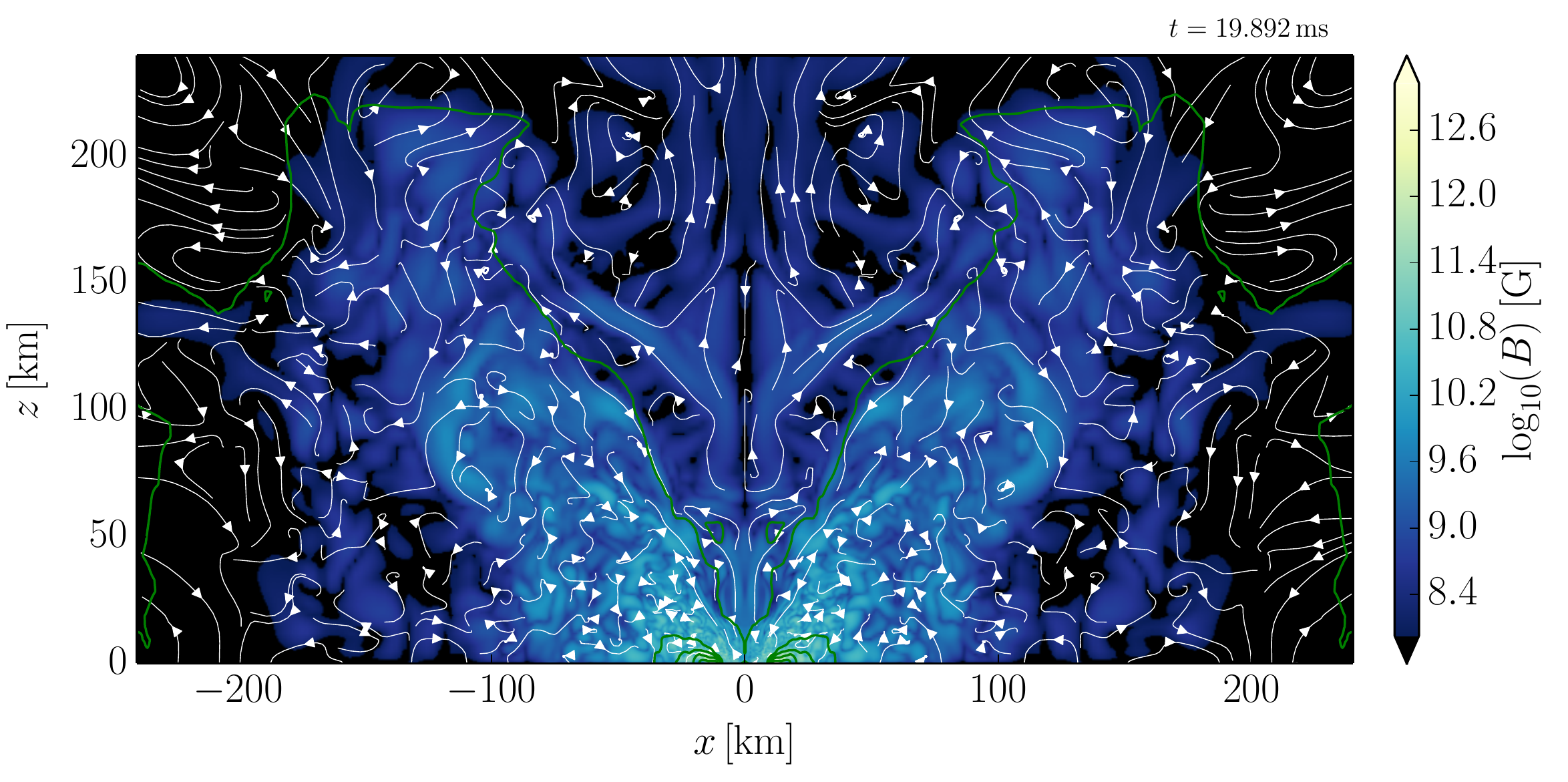}
\caption{The same as Fig. \ref{fig:rho_and_B_xz_large}, but for the IMHD
  simulation. The two columns refer to $t=10.25\,\ms$ (left column), when
  the HMNS has not yet collapsed, and to $t=18.92\,\ms$ {(right column)},
  when a black hole has already been formed. Note again the formation of
  a magnetic-jet structure around the black-hole rotation axis, which
  however is far less regular on large scales than the one produced in
  the RMHD simulation. See also Fig. \ref{fig:rho_and_B_xz.imhd} for a
  view on smaller scales.}
  \label{fig:rho_and_B_xz_large.imhd}
\end{figure*}

Although the IMHD simulations are not the focus of this paper, for
completeness we provide in this Appendix a rapid overview of the
properties of the magnetic-jet structure as obtained within this
approximation. The essence of the results is shown in
Fig. \ref{fig:rho_and_B_xz_large.imhd}, which represents the equivalent
of Fig. \ref{fig:rho_and_B_xz_large} but within IMHD. The different
panels show large-scale, two-dimensional snapshots on the $(x,z)$ planes
of the rest-mass density (top row), and of the magnetic field (bottom
row). The left column of Fig. \ref{fig:rho_and_B_xz_large.imhd} refers to
$t=10.25\,\ms$ (left column), when the HMNS has not yet collapsed to a
black hole, while the right column refers to $t=18.89\,\ms$, when a black
hole has already been formed. Because in the IMHD approximation the
magnetic fields are tightly locked with the matter, it does not come as a
surprise that no ordered magnetic field structure seems to develop before
the HMNS collapses and forms a black hole. This is because the dynamics
of the plasma is quite turbulent at the merger and during the HMNS
stage. However, after a black hole is formed, a well-ordered magnetic-field
structure appears as the system reaches a quasistationary state. We
note again that the formation of a magnetic-jet structure occurs around
the black-hole rotation axis.

\begin{figure*}
\centering
\includegraphics[width=0.48\textwidth]{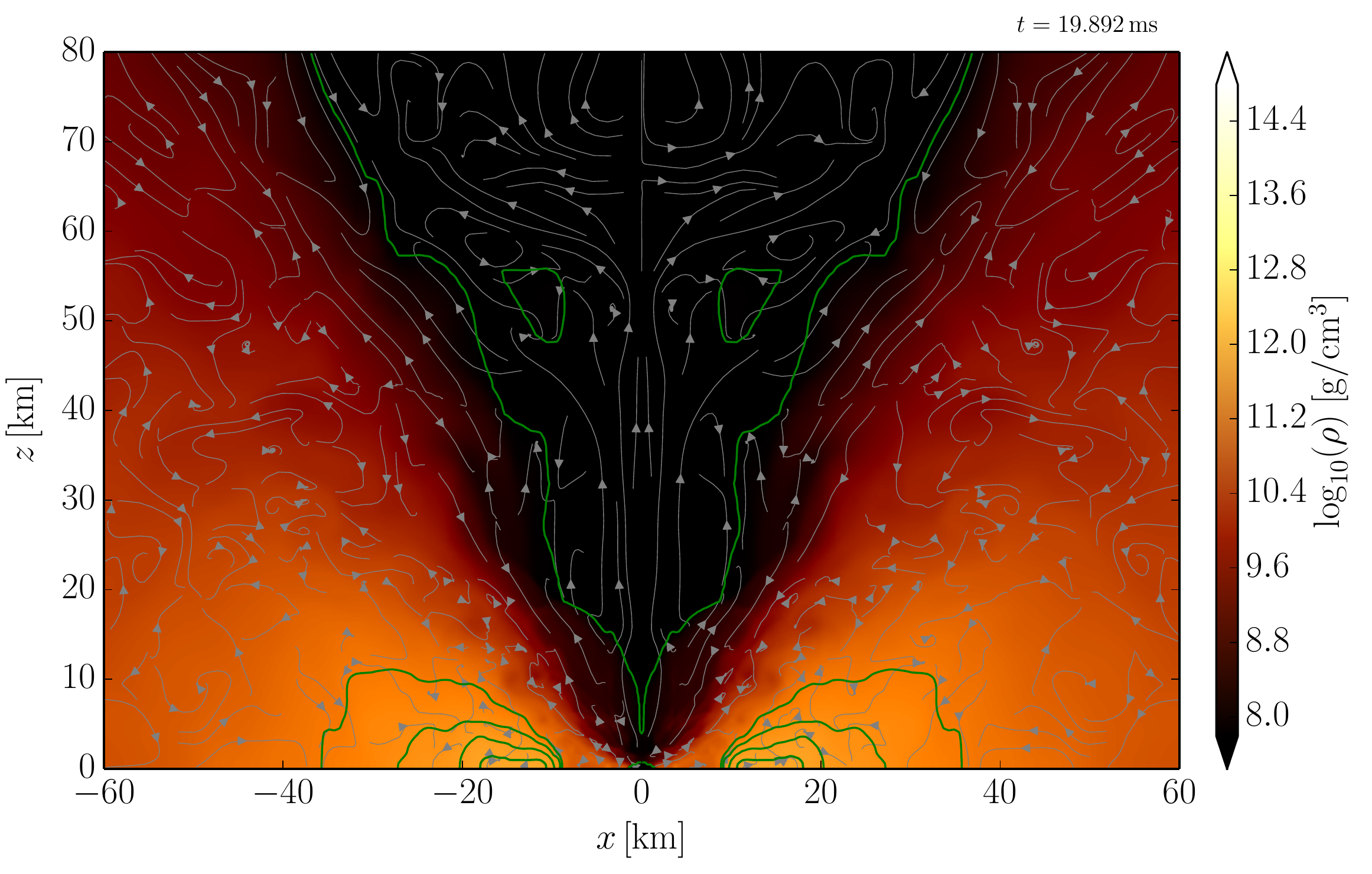}
\hskip 0.5cm
\includegraphics[width=0.48\textwidth]{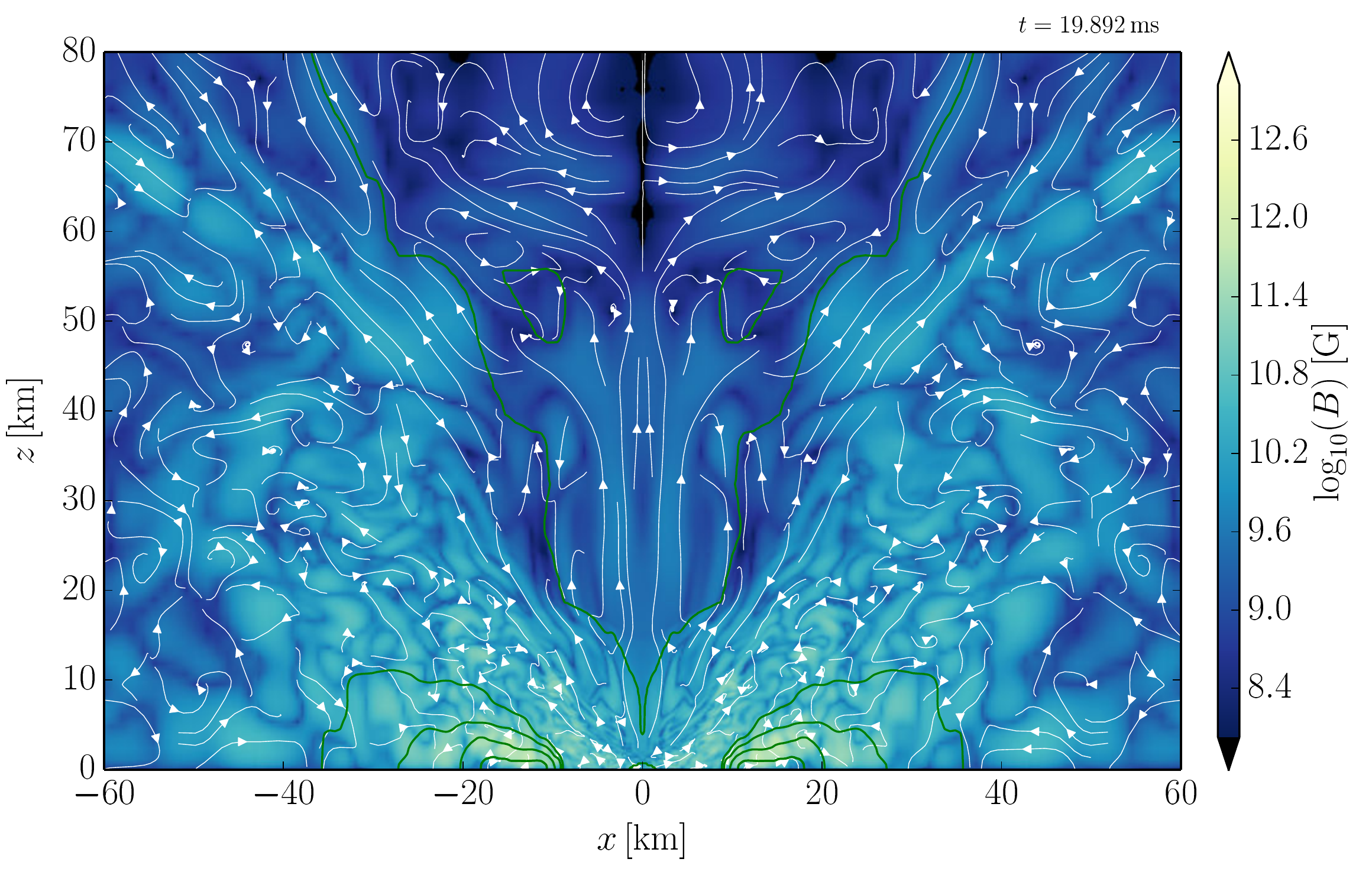}
\caption{The same as Fig. \ref{fig:rho_and_B_xz_large.imhd}, but on a
  scale of $[-60,60]\,\km$ on the $x$-axis and of $[0,80]\,\km$ on the
  $z$-axis. Note that the magnetic-jet structure becomes more evident on
  these scales. See also Fig. \ref{fig:bns:rhoB6} for the corresponding
  quantities in the RMHD simulation.}
  \label{fig:rho_and_B_xz.imhd}
\end{figure*}

Differently from the corresponding RMHD simulation, the magnetic-jet
structure here is not very regular on large scales, and it is necessary to
go down to the length scale of the torus, as shown in
Fig. \ref{fig:rho_and_B_xz.imhd}, for the magnetic-jet structure to
become evident. Note that the magnetic-field lines are almost parallel to
the $z$-axis, in analogy with what was shown in Ref. \cite{Rezzolla:2011}.
Finally, although we are here using only the projection of the magnetic-field lines on the $(x,z)$ plane, the magnified view in
Fig. \ref{fig:rho_and_B_xz.imhd} reveals that the magnetic field in the
low-density funnel is still predominantly poloidal, although not as
ordered as in the RMHD simulation (\cf Fig. \ref{fig:bns:rhoB6}, left and
right panels). At the same time, and not shown here for compactness, the
magnetic field is essentially toroidal in the torus.

\bibliographystyle{apsrev4-1-noeprint}

\end{document}